\documentclass[journal]{IEEEtran}
\usepackage{graphicx}
\usepackage{cite}
\usepackage{picinpar}
\usepackage{amssymb}
\usepackage{amsmath}
\usepackage{amsfonts}
\usepackage{amsthm}
\usepackage{algorithm} 
\usepackage{algorithmic} 
\usepackage{stfloats}
\usepackage{url}
\usepackage{flushend}
\usepackage{colortbl}
\usepackage{soul}
\usepackage{multirow}
\usepackage{pifont}
\usepackage{color}
\usepackage{alltt}
\usepackage{verbatim}
\usepackage{hyperref}
\hypersetup{
	colorlinks = true,
	linkcolor = blue,
	citecolor = blue,
}
\usepackage{enumerate}
\usepackage{siunitx}
\usepackage{breakurl}
\usepackage{epstopdf}
\usepackage{pbox}
\usepackage{color}
\usepackage[table]{xcolor}
\usepackage{subfigure}
\usepackage{verbatim}
\usepackage{pdflscape}
\usepackage{pdfpages}
\newcommand\figref[1]{Fig.~\ref{#1}}
\begin{document}
\title{Applications of Unsupervised Deep Transfer Learning to Intelligent Fault Diagnosis: A Survey and Comparative Study}

\author{
	\vskip 1em	
	Zhibin Zhao,
	Qiyang Zhang,
	Xiaolei Yu,
	Chuang Sun,
	Shibin Wang,
	Ruqiang Yan,~\IEEEmembership{Senior Member,~IEEE}
	and Xuefeng Chen,~\IEEEmembership{Senior Member,~IEEE}

	\thanks{
		Z. Zhao, Q. Zhang, X. Yu, C. Sun, S. Wang, R. Yan and X. Chen are with the State Key Laboratory for Manufacturing Systems
		Engineering, Xi’an Jiaotong University, Xi’an 710049, China.
		E-mail: (zhaozhibin@xjtu.edu.cn; zhangqiyang@stu.xjtu.edu.cn; yxl007@stu.xjtu.edu.cn; ch.sun@xjtu.edu.cn; wangshibin2008@gmail.com; yanruqiang@xjtu.edu.cn;\quad chenxf@mail.xjtu.edu.cn) 
		
		This work was supported by the Natural Science Foundation of China (No. 52105116) and by the China Postdoctoral Science Foundation (No. 2021M692557 and No. 2021TQ0263).
		
		R. Yan is the corresponding author.	
	}
}

\maketitle

\begin{abstract} 
Recent progress on intelligent fault diagnosis (IFD) has greatly depended on deep representation learning and plenty of labeled data.
However, machines often operate with various working conditions or the target task has different distributions with the collected data used for training (the domain shift problem).
Besides, the newly collected test data in the target domain are usually unlabeled, leading to unsupervised deep transfer learning based (UDTL-based) IFD problem.
Although it has achieved huge development, a standard and open source code framework as well as a comparative study for UDTL-based IFD are not yet established.
In this paper, we construct a new taxonomy and perform a comprehensive review of UDTL-based IFD according to different tasks.
Comparative analysis of some typical methods and datasets reveals some open and essential issues in UDTL-based IFD which are rarely studied, including transferability of features, influence of backbones, negative transfer, physical priors, etc.
To emphasize the importance and reproducibility of UDTL-based IFD, the whole test framework will be released to the research community to facilitate future research.
In summary, the released framework and comparative study can serve as an extended interface and basic results to carry out new studies on UDTL-based IFD.
The code framework is available at \url{https://github.com/ZhaoZhibin/UDTL}.
\end{abstract}

\begin{IEEEkeywords}
Intelligent fault diagnosis; Unsupervised deep transfer learning; Taxonomy and survey; Comparative study; Reproducibility
\end{IEEEkeywords}

\markboth{IEEE Trans.\ on Instrum.\ Meas.}%
{}

\definecolor{limegreen}{rgb}{0.2, 0.8, 0.2}
\definecolor{forestgreen}{rgb}{0.13, 0.55, 0.13}
\definecolor{greenhtml}{rgb}{0.0, 0.5, 0.0}

\section{Introduction}
\label{S:1}
\IEEEPARstart{W}{ith} the rapid development of industrial big data and Internet of Things, Prognostic and Health Management (PHM) for industrial equipments, such as aero-engine, helicopter and high-speed train, is becoming increasingly popular, bringing out many intelligent maintenance systems.
Intelligent fault diagnosis (IFD) is becoming an essential branch among PHM systems.
IFD based on traditional machine learning methods \cite{kankar2011fault}, including random forest \cite{cerrada2016fault} and support vector machine \cite{widodo2007support}, has been widely applied in research and industry scenarios.
However, these methods often need to extract features manually or to combine with other advanced signal processing techniques, such as time frequency analysis \cite{tong2020ridge} and sparse representation \cite{zhao2018enhanced,zhao2021fast}.
While, with the increment of available data, data-driven methods with the representation learning ability are also becoming more and more important.
Thus, Deep Learning (DL) \cite{lecun2015deep}, which can extract useful features automatically from original signals, gradually becomes a hot research topic for many fields \cite{ravi2016deep,min2017deep,yang2019robust,yang2020particle} as well as PHM \cite{zhu2020stacked, zhang2020knowledge,xu2018roller}.
Effective DL models, such as Convolutional Neural Network (CNN) \cite{krizhevsky2012imagenet}, Sparse Autoencoder (SAE) \cite{ng2011sparse}, etc., for tasks in PHM have been validated successfully in current research, and a benchmark study is also given in \cite{zhao2020deep} for better comparison and development.

Behind the effectiveness of DL-based IFD, there exist two necessary assumptions: 1) samples from the training dataset (source domain) should have the same distribution with that from the test dataset (target domain); 2) plenty of labeled data are available during the training phase.
Although the labeled data might be generated by dynamic simulations or fault seeding experiments, the generated data are not strictly consistent with the test data in the real scenario.
That is, DL models based on the training dataset only possess a weak generalization ability, when deployed to the test dataset from real applications.
In addition, rotating machinery often operates with varying working conditions, such as loads and speeds, which also requires that trained models using the dataset from one working condition can successfully transfer to the test dataset from another working condition.
In short, these factors make models trained in the source domain hard to be generalized or transferred to the target domain, directly.

Shared features existing in these two domains due to the intrinsic similarity in different application scenarios or different working conditions allow this domain shift manageable.
Hence, to let DL models trained in the source domain be able to be transferred well to the target domain, a new paradigm, called deep transfer learning (DTL) should be introduced into IFD.
One of the effective and direct DTL is to fine-tune DL models with a few labeled data in the target domain, and then the fine-tuned model can be used to diagnose the test samples.
However, the newly collected data or the data under different working conditions are usually unlabeled and it is sometimes very difficult, or even impossible to label these data.
Therefore, in this paper, we investigate the unsupervised version of DTL,  called unsupervised deep transfer learning-based (UDTL-based) IFD, which is to make predictions for unlabeled data on a target domain given labeled data on a source domain.
It is worth mentioning that UDTL is sometimes called unsupervised domain adaptation, and in this paper, we do not make a strict distinction between two concepts.

UDTL is widely used and has achieved tremendous success in the field of computer vision and natural language processing, due to the application value, open source codes, and the baseline accuracy.
However, there are few open source codes or the baseline accuracy in the field of UDTL-based IFD, plenty of research has been published for UDTL-based IFD via simply using models that already have been published in other fields.
Due to the lack of open source codes, results in these papers are very hard to repeat for further comparisons.
This is not beneficial to identify the state-of-the-art methods, and furthermore, it is unfavorable to the advancement of this field on a long view.
Hence, it is very important to perform a comparative study, provide a baseline accuracy, and release open source codes of UDTL-based algorithms.
For testing UDTL-based algorithms, the unified test framework, parameter settings, and datasets are three important aspects to affect fairness and effectiveness of comparisons.
While, due to the inconsistency of these factors, there are a lot of unfair and unsuitable comparisons.
It seems that scholars are continuing to combine new techniques, and the proposed algorithms always have better performance than other former algorithms, which comes to the question: Is the improvement beneficial to IFD or just depends on the excessive parameter adjustment?
However, the open and essential issues in UDTL-based IFD are rarely studied, such as transferability of the features, influence of backbones, etc.

There are already some good review papers about transfer learning in IFD.
Zheng et al. \cite{zheng2019cross} summarized the cross-domain fault diagnosis using the knowledge transfer strategy based on transfer learning and presented some open source datasets, which could be used to verify the performance of diagnosis methods.
Yan et al. \cite{yan2019knowledge} reviewed recent development of knowledge transfer for rotary machine fault diagnosis via using different transfer learning methods and provided four case studies to compare the performance of different methods.
Lei et al. \cite{lei2020applications} reviewed IFD based on machine learning methods with the emphasis on transfer learning theories, which adopt diagnosis knowledge from one or multiple datasets to other related ones, and also pointed out that transfer learning theories might be the essential way to narrow the gap between experimental verification and real applications.
However, all above review papers did not focus on UDTL-based IFD and provide the open source test framework for fair and suitable comparisons.
They all payed more attention to label-consistent (also called closed set) UDTL-based IFD, which assumes that the source domain has the same label space with the target domain, but many recent research papers focused on label-inconsistent or multi-domain UDTL, which is closer to the engineering scenarios.
Thus, a comprehensive review is still required to cover the advanced development of UDTL-based IFD from the cradle to the bloom and to guide the future development. 

In this paper, to fill in this gap, commonly used UDTL-based settings and algorithms are discussed and a new taxonomy of UDTL-based IFD is constructed.
In each separate category, we also give a comprehensive review about recent development of UDTL-based IFD.
Some typical methods are integrated into a unified test framework, which is tested on five datasets.
This test framework with source codes will be released to the research community to facilitate the research on UDTL-based IFD.
With this comparative study and open source codes, the authors try to give a depth discussion (it is worth mentioning that results are just a lower bound of the accuracy) of current algorithms and attempt to find the core that determines the transfer performance.

The main contributions of this paper are summarized as follows:
\begin{enumerate}[1)]
	\item \textit{New taxonomy and review:} we establish a new taxonomy of UDTL-based IFD according to different tasks of UDTL.
	The hierarchical order follows the number of source domains, the usage of target data in the training phase, the label consistence of source and target domains, inclusion relationship between label sets of source and target domains, and a transfer methodological level.
	We also provide the most comprehensive overview of UDTL-based IFD for each type of categories.
	\item \textit{Various datasets and data splitting:} We collect most of the publicly available datasets suitable for UDTL-based IFD and provide a detailed discussion about its adaptability.
	We also discuss the way of data splitting and explain that it is more appropriate to split data into training and test datasets regardless of whether they are in source or target domains.
	\item \textit{Comparative study and further discussion:} We evaluate various UDTL-based IFD methods and provide a systematic and comparative analysis from several perspectives to make the future studies more comparable and meaningful.
	We also discuss the transferability of features, influence of backbones, negative transfer, etc.
	\item \textit{Open source codes:} To emphasize the importance and reproducibility of UDTL-based IFD, we release the whole evaluation code framework that implements all UDTL-based methods discussed in this paper.
	Meanwhile, this is an extensible framework that retains an extended interface for everyone to combine different algorithms and load their own datasets to carry out new studies.
	The code framework is available at \url{https://github.com/ZhaoZhibin/UDTL}.
\end{enumerate}

The rest of this paper is organized as follows: Section \ref{S:2} provides background and definition of UDTL-based IFD.
Basic concepts, evaluation algorithms and comprehensive review of UDTL-based IFD are introduced in Section \ref{S:3} to \ref{S:5}.
After that, in Section \ref{S:6} to \ref{S:8}, datasets, evaluation results and further discussions are investigated, followed by the conclusion part in Section \ref{S:9}.

\section{Background and Definition}
\label{S:2}
\subsection{The Definition of UDTL}
To briefly describe the definition of UDTL, we introduce some basic symbols.
It is assumed that labels in the source domain are all available, and the source domain can be defined as follows:
\begin{align}
\label{eq1}
\mathcal{D}_s = \left\{ {\left( {x_i^s,y_i^s} \right)} \right\}_{i = 1}^{{n_s}} \quad x_i^s \in {X_s},\; y_i^s \in {Y_s},
\end{align}
where $ \mathcal{D}_s $ represents the source domain, $ x_i^s \in \mathbb{R}^d $ is the $ i $-th sample, $ X_s $ is the union of all samples, $ y_i^s $ is the $ i $-th label of the $ i $-th sample, $ Y_s $ is the union of all different labels, and $ n_s $ means the total number of source samples.
Besides, it is assumed that labels in the target domain are unavailable, and thus the target domain can be defined as follows:
\begin{align}
\label{eq2}
\mathcal{D}_t = \left\{ {\left( {x_i^t} \right)} \right\}_{i = 1}^{{n_t}} \quad x_i^t \in {X_t},
\end{align}
where $ \mathcal{D}_t $ represents the target domain, $ x_i^t  \in \mathbb{R}^d $ is the $ i $-th sample, $ X_t $ is the union of all samples, and $ n_t $ means the total number of target samples.

The source and target domains follow the probability distributions $ P $ and $ Q $, respectively.
We hope to build a model $ \beta(\cdot) $ which can classify unlabeled samples $x$ in the target domain:
\begin{align}
\label{eq3}
\hat y = \beta \left( x \right),
\end{align}
where $ \hat y $ is the prediction.
Thus, UDTL is aimed to minimize the target risk $ {\varepsilon}_t \left(  \beta  \right) $ using source data supervision \cite{long2015learning}:
\begin{align}
\label{eq4}
{\varepsilon}_t \left(  \beta  \right) =  \Pr _{\left( {x,y} \right)\sim Q} \left[ {\beta \left( x \right) \ne y} \right].
\end{align}

Also, the total loss of UDTL can be written as:
\begin{align}
\label{total-loss}
\mathcal{L} = \mathcal{L}_{c} + \lambda \mathcal{L}_{\text{UDTL}},
\end{align}
where $\mathcal{L}_{c} $ is the Softmax cross-entropy loss shown in \eqref{cross-entropy}, $ \lambda $ is the trade-off parameter, and $ \mathcal{L}_{\text{UDTL}} $ represents the partial loss to reduce the feature difference between source and target domains.
\begin{align}
\label{cross-entropy}
\mathcal{L}_{c} = 
- \mathbb{E}_{\left(x_i^s, y_i^s\right) \in \mathcal{D}_s} 
\sum_{c=0}^{C-1} \mathbf{1}_{[y_i^s=c]} \log\left[\beta\left(x_i^s\right)\right],
\end{align}
where $ C $ is the number of all possible classes, $ \mathbb{E} $ denotes the mathematical expectation, and $\mathbf{1}$ is the indicator function.

\subsection{Taxonomy of UDTL-based IFD}
\begin{figure*}[!t]
	\centering
	\subfigure{\includegraphics[scale = 0.9]{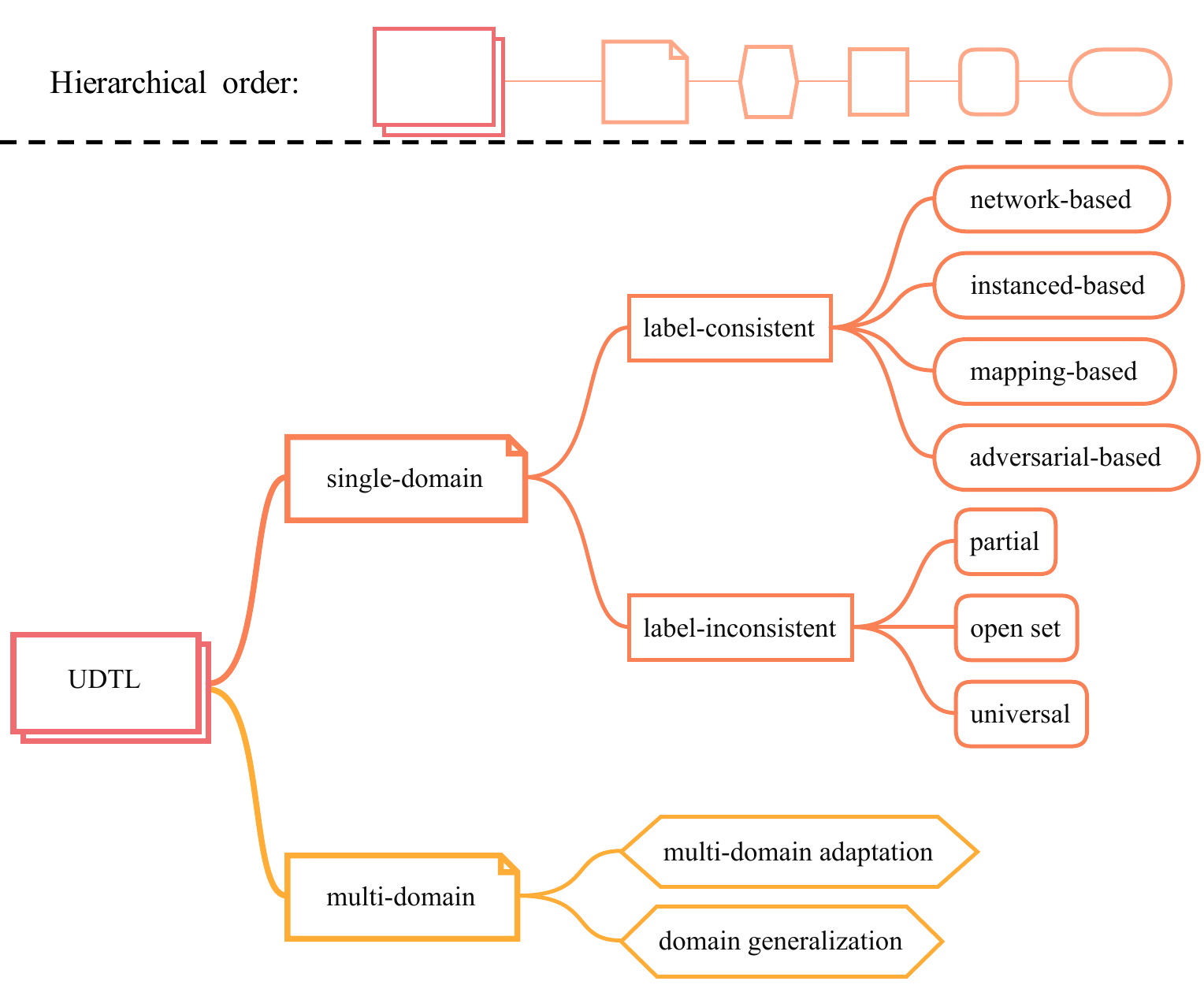}}
	\\ [-10pt]
	\caption{A taxonomy of UDTL-based methods.}
	\label{figure-UDTL}
\end{figure*}

In this section, we present our taxonomy of UDTL-based IFD, as shown in \figref{figure-UDTL}.
We categorize UDTL-based IFD into single-domain and multi-domain UDTL according to the number of source domains from a macro perspective.
In the following, we give a brief introduction of each category and detailed description is given in the next part.

\begin{figure}[!t]
	\centering
	\subfigure{\includegraphics[scale = 0.4]{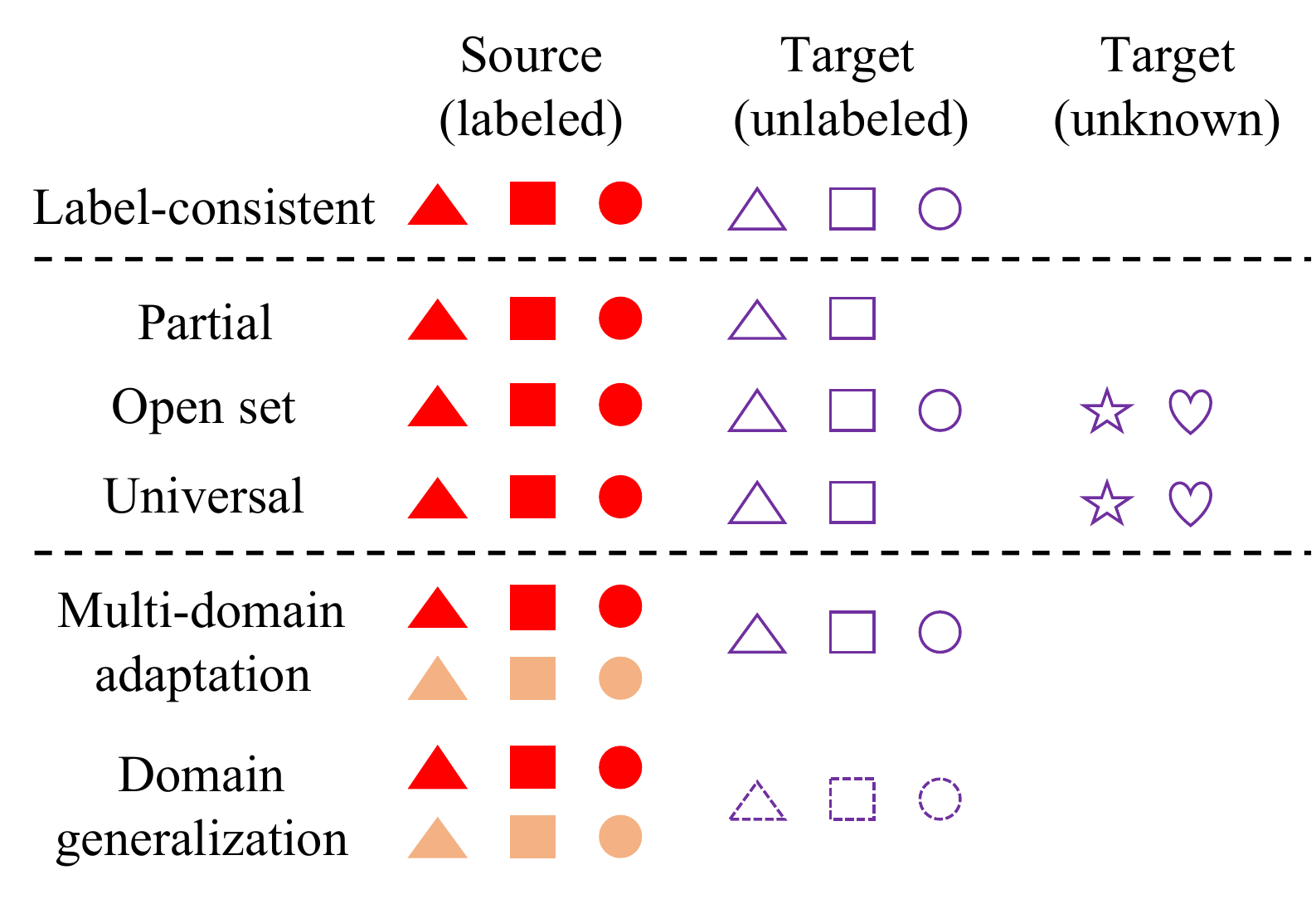}}
	\\ [-10pt]
	\caption{Visualization explanation of different transfer settings. Additionally, different colors represent different domains and dotted lines denote that this domain does not participate in training.}
	\label{figure-UDTLExamples}
\end{figure}

\textit{1) Single-domain UDTL:} These can be further categorized into label-consistent (closed set) and label-inconsistent UDTL.
As shown in \figref{figure-UDTLExamples}, label-consistent UDTL represents the label sets of source and target domains are consistent.
According to Tan et al. \cite{tan2018survey}, label-consistent UDTL can be classified into four categories: network-based, instanced-based, mapping-based, and adversarial-based methods from a methodological level.
Additionally, We categorize label-inconsistent UDTL into partial, open set, and universal tasks based on the inclusion relationship between label sets.
As shown in \figref{figure-UDTLExamples}, partial UDTL means that the target label set is a subspace of the source label set; open set UDTL means that the target label set contains unknown labels; universal UDTL is a combination of the first two conditions.
It is worth mentioning that three tasks can be further divided into the above four methods from a methodological level.

\textit{2) Multi-domain UDTL:} These can be further categorized into multi-domain adaptation and domain generalization (DG) based on the usage of the target data in the training phase.
Multi-domain adaptation means that the unlabeled samples from the target domain participate into the training phase, and DG is the opposite.
Besides, these two conditions can also be further categorized into label-consistent and label-inconsistent UDTL.

\subsection{Motivation of UDTL-based IFD}
Distributions of training and test samples are often different, due to the influence of working conditions, fault sizes, fault types, etc.
Consequently, UDTL-based IFD has been introduced recently to tackle this domain shift problem since there are some shared features in the specific space.
Using these shared features, applications of UDTL-based IFD can be mainly classified into four categories: different working conditions, different types of faults, different locations, and different machines.

\textit{1) Different working conditions:} Due to the influence of speed, load, temperature, etc., working conditions often vary during the monitoring period.
Collected signals may contain domain shift, which means that the distribution of data may differ significantly under different working conditions \cite{li2019multi}.
The aim of UDTL-based IFD is that the model trained using signals under one working condition can be transferred to signals under another different working condition.

\textit{2) Different types of faults:} Label difference between source and target domains may exist since different types of faults would happen on the same component.
Therefore, there are three cases in UDTL-based IFD.
The first one is that unknown fault types appear in the target domain (open set transfer).
The second one is that partial fault types of the source domain appear in the target domain (partial transfer).
The third one is that the first two cases occur at the same time (universal transfer).
The aim of UDTL-based IFD is that the model trained with some types of faults can be transferred to the target domain with different types of faults.

\textit{3) Different locations:} Because sensors installed on the same machine are often responsible for monitoring different components, and sensors located near the fault component are more suitable to indicate the fault information.
However, key components have different probabilities of failure rates, leading to the situation where signals from different locations have different numbers of labeled data.
The aim of UDTL-based IFD is that the model trained with plenty of labeled data from one location can be transferred to the target domain with unlabeled data from other locations.

\textit{4) Different machines:} Enough labeled fault samples of real machines are difficult to collect due to the test cost and security.
Besides, enough labeled data can be generated from dynamic simulations or fault seeding experiments.
However, distributions of data from dynamic simulations or fault seeding experiments are different but similar to those from real machines, due to the similar structure and measurement situations.
Thus, the aim of UDTL-based IFD is that the model can be transferred to test data gathered from real machines.

\subsection{The structure of backbone}
One of the most important parts of UDTL-based IFD is the structure of the backbone, which acts as feature extraction and has a huge impact on the test accuracy.
For example, in the field of image classification, different backbones, such as VGG \cite{simonyan2014very}, ResNet \cite{he2016deep}, etc., have different abilities of feature extraction, leading to different classification performance.

However, for UDTL-based IFD, different studies have their own backbones, and it is difficult to determine whose backbone is better.
Therefore, direct comparisons with the results listed in other published papers are unfair and unsuitable due to different representative capacities of backbones.
In this paper, we try to verify the performance of different UDTL-based IFD methods using the same CNN backbone to ensure a fair comparison.

\begin{figure}[!t]
	\centering
	\subfigure{\includegraphics[scale = 0.3]{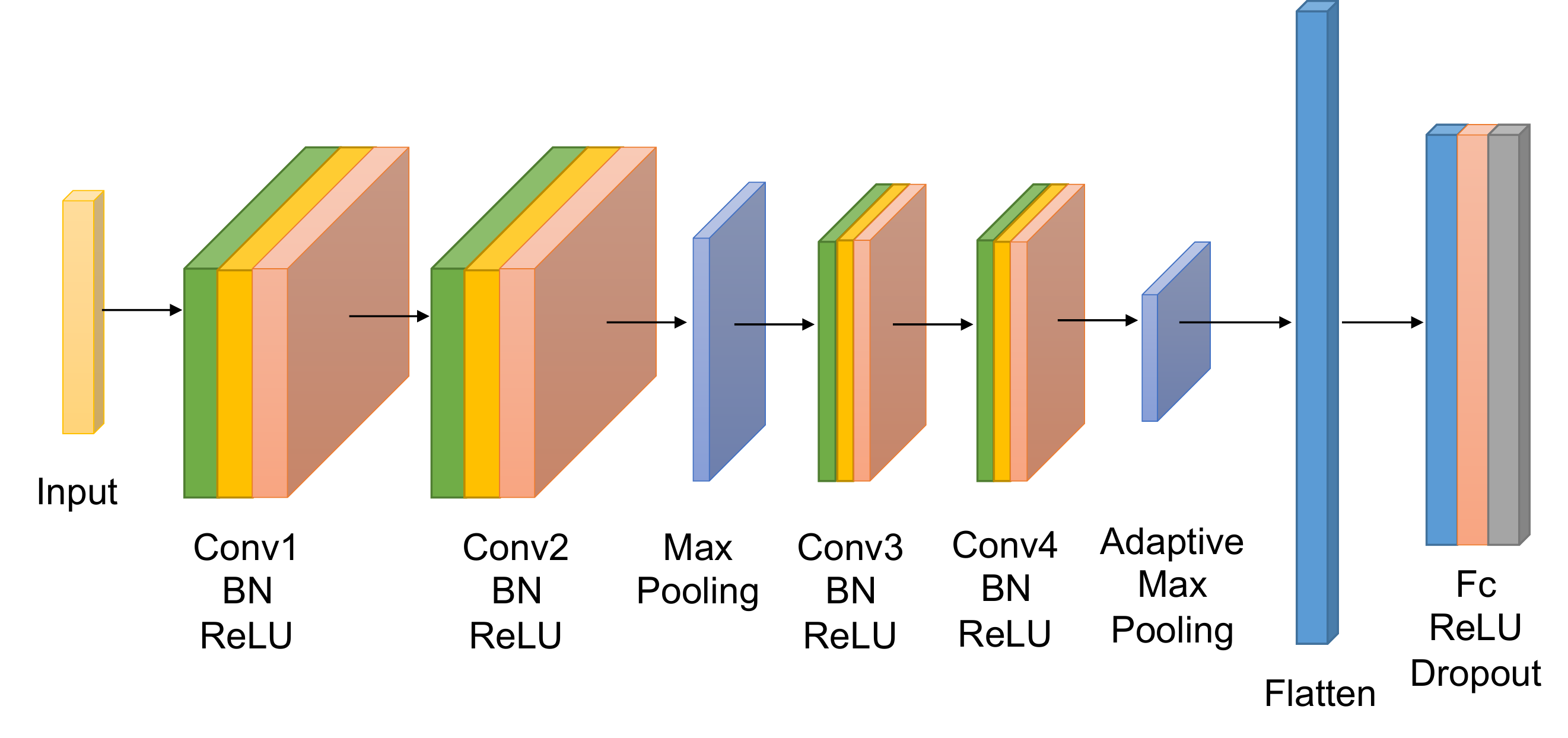}}
	\\ [-5pt]
	\caption{The structure of the backbone.}
	\label{figure-backbone}
\end{figure}

As shown in \figref{figure-backbone}, the CNN backbone consists of four one dimension (1D) convolutional layers that come with an 1D Batch Normalization (BN) layer and a ReLU activation function.
Besides, the second combination comes with an 1D Max Pooling layer, and the fourth combination comes with an 1D Adaptive Max Pooling layer to realize the adaptation of the input length.
The convolutional output is then flattened and passed through a fully-connected (Fc) layer, a ReLU activation function, and a Dropout layer. The detailed parameters are listed in Table \ref{Tab-backbone}.

\begin{table}[!htbp]
	\caption{Parameters of the backbone.} 
	\centering
	\label{Tab-backbone}  
	\begin{tabular}{cc}
		\hline
		Layers      & Parameters \\ \hline
		Conv1    & out\_channels=16, kernel\_size=15 \\ \hline
		Conv2 & out\_channels=32, kernel\_size=3   \\ \hline
		Max Pooling & kernel\_size=2, stride=2  \\ \hline
		Conv3 & out\_channels=64, kernel\_size=3   \\ \hline
		Conv4 & out\_channels=128, kernel\_size=3   \\ \hline
		Adaptive Max Pooling & output\_size=4  \\ \hline
		Fc & out\_features=256  \\ \hline
		Dropout & p=0.5   \\ \hline
	\end{tabular}
\end{table}

\section{Label-consistent UDTL}
\label{S:3}
Label-consistent (also called closed set) UDTL-based IFD assumes that the source domain has the same label space with the target domain.
In this section, we categorize label-consistent UDTL into network-based, instanced-based, mapping-based, and adversarial-based methods from a methodological level.
\subsection{Network-based UDTL}
\subsubsection{Basic concepts}
\begin{figure}[!t]
	\centering
	\subfigure{\includegraphics[scale = 0.4]{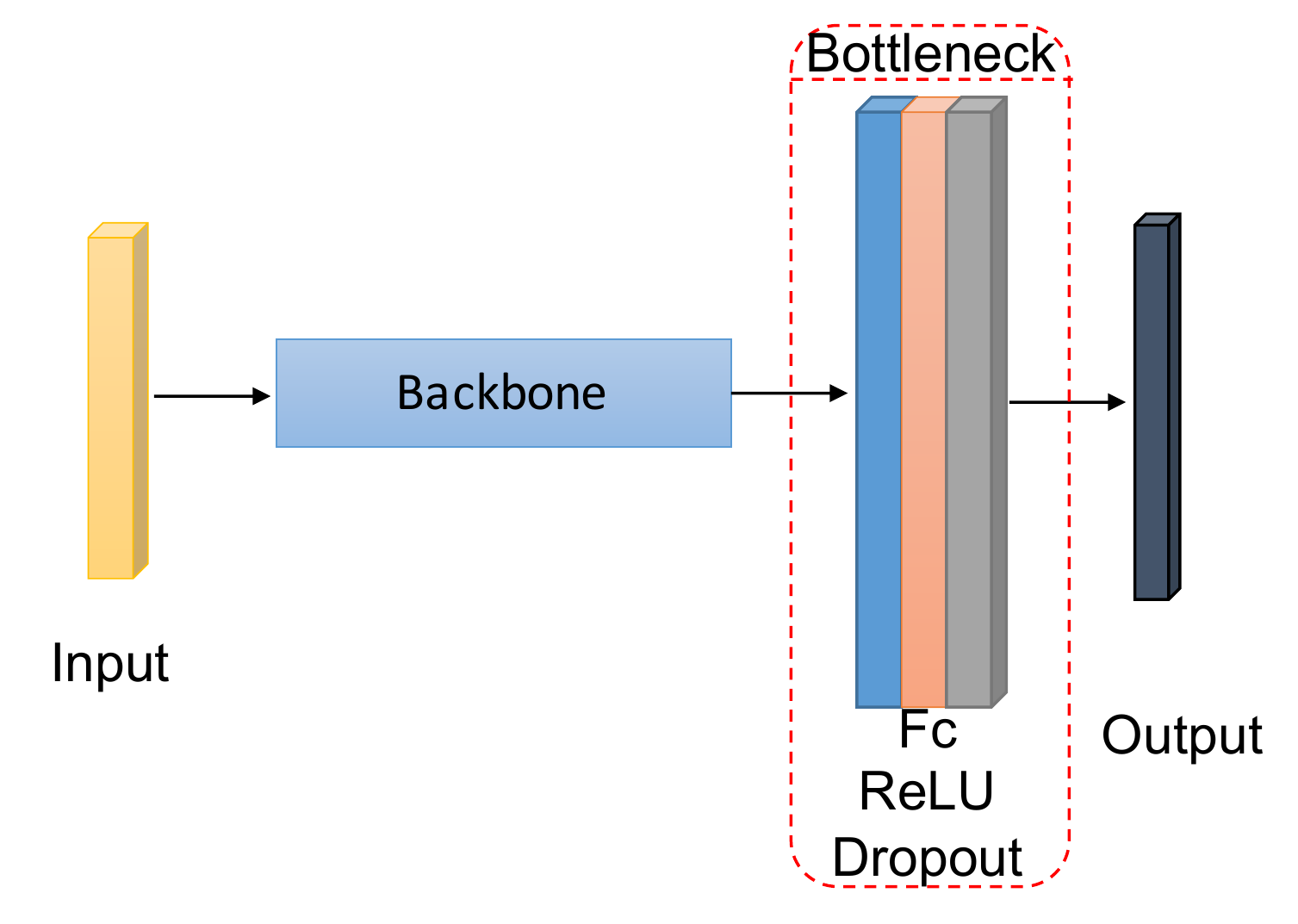}}
	\\ [-10pt]
	\caption{The structure of the basic model.}
	\label{figure-bottleneck}
\end{figure}

Network-based DTL means that partial network parameters pre-trained in the source domain are transferred directly to be partial network parameters of the test procedure or network parameters are fine-tuned with a few labeled data in the target domain.
The most popular network-based DTL method is to fine-tune the trained model utilizing a few labeled data in the target domain.
However, for UDTL-based IFD, labels in the target domain are unavailable.
We use the backbone coming with a bottleneck layer, consisting of a Fc layer (out\_features=256), a ReLU activation function, a Dropout layer ($p=0.5$), and a basic Softmax classifier to construct our basic model (we call it Basis), which is shown in \figref{figure-bottleneck}.
The trained model is used to test samples in the target domain directly, which means that source and target domains share the same model and parameters.

\subsubsection{Applications to IFD}
Pre-trained deep neural networks using the source data were used in \cite{zhang2017transfer,zhang2018method,chen2018incipient,hasan20181d,kim2019new,hasan2019acoustic,sun2018deep,shao2018highly,chen2019transfer,mao2019online,he2020deep,li2020cross} via frozing their partial parameters, and then part of network parameters were transferred to the target network and other parameters were fine-tuned with a small amount of target data.
Pre-trained deep neural networks on ImageNet were used in \cite{sharaf2018beam,cao2018preprocessing,chen2019intelligent,iba2019vibration,ma2019novel} and were fine-tuned with limited target data to adapt the domain of engineering applications.
Ensemble techniques and multi-channel signals were used in \cite{he2020ensemble, di2021ensemble} to initialize the target network which was fine-tuned by a few training samples from the target domain.
Two-dimensional images, such as grey images \cite{wang2020deformable}, time-frequency images \cite{wang2019deep}, and thermal images \cite{shao2020intelligent}, were used to pre-train the specific-designed networks, which were transferred to the target tasks via fine-tuning.

Qureshi et al. \cite{qureshi2017wind} pre-trained nine deep sparse auto-encoders on one wind farm, and predictions on another wind farm were taken by fine-tuning the pre-trained networks.
Zhong et al. \cite{zhong2019novel} trained a CNN on enough normal samples and then replaced Fc layers with support vector machine as the target model.
Han et al. \cite{han2019learning} discussed and compared three fine-tuning strategies: only fine-tuning the classifier, fine-tuning the feature descriptor, and fine-tuning both the feature descriptor and the classifier for diagnosing unseen machine conditions.
Xu et al. \cite{xu2019online} pre-trained the offline CNN on the source domain and directly transferred them to the shallow layers of the online CNN via fine-tuning the online CNN on the target domain for online IFD.
Zhao et al. \cite{zhao2020deepmultiscale} proposed a multi-scale convolutional transfer learning network pre-trained on the source domain, and then the model was transferred to the other different but similar domains with proper fine-tuning.

\subsection{Instanced-based UDTL}
\subsubsection{Basic concepts}
Instanced-based UDTL refers to re-weight instances in the source domain to assist the classifier to predict labels or use statistics of instances to help align the target domain, such as TrAdaBoost \cite{dai2007boosting} and adaptive Batch Normalization (AdaBN) \cite{li2016revisiting}.
In this paper, we use AdaBN to represent one of instanced-based UDTL methods, which does not require labels from the target domain.

BN, which can be used to avoid the issue of the internal covariate shifting, is one of the most important techniques.
BN can promote much faster training speed since it makes the input distribution more stable.
Detailed descriptions and properties can be referred to \cite{ioffe2015batch}.
It is worth mentioning that BN layers are only updated in the training procedure and the global statistics of training samples are used to normalize test samples during the test procedure.

AdaBN, which is a simple and parameter-free technique for the domain shift problem, was proposed in \cite{li2016revisiting} to enhance the generalization ability.
The main idea of AdaBN is that the global statistics of each BN layer are replaced with statistics in the target domain during the test phase.
In our AdaBN realization, after training, we provide two updating strategies to fine-tune statistics of BN layers using target data, including updating via each batch and the whole data.
In this paper, we update statistics of BN layers via each batch considering the memory limit.

\subsubsection{Applications to IFD}
Xiao et al. \cite{xiao2019transfer} used TrAdaBoost to enhance the diagnostic capability of the fault classifier by adjusting the weight factor of each training sample.
Zhang et al. \cite{zhang2017new} and Qian et al. \cite{qian2018new} used AdaBN to improve the domain adaptation ability of the model by ensuring that each layer receives data from a similar distribution.

\subsection{Mapping-based UDTL}
\subsubsection{Basic concepts}
Mapping-based UDTL refers to map instances from both source and target domains to the feature space via a feature extractor.
There are many methods belonging to mapping-based UDTL, such as Euclidean distance, Minkowski distance, Kullback-Leibler, correlation alignment (CORAL) \cite{sun2016deep}, maximum mean discrepancy (MMD) \cite{borgwardt2006integrating,sejdinovic2013equivalence}, multi kernels MMD (MK-MMD) \cite{gretton2012optimal,long2015learning},  joint distribution adaptation (JDA) \cite{long2013transfer}, balanced distribution adaptation (BDA) \cite{wang2017balanced}, and Joint Maximum Mean Discrepancy (JMMD) \cite{long2017deep}.
In this paper, we use MK-MMD, JMMD, and CORAL to represent mapping-based methods and test their performance.

\textbf{\textit{MK-MMD:}}
To introduce the definition of MK-MMD, we briefly explain the concept of MMD.
MMD was first proposed in \cite{borgwardt2006integrating} and was used in transfer learning by many other scholars \cite{pan2010domain,tzeng2014deep}. 
MMD defined in Reproducing Kernel Hilbert Space (RKHS) is a squared distance between the kernel embedding of marginal distributions $ P(X_s) $ and $ Q(X_t) $.
RKHS is a Hilbert space of functions in which point evaluation is a continuous linear functional, and some examples can be found in \cite{zynda2020weights}.
The formula of MMD can be written as follows:
\begin{align}
\label{eq5}
\mathcal{L}_\text{MMD}\left( P, Q \right) = \left\| \mathbb{E}_P \left({\phi \left( {x^s} \right)} \right) - \mathbb{E}_Q \left({\phi \left( {x^t} \right)}\right)  \right\|^2_{\mathcal{H}_k},
\end{align}
where $ \mathcal{H}_k $ is RKHS using the kernel $ k $ (in general, Gaussian kernel is used as the kernel), and $ \phi (\cdot) $ is the mapping to RKHS.

Parameter selection of each kernel is crucial to the final performance.
To tackle this problem, MK-MMD, which could maximize the two-sample test power and minimize the Type \uppercase\expandafter{\romannumeral2} error jointly, was proposed by Gretton et al \cite{gretton2012optimal}.
For MK-MMD, scholars often use the convex combination of $ m $ kernels $ \left\{ k_u \right\} $ to provide effective estimations of the mapping.
\begin{align}
\label{eq6}
K \mathop  = \limits^\Delta  \left\{ {k = \sum\limits_{u = 1}^m {{\alpha_u}{k_u}:\sum\limits_{u = 1}^m {{\alpha_u} = 1} ,\alpha  \ge 0,\forall u} } \right\},
\end{align}
where $ \left\{ \alpha_u \right\}$ are weighted parameters of different kernels (in this paper, all $ \alpha_u = \frac{1}{m}$).

\begin{figure}[!t]
	\centering
	\subfigure{\includegraphics[scale = 0.35]{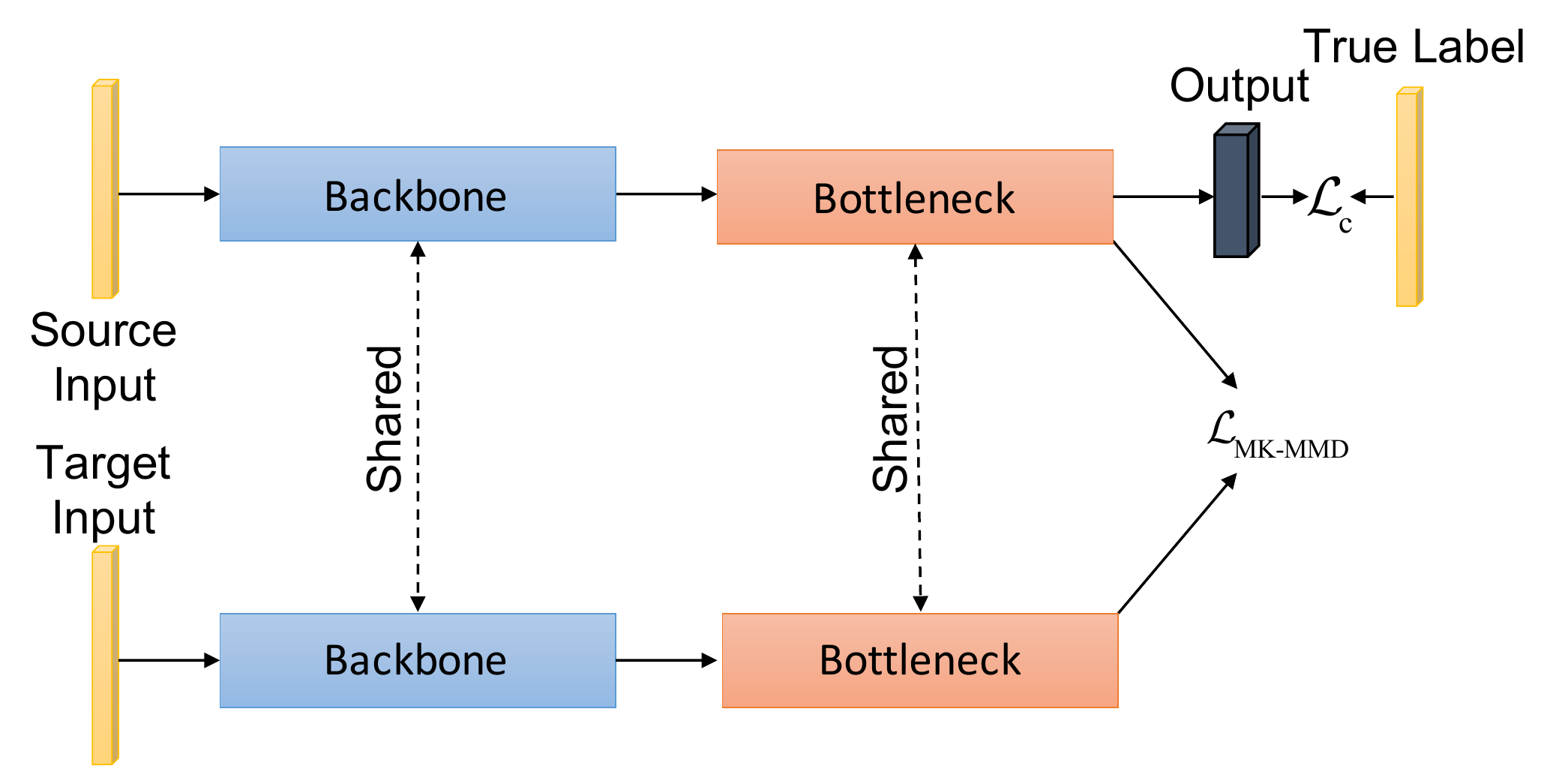}}
	\\ [-10pt]
	\caption{The UDTL-based IFD model based on MK-MMD.}
	\label{figure-MKMMD}
\end{figure}

Inspired by deep adaptation networks (DAN) proposed in \cite{long2015learning}, we design an UDTL-based IFD model by adding MK-MMD into the loss function to realize the feature alignment shown in \figref{figure-MKMMD}.
In addition, the final loss function is defined as follows:
\begin{align}
\label{eq7}
\mathcal{L}  = \mathcal{L}_{c} + \lambda_{\text{MK-MMD}} \mathcal{L}_{\text{MK-MMD}} \left( \mathcal{D}_s, \mathcal{D}_t \right),
\end{align}
where $ \lambda_{\text{MK-MMD}} $ is a trade-off parameter and $ \mathcal{L}_{\text{MK-MMD}} $ means the multi-kernel version of MMD.
Besides, we simply use the Gaussian kernel and the number of kernels is equal to five.
The bandwidth of each kernel is set to be median pairwise distances on training data according to the median heuristic \cite{gretton2012optimal}.

\textbf{\textit{JMMD:}}
MMD and MK-MMD, which are defined to solve the problem $ P(X_s) \neq Q(X_t) $, cannot be used to tackle the domain shift generated by joint distributions (e.g. $P(X_s, Y_s) \neq Q(X_t, Y_t)$).
Thus, JMMD, proposed in \cite{long2017deep}, was designed to measure the distance of empirical joint distributions $ P(X_s, Y_s) $ and $ Q(X_t, Y_t)$.
The formula of JMMD is written as follows \cite{long2017deep}:
\begin{equation}
\label{JMMD}
\begin{aligned}
&\mathcal{L}_\text{JMMD}\left( P, Q \right) = \\
&\left\| \mathbb{E}_P \left({\otimes_{l=1}^{|L|} \phi^l \left( {z_l^{s}} \right)} \right) 
- \mathbb{E}_Q \left(\otimes_{l=1}^{|L|}{\phi^l \left( {z_l^{t}} \right)}\right)  \right\|^2_{\otimes_{l=1}^{|L|}{\mathcal{H}^l}},
\end{aligned}
\end{equation}
where $ \otimes_{l=1}^{|L|} \phi^l \left( {z_{l}} \right)  = \phi^1 \left( {z_{1}} \right) \otimes \dots \otimes \phi^{|L|} \left( {z_{|L|}} \right)$ is the feature mapping in the tensor product Hilbert space, $ L $ is the set of higher network layers, $ |L| $ is the number of layers, $ z_l^{s} $ means the activation of the $l-$th layer generated by the source domain, and $z_l^{t}$ means the activation of the $l-$th layer generated by the target domain.

\begin{figure}[!t]
	\centering
	\subfigure{\includegraphics[scale = 0.35]{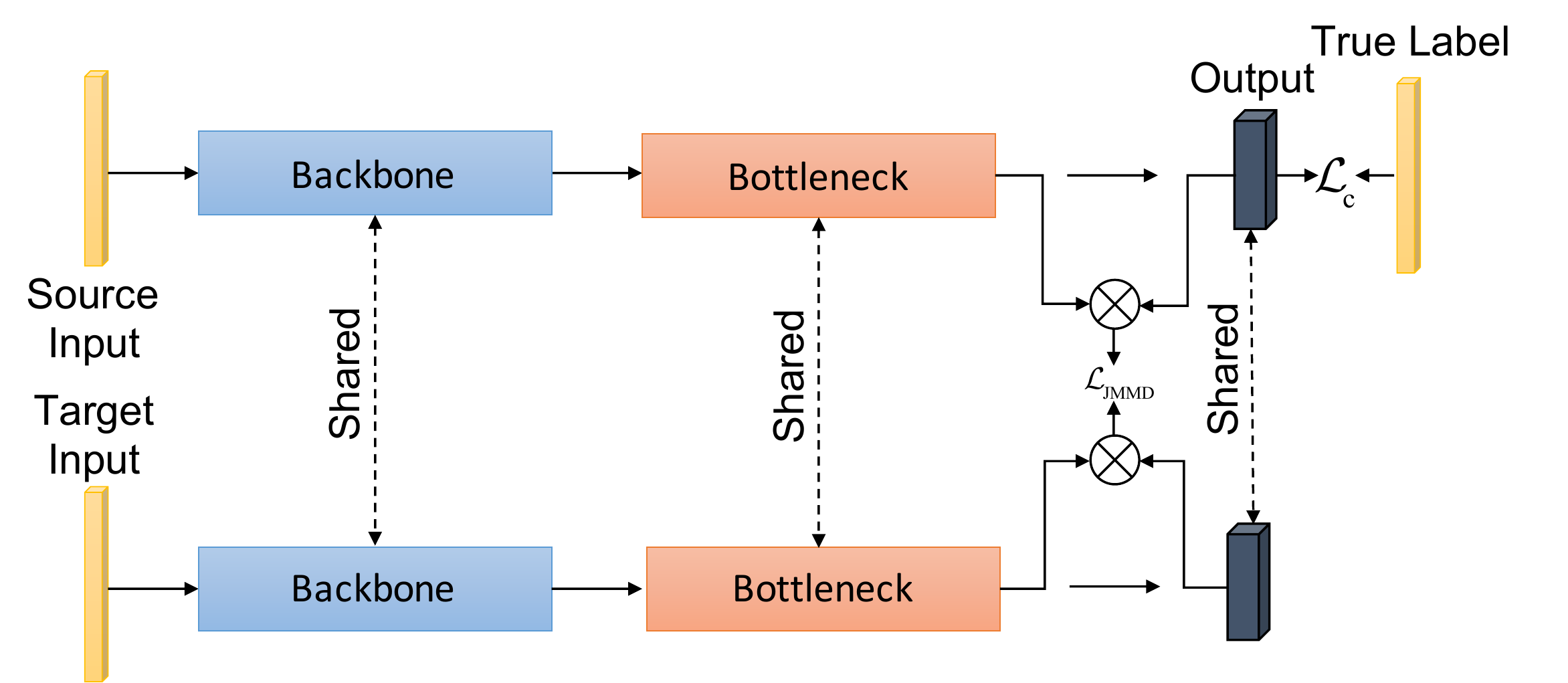}}
	\\ [-10pt]
	\caption{The UDTL-based IFD model based on JMMD.}
	\label{figure-JMMD}
\end{figure}

Inspired by Joint Adaptation Network (JAN) which uses JMMD to align the domain shift \cite{long2017deep}, we design an UDTL-based IFD method by adding JMMD into the loss function to realize feature alignment shown in \figref{figure-JMMD}.
The final loss function is defined as follows:
\begin{align}
\label{JAN}
\mathcal{L}  = \mathcal{L}_{c} + \lambda_{\text{JMMD}} \mathcal{L}_{\text{JMMD}} \left(\mathcal{D}_s, \mathcal{D}_t \right),
\end{align}
where $ \lambda_{\text{JMMD}} $ is a trade-off parameter.
Additionally, the parameter setting of JMMD is the same as that in JAN.

\textbf{\textit{CORAL:}}
The CORAL loss, which aims to align the second-order statistics of source and target distributions, was first proposed in \cite{sun2016return} and was further used in UDTL \cite{sun2016deep}.
First of all, following \cite{sun2016return} and \cite{sun2016deep}, we give the basic definition of the CORAL loss as:
\begin{align}
\label{CORAL_loss}
\mathcal{L}_{\text{CORAL}} \left( \mathcal{D}_s, \mathcal{D}_t  \right) =  \dfrac{1}{4d^2} ||C^s - C^t||_F^2,
\end{align}
where $ ||\cdot||_F $ is the Frobenius norm and $ d $ is the dimension of each sample.
$ C^s $ and $ C^t $ defined in \eqref{covariance} are covariance matrices.
\begin{equation}
\label{covariance}
\begin{aligned}
C^s & = \dfrac{1}{n_s - 1} \left(X_s^TX_s - \dfrac{1}{n_s}(\textbf{1}^TX_s)^T(\textbf{1}^TX_s)\right),\\
C^t & = \dfrac{1}{n_t - 1} \left(X_t^TX_t - \dfrac{1}{n_t}(\textbf{1}^TX_t)^T(\textbf{1}^TX_t)\right),
\end{aligned}
\end{equation}
where $ \textbf{1} $ represents the column vector whose elements are all equal to one.

\begin{figure}[!t]
	\centering
	\subfigure{\includegraphics[scale = 0.35]{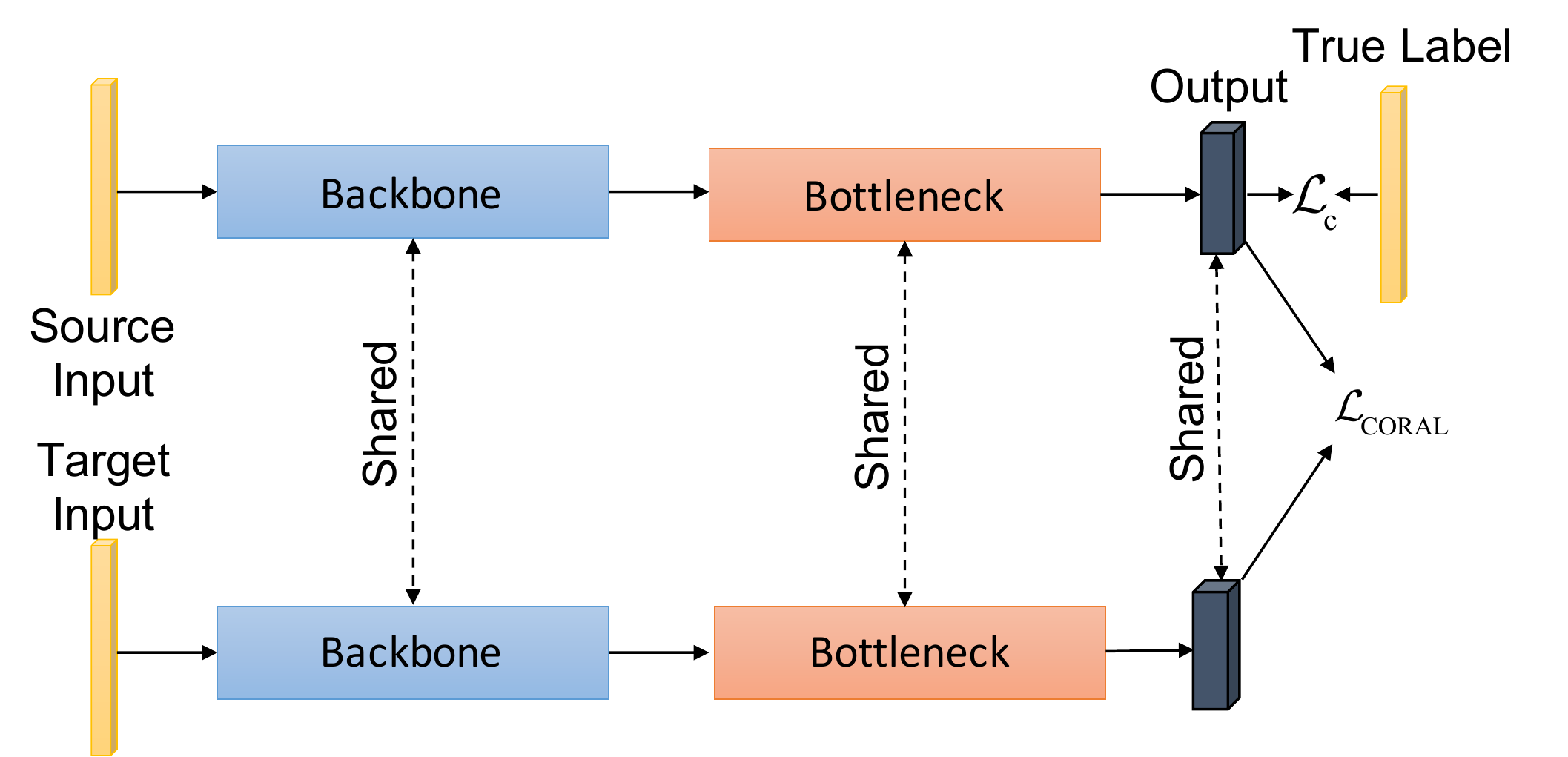}}
	\\ [-10pt]
	\caption{The UDTL-based IFD model based on CORAL.}
	\label{figure-CORAL}
\end{figure}

Inspired by Deep CORAL proposed in \cite{sun2016deep}, we design an UDTL-based IFD method by adding the CORAL loss into the loss function to realize the feature transfer shown in \figref{figure-CORAL}.
Also, the final loss function is defined as follows:
\begin{align}
\label{Deep_CORAL}
\mathcal{L}  = \mathcal{L}_{c} + \lambda_{\text{CORAL}} \mathcal{L}_{\text{CORAL}} \left( \mathcal{D}_s, \mathcal{D}_t  \right),
\end{align}
where $ \lambda_{\text{CORAL}} $ is a trade-off parameter.

\subsubsection{Applications to IFD}
BDA was used in \cite{wang2018power, wang2020network} to adaptively balance the importance of the marginal and conditional distribution discrepancy between feature domains learned by deep neural networks for IFD.
The CORAL loss \cite{wang2019hierarchical,an2020deep} and maximum variance discrepancy (MVD) \cite{zhang2020unsupervised} were used to reduce the distribution discrepancy between different domains.
Qian et al. \cite{qian2018new,qian2019deep} considered the higher-order moments and proposed an HKL divergence to adjust domain distributions for rotating machine fault diagnosis.
The distance designed to measure source and target tensor representations was proposed in \cite{hu2020cross} to align tensor representations into the invariant tensor subspace for bearing fault diagnosis.

Another metric distance, called MMD, was widely used in the field of intelligent diagnosis \cite{lu2016deep,zhang2018intelligent,wen2017new,yang2019intelligent,tang2019transfer,li2018cross,xu2019digital,an2019intelligent,li2020intelligent}.
Tong et al. \cite{tong2018bearing1,tong2018bearing2} reduced marginal and conditional distributions simultaneously across domains based on MMD in the feature space by refining pseudo test labels for bearing fault diagnosis.
Wang et al. \cite{wang2020multi} proposed a conditional MMD based on estimated pseudo labels to shorten the conditional distribution distance for bearing fault diagnosis.
The marginal and conditional distributions were aligned simultaneously in multiple layers via minimizing MMD \cite{li2020deepbalanced,lu2021new}.
Yang et al. \cite{yang2020polynomial} replaced the Gaussian kernel with a polynomial kernel in MMD for better aligning the distribution discrepancy.
Cao et al. \cite{cao2021domain} proposed a pseudo-categorized MMD to narrow the intra-class cross-domain distribution discrepancy.
MMD was also combined with other techniques, such as Grassmann manifold \cite{zhao2020novel}, locality preserving projection \cite{zheng2020new}, and graph Laplacian regularization \cite{zhang2020novel,qian2021discriminative}, to boost the performance of distribution alignment.

MK-MMD was used in \cite{li2018robust,yang2018transfer,an2019generalization,li2019multi,zhu2019new,che2020domain} to better transfer the distribution of learned features in the source domain to that in the target domain for IFD.
Han et al. \cite{han2019deep} and Qian et al. \cite{qian2019novel} used JDA to align both conditional and marginal distributions simultaneously to construct a more effective and robust feature representation for substantial distribution difference. Wu et al. \cite{wu2020adaptive} further used the grey wolf optimization algorithm to learn the parameters of JDA.
Based on JMMD, Cao et al. \cite{cao2020deep} proposed a soft JMMD to reduce both the marginal and conditional distribution discrepancy with the enhancement of auxiliary soft labels.

\subsection{Adversarial-based UDTL}
\subsubsection{Basic concepts}
Adversarial-based UDTL refers to an adversarial method using a domain discriminator to reduce the feature distribution discrepancy between source and target domains produced by a feature extractor. 
In this paper, we use two commonly used methods including domain adversarial neural network (DANN) \cite{ganin2016domain} and conditional domain adversarial network (CDAN) \cite{long2018conditional} to represent adversarial-based methods and test the corresponding accuracy.

\textbf{\textit{DANN:}}
Similar to MMD and MK-MMD,  DANN is defined to solve the problem $ P(X_s) \neq Q(X_t)  $.
It aims to train a feature extractor, a domain discriminator distinguishing source and target domains, and a class predictor, simultaneously to align source and target distributions.
That is, DANN trains the feature extractor to prevent the domain discriminator from distinguishing differences between two domains.
Let $ G_f $ be the feature extractor whose parameters are $ \theta_f $, $ G_c $ be the class predictor whose parameters are $ \theta_c $, and $ G_d $ be the domain discriminator whose parameters are $ \theta_d $.
After that, the prediction loss and the adversarial loss (the binary cross-entropy loss) can be rewritten as follows:
\begin{equation}
\label{A-softmax-loss}
\begin{aligned}
&\mathcal{L}_{c}(\theta_f,\theta_c) = 
\\
&- \mathbb{E}_{\left(x_i^s, y_i^s\right) \in \mathcal{D}_s} 
\sum_{c=0}^{C-1} \mathbf{1}_{[y_i^s=c]} \log\left[G_c\left(G_f\left(x_i^s;\theta_f\right);\theta_c\right)\right],
\end{aligned}
\end{equation}
\begin{equation}
\label{A-binary-loss}
\begin{aligned}
\mathcal{L}_{\text{DANN}}\left(\theta_f,\theta_d\right) = 
&- \mathbb{E}_{x_i^s \in \mathcal{D}_s} 
\log\left[G_d\left(G_f\left(x_i^s;\theta_f\right);\theta_d\right)\right]- 
\\
&\mathbb{E}_{x_i^t \in \mathcal{D}_t} 
\log\left[1- G_d\left(G_f\left(x_i^t;\theta_f\right);\theta_d\right)\right].
\end{aligned}
\end{equation}
To sum up, the total loss of DANN can be defined as:
\begin{align}
\label{DANN_loss}
\mathcal{L}\left(\theta_f,\theta_c,\theta_d\right) = \mathcal{L}_c\left(\theta_f,\theta_c\right) - \lambda_{\text{DANN}}\mathcal{L}_{\text{DANN}}\left(\theta_f,\theta_d\right),
\end{align}
where $ \lambda_{\text{DANN}} $ is a trade-off parameter.

During the training procedure, we need to minimize the prediction loss to allow the class predictor to predict true labels as much as possible.
Additionally, we also need to maximize the adversarial loss to make the domain discriminator difficult to distinguish differences.
Thus, solving the saddle point problem ($ \hat\theta_f,\hat\theta_c,\hat\theta_d $) is equivalent to the following minimax optimization problem:
\begin{equation}
\label{DANN_Optimization}
\begin{aligned}
\left(\hat{\theta_f}, \hat{\theta_c}\right) &= \arg\min\limits_{\theta_f,\theta_c} \mathcal{L}\left(\theta_f,\theta_c,\hat \theta_d\right),\\
\left(\hat{\theta_d}\right) &= \arg\max\limits_{\theta_d} \mathcal{L}\left(\hat\theta_f,\hat\theta_c,\theta_d\right).
\end{aligned}
\end{equation}
Following the statement in \cite{ganin2016domain}, we can simply add a special \textit{gradient reversal layer} (GRL), which changes signs of the gradient from the subsequent level and is parameter-free, to solve the above optimization problem.

\begin{figure}[!t]
	\centering
	\subfigure{\includegraphics[scale = 0.32]{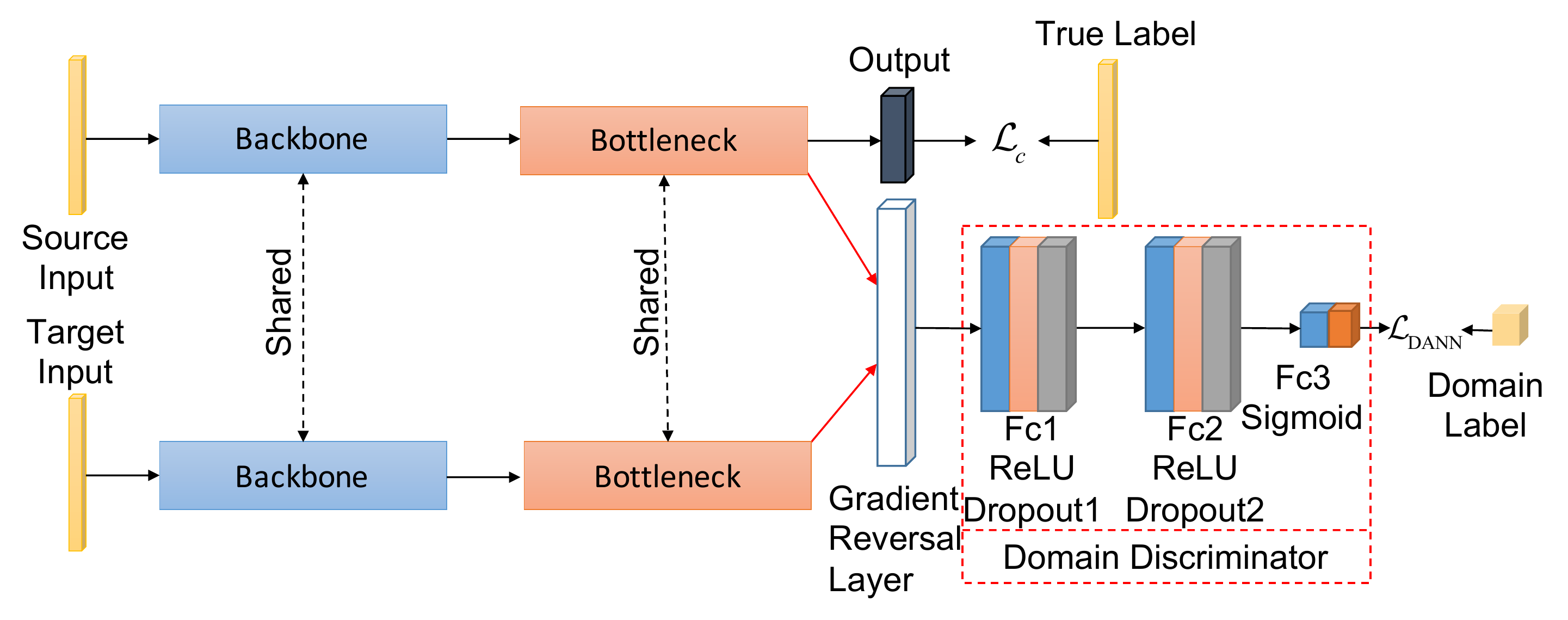}}
	\\ [-10pt]
	\caption{The UDTL-based IFD model based on DANN.}
	\label{figure-DANN}
\end{figure}

We design an UDTL-based IFD model via adding the adversarial idea into the loss function to realize the feature transfer between source and target domains shown in \figref{figure-DANN}.
It can be observed that we use a three-layer Fc binary classifier as our domain discriminator which is the same as \cite{ganin2016domain}.
The output features of these Fc layers are 1024 (Fc1), 1024 (Fc2), and 2 (Fc3), respectively.
The parameter of a Dropout layer is $p=0.5$.

\textbf{\textit{CDAN:}}
Although DANN can align the distributions of two domains efficiently, there may still exist some bottlenecks.
As stated in \cite{long2018conditional}, DANN cannot capture complex multi-modal structures and it is hard to condition the domain discriminator safely.
Based on this statement, Long et al. \cite{long2018conditional} proposed a new adversarial-based UDTL model called CDAN to solve the problem $ P(X_s, Y_s) \neq Q(X_t, Y_t)  $.
To briefly introduce the main idea inside CDAN, we first need to define the multi-linear map $ \otimes $, which means the outer product of multiple random vectors.
If two random vectors $ x $ and $ y $ are given, the mean mapping $ x \otimes y$ can capture the complex multi-modal structures inside the data completely.
Besides, the cross-covariance $ \mathbb{E}_{xy} [\phi(x) \otimes \phi(y)]$ can be used to model the joint distribution $ P(x, y) $ successfully.
Thus, the conditional adversarial loss is defined as follows:
\begin{equation}
\label{CDAN-binary-loss}
\begin{aligned}
\mathcal{L}_{\text{CDAN}}&(\theta_f,\theta_d) = 
- \mathbb{E}_{x_i^s \in \mathcal{D}_s} 
\log\left[G_d\left(G_f(x_i^s) \otimes G_c\left(G_f(x_i^s)\right)\right)\right]
\\
&-\mathbb{E}_{x_i^t \in \mathcal{D}_t} 
\log\left[1- G_d\left(G_f(x_i^t) \otimes G_c\left(G_f(x_i^t)\right)\right)\right],
\end{aligned}
\end{equation}
and the prediction loss is the same as that in \eqref{A-softmax-loss}.

To relax the influence with uncertain predictions, the entropy criterion $ H(p) = - \sum_{c=0}^{C-1} p_c \log p_c $ is used to define the uncertainty of predictions by classifiers, where $ p_c $ is the probability of the predicted result corresponding to the label $ c $.
According to the defined entropy-aware weight function shown in \eqref{entropy-weight}, those hard-to-transfer samples are re-weighted with lower weights in the modified conditional adversarial loss \eqref{modified-CDAN-binary-loss}:
\begin{align}
\label{entropy-weight}
w\left(H\left(p\right)\right) = 1 + e^{-H\left(p\right)}.
\end{align}
\begin{equation}
\label{modified-CDAN-binary-loss}
\begin{aligned}
\mathcal{L}_{\text{CDAN}}(\theta_f,\theta_d) = 
&- \mathbb{E}_{x_i^s \in \mathcal{D}_s} w\left(H\left(p_i^s\right)\right)
\\
&\log\left[G_d\left(G_f(x_i^s) \otimes G_c\left(G_f(x_i^s)\right)\right)\right]
\\
&- \mathbb{E}_{x_i^t \in \mathcal{D}_t} w\left(H\left(p_i^t\right)\right)
\\
&\log\left[1- G_d\left(G_f(x_i^t) \otimes G_c\left(G_f(x_i^t)\right)\right)\right].
\end{aligned}
\end{equation}

\begin{figure}[!t]
	\centering
	\subfigure{\includegraphics[scale = 0.32]{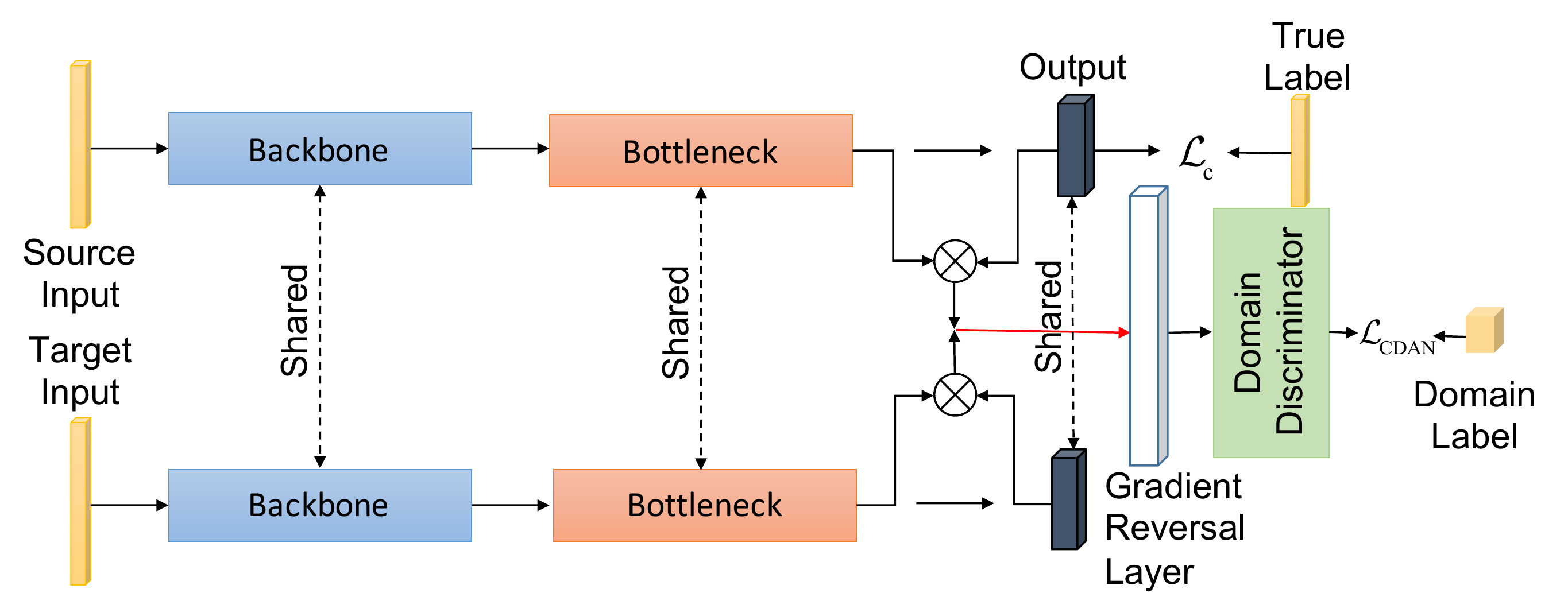}}
	\\ [-10pt]
	\caption{The UDTL-based IFD model based on CDAN.}
	\label{figure-CDAN}
\end{figure}

We design an UDTL-based IFD model via embedding the conditional adversarial idea into the loss function to realize the feature transfer shown in \figref{figure-CDAN}.
Also, the final loss function is defined as follows:
\begin{align}
\label{CDAN-loss}
\mathcal{L}\left(\theta_f,\theta_c,\theta_d\right) = \mathcal{L}_c\left(\theta_f,\theta_c\right) - \lambda_{\text{CDAN}}\mathcal{L}_{\text{CDAN}}\left(\theta_f,\theta_d\right),
\end{align}
where $ \lambda_{\text{CDAN}} $ is a trade-off parameter.

\subsubsection{Applications to IFD}
In \cite{zhang2018adversarial,han2019novel,guo2018deep,wang2019domain,chen2020domain,zou2020adversarial,li2021knowledge,li2021domain}, the feature extractor was pre-trained with the labeled source data and was used to generate target features.
After that, features from source and target domains were trained to maximize the domain discriminator loss, leading to distribution alignment for IFD.
Classifier discrepancy \cite{jiao2019unsupervised, yu2020symmetric,jiao2020double}, which means using separate classifiers for source and target domains, was introduced in UDTL-based IFD via an adversarial training process.
Meanwhile, adversarial training was also combined with other metric distances, such as L1 alignment \cite{li2019deep}, MMD \cite{jia2020novel}, MK-MMD \cite{zhang2020new}, and JMMD \cite{jiao2020residual}, to better match the feature distributions between different domains for IFD.
Li et al. \cite{li2021intelligent} used two feature extractors and classifiers trained using MMD and domain adversarial training, respectively, and meanwhile ensemble learning was further utilized to obtain the final results.
Qin et al. \cite{qin2021multi} proposed a multiscale transfer voting mechanism (MSTVM) to improve the classical domain adaption models and the verified model was trained by MMD and domain adversarial training.
In addition, Qin et al. also proposed the parameter sharing \cite{qin2021parameter} and multiscale \cite{yao2021multiscale} ideologies to reduce the complexity of network structures and extract more domain-invariant features.
The verified models were trained by domain adversarial training embedded with metric distances, like MMD and CORAL.

Wasserstein distance was used in \cite{cheng2020wasserstein, zhang2019deep, wang2020triplet,she2020wasserstein} to guide adversarial training for aligning the discrepancy of distributions for IFD.
Yu et al. \cite{yu2020conditional} combined conditional adversarial DA with a center-based discriminative loss to realize both distribution discrepancy and feature discrimination for locomotive fault diagnosis.
Li et al. \cite{li2021deep} proposed a strategy for bearing fault diagnosis based on minimizing the joint distribution domain-adversarial loss which embedded the pseudo-label information into the adversarial training process.
Besides, another strategy using adversarial-based methods contained adopting GAN to generate samples for the target domain \cite{xie2018transfer,dixit2021intelligent}.

\section{Label-inconsistent UDTL}
\label{S:4}
Considering that the label sets of source and target domains are hard to be consistent in real application, it is significant to study the label-inconsistent UDTL.
In this paper, three label-inconsistent transfer settings, including partial UDTL, open set UDTL, and universal UDTL, are studied.
\subsection{Partial UDTL}
\subsubsection{Basic concepts}
Partial UDTL, which was proposed in \cite{cao2018partial1}, is a transfer learning paradigm where the target label set ${{\cal C}_t}$ is a subspace of the source label set ${{\cal C}_s}$, i.e. ${{\cal C}_t} \subset {{\cal C}_s}$.

\subsubsection{Partial adversarial domain adaptation (PADA)}
One of popular partial UDTL methods named partial adversarial domain adaptation (PADA) was proposed by Cao et. al \cite{cao2018partial2}.
The model of PADA is similar to DANN and further considers that probabilities of assigning target data to the source-private classes would be small, and label predictions on all target data are average to quantify the contribution of each source class.
\begin{align}
\gamma = \frac{1}{n_t}\sum\limits_{i = 1}^{n_t} {y_i^t}.
\end{align}

After normalizing $\gamma$ by its maximum value, $\hat{\gamma}$ is served as a class-level weight:
\begin{align}
\hat{\gamma} = \frac{\gamma}{\max \left( \gamma  \right)}.
\end{align}
Via applying this class-level weight to the loss of the class predictor and domain discriminator, contributions of source samples belonging to source-private classes can be reduced.
The prediction and adversarial losses are rewritten as follows:
\begin{equation}
\begin{aligned}
\mathcal{L}_{c}(\theta_f,\theta_c) = &
- \mathbb{E}_{\left(x_i^s, y_i^s\right) \in \mathcal{D}_s} {\hat{\gamma}_{y_i^s}}
\\
&\sum_{c=0}^{C-1} \mathbf{1}_{[y_i^s=c]} \log\left[G_c\left(G_f\left(x_i^s;\theta_f\right);\theta_c\right)\right].
\end{aligned}
\end{equation}

\begin{equation}
\begin{aligned}
\mathcal{L}_{\text{PADA}}\left(\theta_f,\theta_d\right) =& 
- \mathbb{E}_{x_i^s \in \mathcal{D}_s}{\hat{\gamma}_{y_i^s}} 
\log\left[G_d\left(G_f\left(x_i^s;\theta_f\right);\theta_d\right)\right]
- 
\\
&\mathbb{E}_{x_i^t \in \mathcal{D}_t} 
\log\left[1- G_d\left(G_f\left(x_i^t;\theta_f\right);\theta_d\right)\right],
\end{aligned}
\end{equation}
where $y_i^s$ is the truth label of source sample $x_i^s$, $\hat{\gamma}_{y_i^s}$ is the normalized class weight, and $\lambda_{\text{PADA}}$ is a trade-off parameter.

\begin{figure}[!t]
	\centering
	\subfigure{\includegraphics[scale = 0.32]{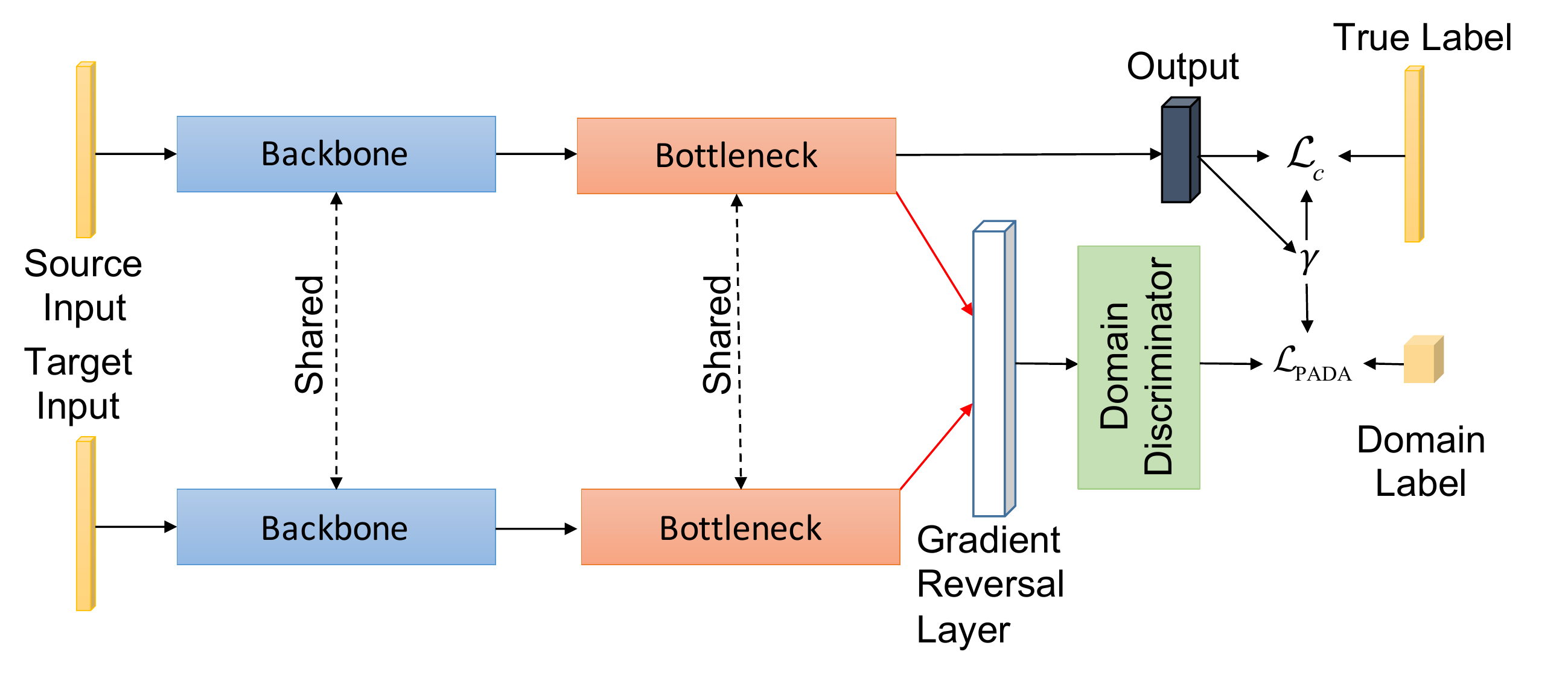}}
	\\ [-10pt]
	\caption{The UDTL-based model based on PADA.}
	\label{figure-PADA}
\end{figure}

 We design an UDTL-based model via applying the class-level weight to the loss function to reduce the influence of outlier source classes as shown in \figref{figure-PADA}. The final loss function is defined as follows:
\begin{align}
\mathcal{L}\left(\theta_f,\theta_c,\theta_d\right) = \mathcal{L}_c\left(\theta_f,\theta_c\right) - \lambda_{\text{PADA}}\mathcal{L}_{\text{PADA}}\left(\theta_f,\theta_d\right).
\end{align}

\subsubsection{Applications to IFD}
In \cite{jiao2019classifier}, two classification networks were constructed and the class-level weights for the source domain were calculated using the target label predictions of two networks, and then weights were applied to the source classification loss to down-weight the influence of outlier source samples.
Li et al. \cite{li2020novel} added weight modules to the adversarial transfer network and a weighting learning strategy was constructed to quantify the transferability of source samples.
Via filtering out irrelevant source samples, the distribution discrepancy across domains in the shared label space could be reduced.
Li et al. \cite{li2020deep} proposed a conditional data alignment technique to align distributions of healthy data and a prediction consistency technique to align the distributions of other classes in two domains.
In \cite{li2020partial}, to facilitate the positive transfer of shared classes and reduce the negative transfer of outlier classes, the average domain prediction loss of each source class was used as the class-level weight.
To avoid the potential negative effect and preserve the inter-class relationships, Wang et al. \cite{wang2020missing} proposed to unilaterally align the target domain to the source domain via adding a consistency loss which forces aligned source features to be close to pre-trained source features.
Deng et al. \cite{deng2021double} constructed sub-domain discriminator for each class to achieve better flexibility.
A double layer attention mechanism was proposed to assign different attentions to sub-domain discriminators and different attentions to samples for selecting relevant samples.
Yang et al. \cite{yang2021deep} proposed to learn domain-asymmetry factors via training a domain discriminator and source samples were weighted in the distribution adaptation to block irrelevant knowledge.

\subsection{Open set UDTL}
\subsubsection{Basic concepts}
Considering that the label space of target domain is uncertain for UDTL, Saito et al. proposed open set domain adaptation (OSDA) that the target domain could contain samples of classes which were absent in the source domain \cite{saito2018open}, i.e. ${{\cal C}_s} \subset {{\cal C}_t}$.
The goal of OSDA is to correctly classify known-class target samples and recognize unknown-class target samples as an additional class.
\subsubsection{Open set back-propagation (OSBP)}
Saito et al. \cite{saito2018open} proposed an adversarial-based UDTL method, named OSBP, which aimed to make a pseudo decision boundary for unknown class.
The model of OSBP is composed of a feature extractor $G_f$ and a $C+1$ classifier $G_c$, where $C$ denotes the number of source classes.
The outputs of $G_c$ are then input into Softmax to obtain class probabilities.
The probability of $x$ being classified into class $c$ is defined as
$p_c^t = \frac{\exp \left( G_c\left( G_f(x) \right) \right)}{\sum_{k=0}^{C} \exp \left( G_k\left( G_f(x) \right) \right)}$.
1$\sim{C}$ and $C+1$ dimensions indicate the probability of known and unknown classes, respectively. 

To correctly classify source samples, the feature extractor and classifier are trained using the prediction loss $\mathcal{L}_{c}$ in \eqref{cross-entropy}.
Moreover, the classifier is trained to recognize target samples as an unknown class via training the classifier to output $p_{C + 1}^t = \tau$, where $\tau$ ranges from 0 to 1.
While the feature extractor is trained to deceive the classifier via training the feature extractor to allow $p_{C + 1}^t$ higher or lower than $\tau$.
In this way, a good boundary between known and unknown target samples can be constructed.
A binary cross-entropy loss is used for the adversarial training:
\begin{align}
{{\cal L}_{\text{OSBP}}}\left( {{\theta _f},{\theta _c}} \right) =  -\tau\log \left( {p_{C + 1}^t} \right) - \left( {1 - \tau} \right)\log \left( {1 - p_{C + 1}^t} \right).
\end{align}

\begin{figure}[!t]
	\centering
	\subfigure{\includegraphics[scale = 0.32]{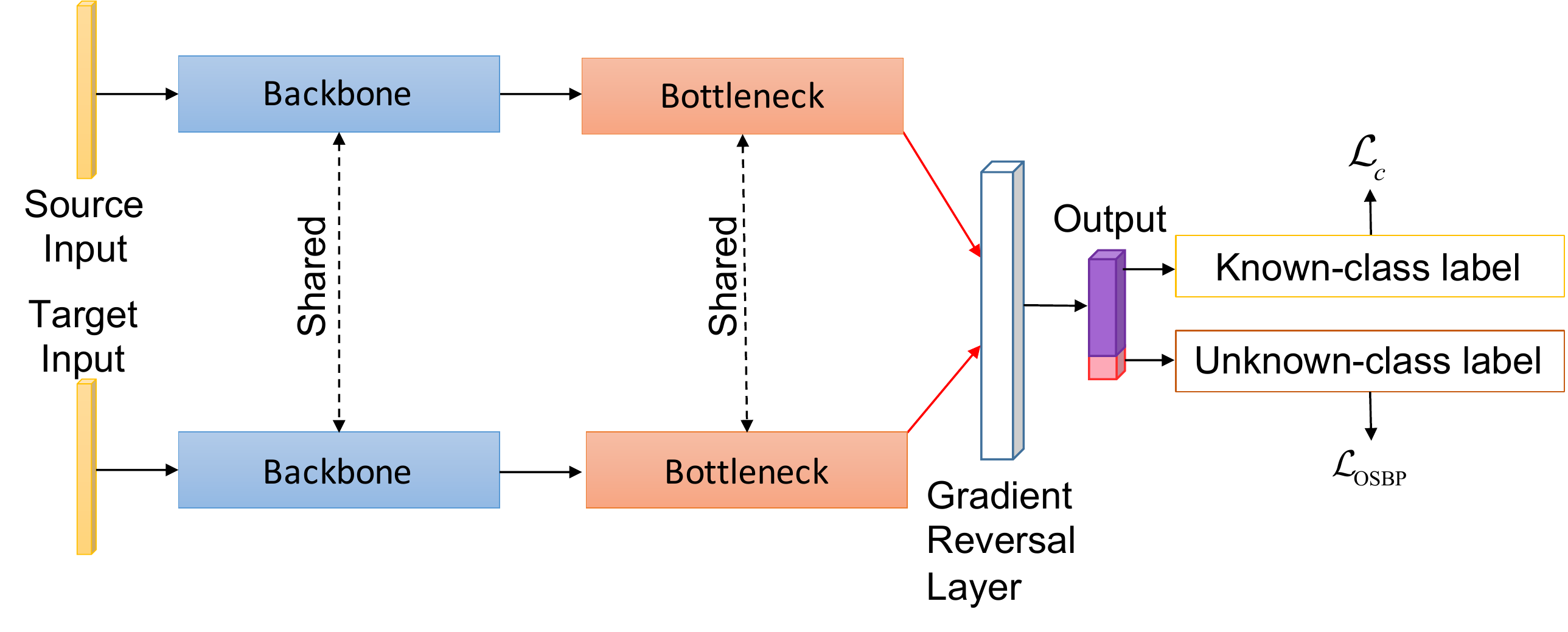}}
	\\ [-10pt]
	\caption{The UDTL-based model based on OSBP.}
	\label{figure-OSBP}
\end{figure}

We design an UDTL-based model via introducing the $C+1$ classifier and adding the adversarial idea to the loss function to make a pseudo decision boundary for the unknown class as shown in \figref{figure-OSBP}.
The saddle point ($ \hat\theta_f,\hat\theta_c$) is solved using the following min-max optimization problem:
\begin{equation}
\begin{aligned}
\left( {\hat {{\theta _f}}} \right) &= \arg \mathop {\max }\limits_{{\theta _f}} {{\cal L}_{\rm{c}}}\left( {{\theta _f},{\hat \theta }_c} \right) - \lambda_{\text{OSBP}}{\cal L}_{\text{OSBP}}\left( \theta _f, {\hat \theta }_c \right),\\
\left( {\hat {{\theta _c}}} \right) &= \arg \mathop {\min }\limits_{{\theta _c}} {{\cal L}_c}\left( {\hat {{\theta _f}},{\theta _c}} \right) + {\lambda _{\text{OSBP}}}{{\cal L}_{\text{OSBP}}}\left( {\hat {{\theta _f}},{\theta _c}} \right).
\end{aligned}
\end{equation}

\subsubsection{Applications to IFD}
Li et al. \cite{li2020two} proposed a new fault classifier to detect the unknown class, and a convolutional auto-encoder model was further built to recognize the number of new fault types in \cite{li2020deep2}.
Zhang et al. \cite{zhang2021open} proposed an instance-level weighted UDTL method to apply similarities of target samples during feature alignment.
To identify target samples with outlier classes, an outlier classifier was trained using target instances with pseudo outlier labels.

\subsection{Universal UDTL}
\subsubsection{Basic concepts}
You et al. \cite{you2019universal} proposed universal domain adaptation (UDA) which imposed no prior knowledge on label sets.
In UDA, for a given source label set and a target label set, they might contain a common label set and hold a private label set, respectively.
UDA requires the model to either classify the target sample correctly if it is associated with a label in the common label set, or mark it as “unknown” otherwise.
Let ${\cal C}_s$ denotes the source label set, ${\cal C}_t$ denotes the target label set, and
${\cal C} = {\cal C}_s \cap {\cal C}_t$ denotes the common label set.
$\overline {\cal C}_s = {\cal C}_s\backslash \cal C$ and $\overline {\cal C} _t = {\cal C}_t\backslash \cal C$ represent the source and target private label sets, respectively.

\subsubsection{Universal adaptation network (UAN)}
You et al. \cite{you2019universal} proposed UAN and designed an instance-level transferability criterion, exploiting the domain similarity and prediction uncertainty.
The model of UAN is similar to that of DANN, while the difference is that UAN adds a non-adversarial domain discriminator $G_d'$.
The non-adversarial domain discriminator $G_d'$ obtains the domain similarity $d'=G_d'(G_f(x))$. They assumed that ${\mathbb{E}_{x\sim{p_{{{\overline {\cal C} }_s}}}}}d' > {\mathbb{E}_{x\sim{p_{_{\cal C}}}}}d' > {\mathbb{E}_{x\sim{q_{_{\cal C}}}}}d' > {\mathbb{E}_{x\sim{q_{{{\overline {\cal C} }_t}}}}}d'$, where ${p_{{{\overline {\cal C} }_s}}}$ is the distribution of source data belonging to the label set ${\overline C _s}$ and ${q_{{{\overline {\cal C} }_t}}}$ is the distribution of target data belonging to label set ${\overline C_t}$.
${p_{\cal C}}$ and ${q_{\cal C}}$ are the distributions of source and target data belonging to ${\cal C}$, respectively.
Considering that entropy can quantify the prediction uncertainty, they assumed that
${\mathbb{E}_{x\sim{q_{{{\overline {\cal C} }_t}}}}}H\left( p \right) > {\mathbb{E}_{x\sim{q_{_{\cal C}}}}}H\left( p \right) > {\mathbb{E}_{x\sim{p_{_{\cal C}}}}}H\left( p \right) > {\mathbb{E}_{x\sim{p_{{{\overline {\cal C} }_s}}}}}H\left( p \right)$.

The instance-level transferability criterion for source and target samples can be defined as follows:
\begin{align}
\omega \left( {x_i^s} \right) = \frac{{H\left( {p\left( {x_i^s} \right)} \right)}}{{\log \left| {{{\cal C}_s}} \right|}} - d'\left( {x_i^s} \right),
\end{align}
\begin{align}
\omega \left( {x_i^t} \right) = d'\left( {x_i^t} \right) - \frac{{H\left( {p\left( {x_i^t} \right)} \right)}}{{\log \left| {{{\cal C}_s}} \right|}},
\end{align}
where $\omega \left( {x_i^s}\right)$ and $\omega \left( {x_i^t}\right)$ indicate the probability of a source sample $x_i^s$ and a target sample $x_i^t$ belonging to the common label set ${\cal C}$.
\begin{figure}[!t]
	\centering
	\subfigure{\includegraphics[scale = 0.32]{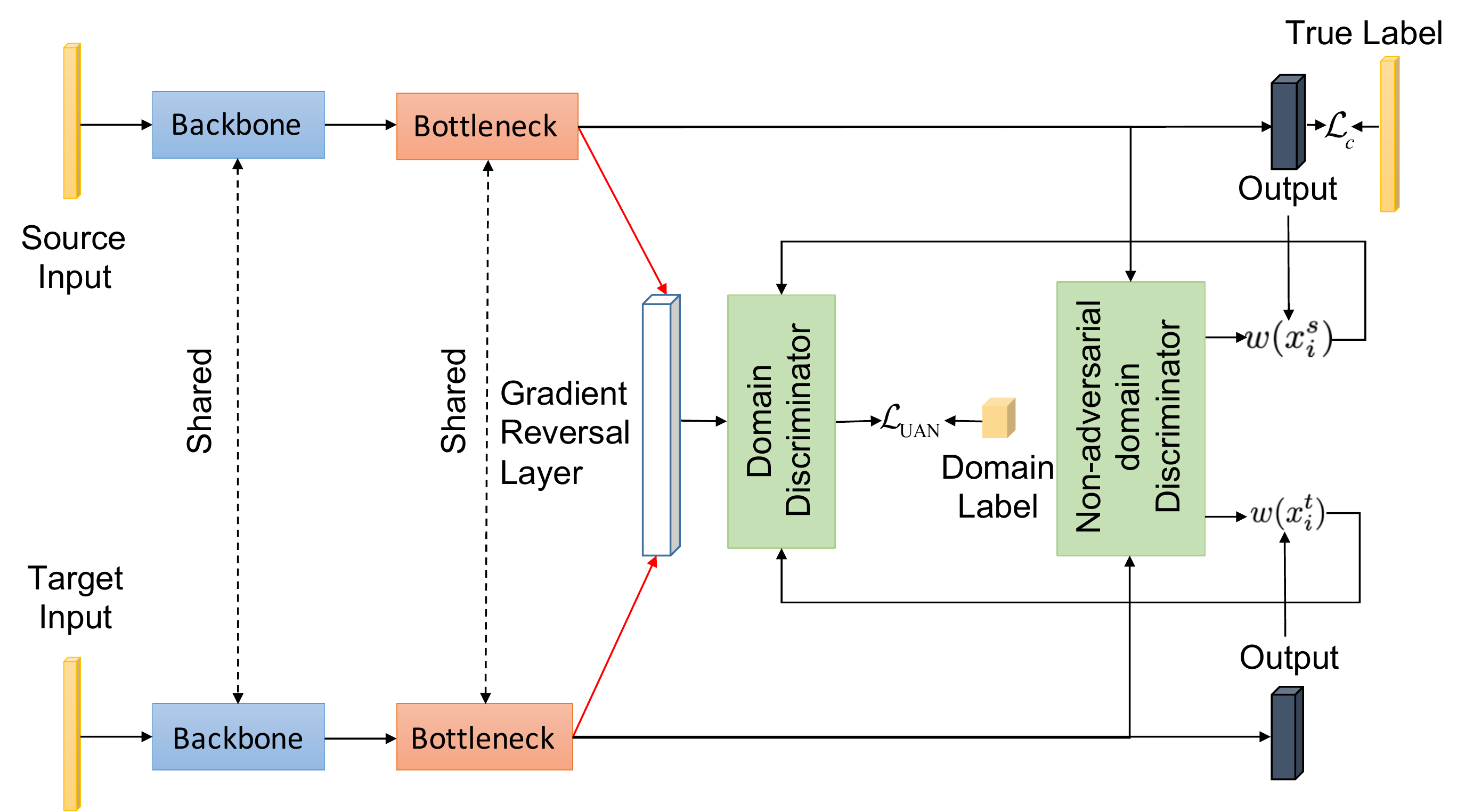}}
	\\ [-10pt]
	\caption{The UDTL-based model based on UAN.}
	\label{UAN}
\end{figure}
The loss of domain discriminator $G_d$ in \eqref{A-binary-loss} is modified to:
\begin{equation}
\begin{aligned}
\mathcal{L}_{\text{UAN}}\left(\theta_f,\theta_d\right) = 
&- \mathbb{E}_{x_i^s \in \mathcal{D}_s} \omega \left( {x_i^s} \right)
\log\left[G_d\left(G_f\left(x_i^s;\theta_f\right);\theta_d\right)\right]
-
\\
&\mathbb{E}_{x_i^t \in \mathcal{D}_t} \omega \left( {x_i^t} \right)
\log\left[1- G_d\left(G_f\left(x_i^t;\theta_f\right);\theta_d\right)\right].
\end{aligned}
\end{equation}

The loss of non-adversarial domain discriminator $G_{d'}$ is:
\begin{equation}
\begin{aligned}
\mathcal{L}_{d'}\left(\theta_f,\theta_{d'}\right)
= &
- \mathbb{E}_{x_i^s \in \mathcal{D}_s}
\log \left[G_{d'}\left(G_f\left(x_i^s;\theta_f\right);\theta_{d'}\right)\right]
- 
\\
&\mathbb{E}_{x_i^t \in \mathcal{D}_t}
\log\left[1 - G_{d'}\left(G_f\left(x_i^t;\theta_f\right);\theta_{d'}\right)\right].
\end{aligned}
\end{equation}
where $\theta_{d'}$ is the parameters of non-adversarial domain discriminator $G_{d'}$.
The saddle point ($ \hat\theta_f,\hat\theta_c,\hat\theta_d,\hat\theta_{d'} $) can be solved using the following min-max optimization problem:
\begin{equation}
\begin{aligned}
\left( \hat {\theta_f}, \hat {\theta_c} \right) 
&= 
\arg \min \limits_{\theta_f, \theta_c} {\cal L}_c \left( \theta_f, \theta_c \right) - \lambda_{\text{UAN}}{\cal L}_{\text{UAN}}\left( \theta_f, {\hat \theta}_d \right),
\\
\left( \hat {\theta_d} \right) 
&= 
\arg \min \limits_{\theta_d} {\cal L}_{\text{UAN}} \left({\hat \theta}_f, {\hat \theta}_c, {\theta_d} \right),
\\
\left( \hat {\theta_{d'}} \right) 
&= 
\arg \min \limits_{\theta_{d'}} {\cal L}_{d'} \left( {\hat \theta}_f, \theta_{d'} \right).
\end{aligned}
\end{equation}

Via training UAN, distributions of source and target data in the shared label set can be maximally aligned and the category gap can be reduced. 
In the test phase, for a target sample $x_i^t$, if its $\omega(x_i^t)$ is higher than the threshold $\omega_0$, it is regarded as the unknown class, otherwise it is predicted by its label prediction. 

\subsubsection{Applications to IFD}
Zhang et al. \cite{zhang2021universal} proposed a selective UDTL method. Class-wise weights were applied to the source domain and instance-wise weights were applied to the target domain.
An outlier identifier was trained to recognize unknown fault modes.
Yu et al. \cite{9394793} proposed a bilateral weighted adversarial network to align feature distributions of shared-class source and target samples, and to disentangle shared-class and outlier-class samples.
After model training, the extreme value theory (EVT) model was established on the feature representation of source samples and was further used to detect unknown-class samples in the target domain.
\section{Multi-domain UDTL}
\label{S:5}
Considering that a single source domain might not be enough for UDTL in real applications, it is also important to consider multi-domain UDTL, which can help learn domain-invariant features.
In this paper, two kinds of multi-domain UDTL settings, including multi-domain adaptation (using the target data in the training phase) and domain generalization (not using the target data in the training phase), are studied.
Because there are multiple source domains in multi-domain UDTL, we first need to redefine some basic symbols.
Let $\left\{ \mathcal{D}_{s,n} \right\}_{0}^{n_{sd}-1}$ denote the source domains, where $n_{sd}$ denotes the number of source domains. 
$\mathcal{D}_t$ denotes the target domain.
$\mathcal{D}_{s,n}$ means the $n$-th source domain.
$x_{i,n}^{s}$ and $y_{i,n}^{s}$ are the $i$-th sample and its corresponding label. 
Besides, $d_{i,n}^{s}$ is the domain label of $x_{i,n}^{s}$ and $d_i^t$ ($d_{i}^t={n}_{sd}$) is the domain label of $x_i^t$. 

\subsection{Multi-domain adaptation}
\subsubsection{Basic concepts}
The traditional UDTL based on one single source domain cannot make full use of the data from multi-source domains, which might fail to find the private relationship and domain-invariant features.
Thus, multi-domain adaptation aims to utilize labeled multi-source domains and unlabeled target domains to dig the relationship and domain-invariant features.

\subsubsection{Multi-source unsupervised adversarial domain adaptation (MS-UADA)}
\begin{figure}[!t]
	\centering
	\subfigure{\includegraphics[scale = 0.35]{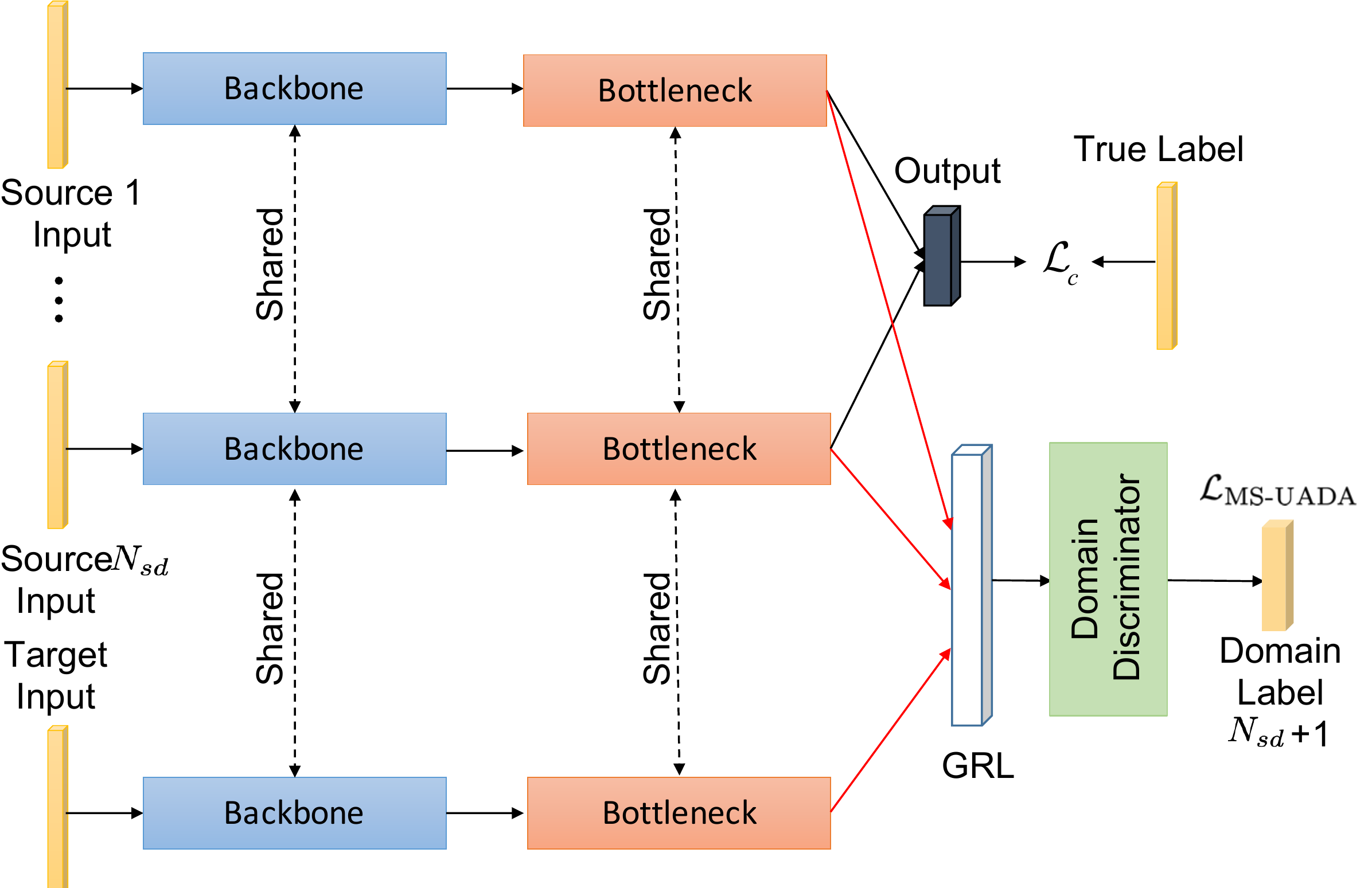}}
	\\ [-10pt]
	\caption{The UDTL-based model based on MS-UADA.}
	\label{multi_DA_structure}
\end{figure}

There are mainly two ways to realize multi-domain adaptation.
One is that features should be domain-invariant \cite{dai2020adversarial}, that is, the gap between different domains, including source and target domains in the feature space should be as small as possible.
The other way is to find a source domain, which is the most similar to the target domain \cite{zhao2020multi,ANewPenaltyYAN}.
The second way requires a distance to measure the similarity among domains.
In this paper, we used the method proposed in \cite{dai2020adversarial} called multi-source unsupervised adversarial adaptation (MS-UADA) to realize multi-domain adaptation, and the structure is shown in \figref{multi_DA_structure}.
The loss of MS-UADA for the domain discriminator $G_d$ is defined as follows: 
\begin{align}
\label{ad_MDA}
\begin{split}
&\mathcal{L}{_{\text{MS-UADA}}}= -{{\mathbb{E}}_{x_{i}^{t}\in {\mathcal{D}_t}}} \mathbf{1}_{[d_i^t=n_{sd}]} \log \left[G_d\left(G_f\left(x_i^t;\theta_f\right);\theta_d\right)\right]
\\&-\sum\limits_{n=0}^{{{n}_{sd}}-1}{{\mathbb{E}}_{x_{i,n}^{s}\in {\mathcal{D}_{s,n}}}}
\sum\limits_{d=0}^{{{n}_{sd}}-1}{{\mathbf{1}_{[ d_{i,n}^{s}=d ]}} \log \left[G_d\left(G_f\left(x_{i,n}^s;\theta_f\right);\theta_d\right)\right]}.
\end{split}
\end{align}
To reduce the gap, the features from ${{G}_{f}}$ should confuse ${{G}_{d}}$, which means ${{G}_{d}}$ cannot realize domain classification.
Thus the training processing can be seen as a minimax game, and the total loss is:
\begin{align}
\label{total_loss_MDA}
\mathcal{L}\left( {{\theta }_{f}},{{\theta }_{c}},{{\theta }_{d}} \right)=\mathcal{L}_c-{{\lambda }_{\text{MS-UADA}}}\mathcal{L}{_{\text{MS-UADA}}},
\end{align}
where ${{\lambda }_{\text{MS-UADA}}}$ is the trade-off parameter.
The way to optimize $\mathcal{L}\left( {{\theta }_{f}},{{\theta }_{c}},{{\theta }_{d}} \right)$ is consistent with the DANN.

\subsubsection{Applications to IFD}
Zhu et al. \cite{ANewMultipleSourceDomainAdaptation} proposed an adversarial learning strategy in multi-domain adaptation to capture the fault feature representation.
Rezaeianjouybari et al. \cite{REZAEIANJOUYBARI2021109359} proposed a novel multi-source adaptation framework, which could realize the alignment in both feature and task levels.
Zhang et al. \cite{Zhang_2020} proposed an adversarial multi-domain adaptation according to a classifier alignment method to capture domain-invariant features from multiple source domains.
He et al. \cite{He2020access} proposed a method based on K-means and space transformation in multi-source domains.
Wei et al. \cite{WEI2021107744} proposed a multi-source adaptation framework to learn domain-invariant features on the basis of distributional similarities.
Zhang et al.  \cite{zhang2020unsupervised} proposed an enhanced transfer joint matching approach based on MVD and MMD for multi-domain adaptation.
Li et al. \cite{DiagnosingRotating2020} proposed a multi-domain adaptation method to learn the diagnostic knowledge via domain adversarial training.
Huang et al. \cite{huang2021multi} proposed a multi-source dense adaptation adversarial network to realize fault diagnosis from various working conditions.

\subsection{Domain generalization (DG)}
\subsubsection{Basic concepts}
Domain generalization (DG) is to learn shared knowledge from multiple source domains and generalize knowledge to the target domain which is unseen in the training phase.
The biggest difference of DG is that unlabeled samples in the target domain only appear in the test phase.
Based on the discussion in \cite{DG2018CVPR}, the core idea of DG is that learned domain-invariant features should satisfy the following two properties: 1) Features extracted by ${{G}_{f}}$ should be discriminative.
2) Features extracted from different source domains should be domain-invariant.
More detailed information can be referred to \cite{DG2018CVPR}.

\subsubsection{Invariant adversarial network (IAN)}
According to the above description, the performance of DG depends on discriminative and domain-invariant features.
Domain-invariant features require diagnosis models to reduce the feature gap among different domains.
As described in the previous section, adversarial training can reduce the gap of features among different domains.
In this paper, a simple adversarial training method called invariant adversarial network (IAN) \cite{LI2020409,DG2018ECCV} based on DANN is used in DG to help ${{G}_{f}}$ extract domain-invariant features via aligning the marginal distribution.
The structure of IAN is shown in \figref{multi_DA_structure} and the loss of IAN for the domain discriminator $G_d$ is defined as follows:
\begin{align}
\label{ad_DG}
\begin{split}
\mathcal{L}_{\text{IAN}}
=&
-\sum\limits_{n=0}^{{{n}_{sd}}-1}
{{\mathbb{E}}_{x_{i,n}^{s}\in {\mathcal{D}_{s,n}}}}
\sum\limits_{d=0}^{{{n}_{sd}}-1}
\\&{{\mathbf{1}_{[ d_{i,n}^{s}=d ]}} \log \left[G_d\left(G_f\left(x_{i,n}^s;\theta_f\right);\theta_d\right)\right]}.
\end{split}
\end{align}
${{G}_{f}}$ should confuse ${{G}_{d}}$.
Thus the total loss of IAN is a minimax game:
\begin{align}
\label{DG_final_loss}
\mathcal{L}\left( \theta_f,\theta_c,\theta_d \right)=\mathcal{L}_c-\lambda_\text{IAN}\mathcal{L}_\text{IAN},
\end{align}
where $\lambda_\text{IAN}$ is the trade-off parameter.
The way to optimize $\mathcal{L}\left(\theta_f,\theta_c,\theta_d \right)$ is consistent with DANN.
\begin{figure}[!t]
	\centering	\subfigure{\includegraphics[scale = 0.35]{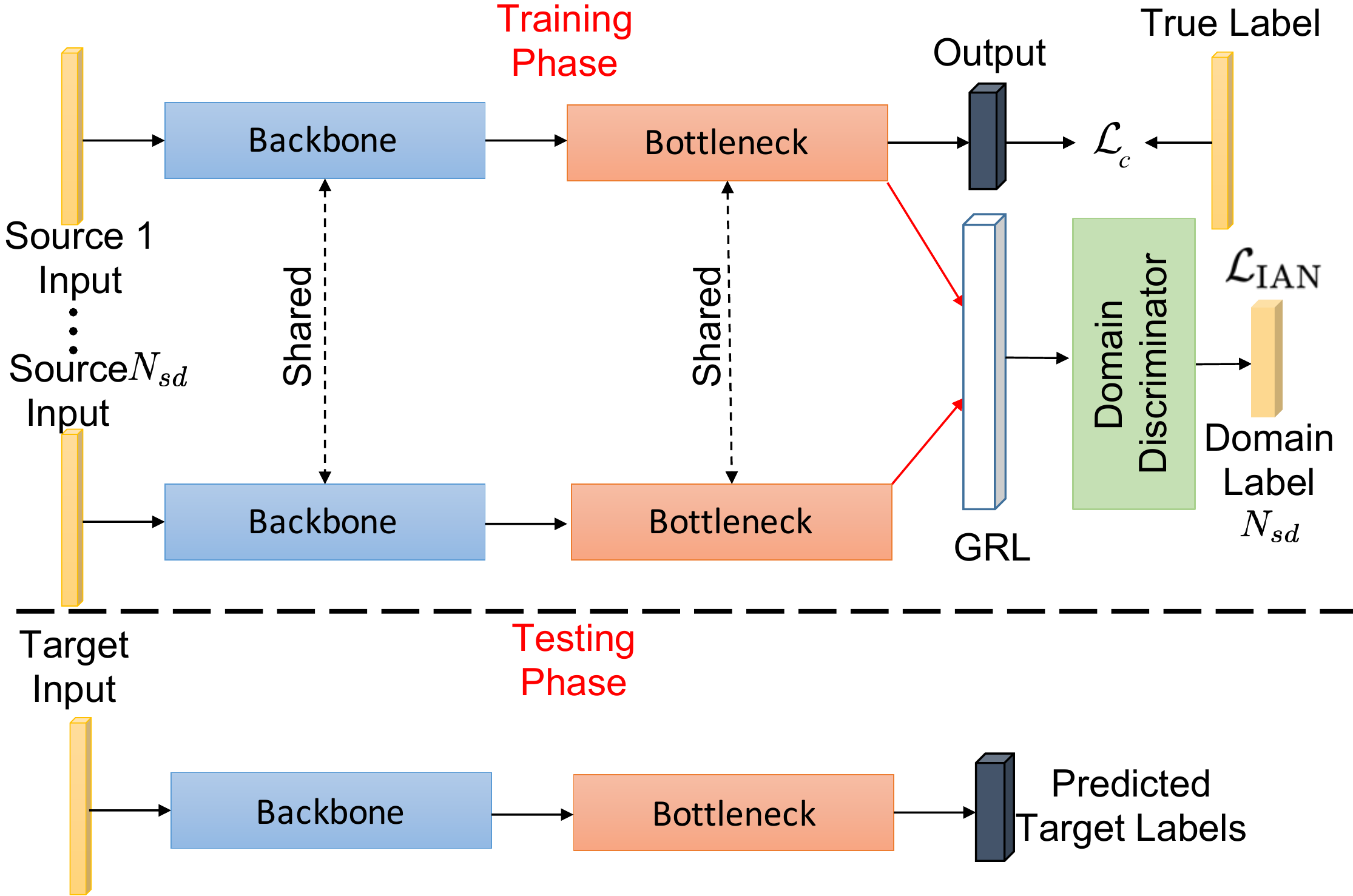}}
	\\ [-10pt]
	\caption{The UDTL-based model based on IAN.}
	\label{DGstructure}
\end{figure}

\subsubsection{Applications to IFD}
Zheng et al. \cite{DGPriori} proposed a DG network based on a \textit{priori} diagnosis knowledge and preprocessing techniques for IFD.
Liao et al. \cite{liao2020deep} proposed a deep semi-supervised DG network to use unlabeled and labeled source data by the Earth-mover distance.
Li et al. \cite{LI2020409} proposed a DG method in IFD by a combination of data augmentation, adversarial training, and distance-based metric.
Yang et al. \cite{centerlossDG2020} used a center loss to learn domain-invariant features across various source domains to realize DG.
Zhang et al. \cite{zhang2021ConditionalAdversarialDG} proposed a conditional adversarial DG method based on a single discriminator for better transfer and low computational complexity.
Han et al. \cite{han2021hybrid} proposed a DG based hybrid diagnosis network for deploying to unseen working conditions via the triplet loss and adversarial training.

\section{Datasets}
\label{S:6}
\subsection{Open source Datasets}
Open source datasets are very important for development, comparisons, and evaluation of different algorithms.
In this comparative study, we mainly test five datasets to verify the performance of different UDTL methods.
The detailed description of five datasets is given as follows: 
\subsubsection{Case Western Reserve University (CWRU) dataset}
The CWRU dataset provided by Case Western Reserve University Bearing Data Center \cite{CWRU} is one of the most famous open source datasets in IFD and has been already used by tremendous published papers.
Following other papers, this paper also uses the drive end bearing fault data whose sampling frequency is equal to 12 kHz and ten bearing conditions are listed in Table \ref{Tab_1}.
In Table \ref{Tab_1}, one normal bearing (NA) and three fault types including inner fault (IF), ball fault (BF) and outer fault (OF) are classified into ten categories (one health state and nine fault states) according to different fault sizes.

Besides, as shown in Table \ref{Tab_2}, CWRU consists of four motor loads corresponding to four operating speeds.
For the transfer learning task, this paper considers these working conditions as different tasks including 0, 1, 2, and 3.
For example, task 0 $ \longrightarrow $ 1 means that the source domain with a motor load 0 HP transfers to the target domain with a motor load 1 HP.
In total, there are twelve transfer learning tasks.

\begin{table}[!t]
	\caption{The description of class labels of CWRU.} 
	\centering
	\label{Tab_1}  
	\begin{tabular}{cccccc}
		\hline
		Class Label       & 0  & 1  & 2  & 3  & 4  \\ \hline
		Fault Location    & NA & IF & BF & OF & IF \\ \hline
		Fault Size (mils) & 0  & 7  & 7  & 7  & 14  \\ \hline
		Class Label       & 5  & 6  & 7  & 8  & 9  \\ \hline
		Fault Location    & BF & OF & IF & BF & OF \\ \hline
		Fault Size (mils) & 14 & 14 & 21 & 21 & 21 \\ \hline
	\end{tabular}
\end{table}

\begin{table}[!t]
	\caption{The transfer learning tasks of CWRU.} 
	\centering
	\label{Tab_2}  
	\begin{tabular}{ccccc}
		\hline
		Task        & 0    & 1    & 2    & 3    \\ \hline
		Load (HP)   & 0    & 1    & 2    & 3    \\ \hline
		Speed (rpm) & 1797 & 1772 & 1750 & 1730 \\ \hline
	\end{tabular}
\end{table}

\subsubsection{Paderborn University (PU) dataset}
The PU dataset acquired from Paderborn University is a bearing dataset \cite{lessmeier2016condition,PU} which consists of artificially induced and real damages.
The sampling frequency is equal to 64 kHz.
Via changing the rotating speed of the drive system, the radial force onto the test bearing and the load torque on the drive train, the PU dataset consists of four operating conditions as shown in Table \ref{PU_Tab_1}. 

Thirteen bearings with real damages caused by accelerated lifetime tests \cite{lessmeier2016condition} are used to study transfer learning tasks among different working conditions (twenty experiments were performed on each bearing code, and each experiment sustained four seconds).
The categorization information is presented in Table \ref{PU_Tab_2} (the meaning of contents is explained in \cite{lessmeier2016condition}).
In total, there are twelve transfer learning settings.

\begin{table}[!t]
	\caption{The transfer learning tasks and operating parameters of PU.} 
	\centering
	\label{PU_Tab_1}  
	\begin{tabular}{ccccc}
		\hline
		Task        & 0    & 1    & 2    & 3    \\ \hline
		Load Torque (Nm)   & 0.7    & 0.7    & 0.1    & 0.7    \\ \hline
		Radial Force (N)   & 1000    & 1000    & 1000    & 400    \\ \hline
		Speed (rpm) & 1500 & 900 & 1500 & 1500 \\ \hline
	\end{tabular}
\end{table}

\begin{table*}[!t]
	\caption{The information of bearings with real damages.} 
	\centering
	\label{PU_Tab_2}  
	\begin{tabular}{cccccc}
	\hline
	\begin{tabular}[c]{@{}c@{}}Bearing\\ Code\end{tabular}  & Damage                        & Bearing Element  & Combination & Characteristic of Damage & Label \\ \hline
	KA04                                                    & fatigue: pitting              & OR               & S           & single point             & 0     \\ \hline
	KA15                                                    & plastic deform: indentations  & OR               & S           & single point             & 1     \\ \hline
	KA16                                                    & fatigue: pitting              & OR               & R           & single point             & 2     \\ \hline
	KA22                                                    & fatigue: pitting              & OR               & S           & single point             & 3     \\ \hline
	KA30                                                    & plastic deform: indentations  & OR               & R           & distributed              & 4     \\ \hline
	KB23                                                    & fatigue: pitting              & IR(+OR)          & M           & single point             & 5     \\ \hline
	KB24                                                    & fatigue: pitting              & IR(+OR)          & M           & distributed              & 6     \\ \hline
	KB27                                                    & plastic deform: indentations  & OR+IR            & M           & distributed              & 7     \\ \hline
	KI14                                                    & fatigue: pitting              & IR               & M           & single point             & 8     \\ \hline
	KI16                                                    & fatigue: pitting              & IR               & S           & single point             & 9     \\ \hline
	KI17                                                    & fatigue: pitting              & IR               & R           & single point             & 10    \\ \hline
	KI18                                                    & fatigue: pitting              & IR               & S           & single point             & 11    \\ \hline
	KI21                                                    & fatigue: pitting              & IR               & S           & single point             & 12    \\ \hline
	\multicolumn{6}{l}{\begin{tabular}[c]{@{}l@{}}OR: outer ring; IR: inner ring;\\ S: single damage; R: repetitive damage; M: multiple damages\end{tabular}} \\ \hline
	\end{tabular}
\end{table*}

\subsubsection{JiangNan University (JNU) dataset}
The JNU dataset is a bearing dataset acquired by Jiang Nan University, China.
JNU can be downloaded from \cite{JNU} and scholars can refer to \cite{li2013sequential} for more detailed information.
Four kinds of health conditions, including NA, IF, OF, and BF, were carried out.
Vibration signals were sampled under three rotating speeds (600 rpm, 800 rpm, and 1000 rpm) with the sampling frequency 50 kHz.
Four rotating speeds set to be 600 rpm, 800 rpm, and 1000 rpm are considered as different tasks denoted as task 0, 1, and 2.
In total, there are six transfer learning settings. 

\subsubsection{PHM Data Challenge on 2009 (PHM2009) dataset}
The PHM2009 dataset is a generic industrial gearbox dataset provided by the PHM Data Challenge competition \cite{PHM}.
The sampling frequency is set to 200 KHz/3.
Fourteen experiments (eight for spur gears and six for helical gears) were performed.

In this paper, we utilize the helical gears dataset (six conditions) collected from accelerometers mounted on input shaft retaining plates.
PHM2009 contains five rotating speeds and two loads, but only data collected from the former four shaft speeds under a high load are considered.
Four rotating speeds set to be 30 Hz, 35 Hz, 40 Hz, and 45 Hz are considered as different tasks denoted as task 0, 1, 2, and 3.
In total, there are twelve transfer learning settings.

\subsubsection{Southeast University (SEU) dataset}
The Southeast University (SEU) dataset is a gearbox dataset provided by Southeast University, China \cite{shao2018highly,SEU}.
This dataset consists of two sub-datasets, including the bearing and gear datasets, which were both collected from Drivetrain Dynamics Simulator.
Eight channels were collected, and we use the data from the channel 2.
As shown in Table \ref{SEU_Tab_1}, each sub-dataset consists of five conditions: one health state and four fault states.

Two kinds of working conditions with rotating speed - load configuration set to be 20 Hz - 0 V and 30 Hz - 2 V are considered as different tasks denoted as task 0 and 1.
In total, there are two transfer learning settings.

\begin{table}[!t]
	\caption{The transfer learning tasks of SEU.} 
	\centering
	\label{SEU_Tab_1}  
	\begin{tabular}{cccc}
		\hline
		Label        & Location & Type                    & Description                                   \\ \hline
		\multirow{2}{*}{0} & Gear     & \multirow{2}{*}{Health} & \multirow{2}{*}{}                             \\ \cline{2-2}
		& Bearing  &                         &                                               \\ \hline
		1                  & Bearing  & Ball                    & Crack in the ball                      \\ \hline
		2                  & Bearing  & Outer                   & Crack in the outer ring                \\ \hline
		3                  & Bearing  & Inner                   & Crack in the inner ring                \\ \hline
		4                  & Bearing  & Combination             & Crack in the inner and outer rings \\ \hline
		5                  & Gear     & Chipped                 & Crack in the gear feet                 \\ \hline
		6                  & Gear     & Miss                    & Missing feet in the gear           \\ \hline
		7                  & Gear     & Surface                 & Wear in the surface of the gear        \\ \hline
		8                  & Gear     & Root                    & Crack in the root of the gear feet     \\ \hline
	\end{tabular}
\end{table}

\subsection{Data preprocessing and splitting}
Data preprocessing and splitting are two important aspects in terms of performance of UDTL-based IFD.
Although UDTL-based methods often possess automatic feature learning capabilities, some data processing steps can help models achieve better performance, such as Short-time Fourier Transform (STFT) in speech signal classification and the data normalization in image classification.
Besides, there often exist some pitfalls in the training phase, especially test leakage.
That is, test samples are unheedingly used in the training phase.

\subsubsection{Input types}
There are two kinds of input types tested in this paper, including the time domain input and the frequency domain input.
For the former one, signals are used as the input directly and the sample length is 1024 without any overlapping.
For the latter one, signals are first transformed into the frequency domain and the sample length is 512 due to the symmetry of spectral coefficients.

\subsubsection{Normalization}
Data normalization is the basic procedure in UDTL-based IFD, which can keep input values into a certain range.
In this paper, we use the Z-score normalization.

\subsubsection{Data splitting}
\begin{figure}[!t]
	\centering
	\subfigure{\includegraphics[scale = 0.3]{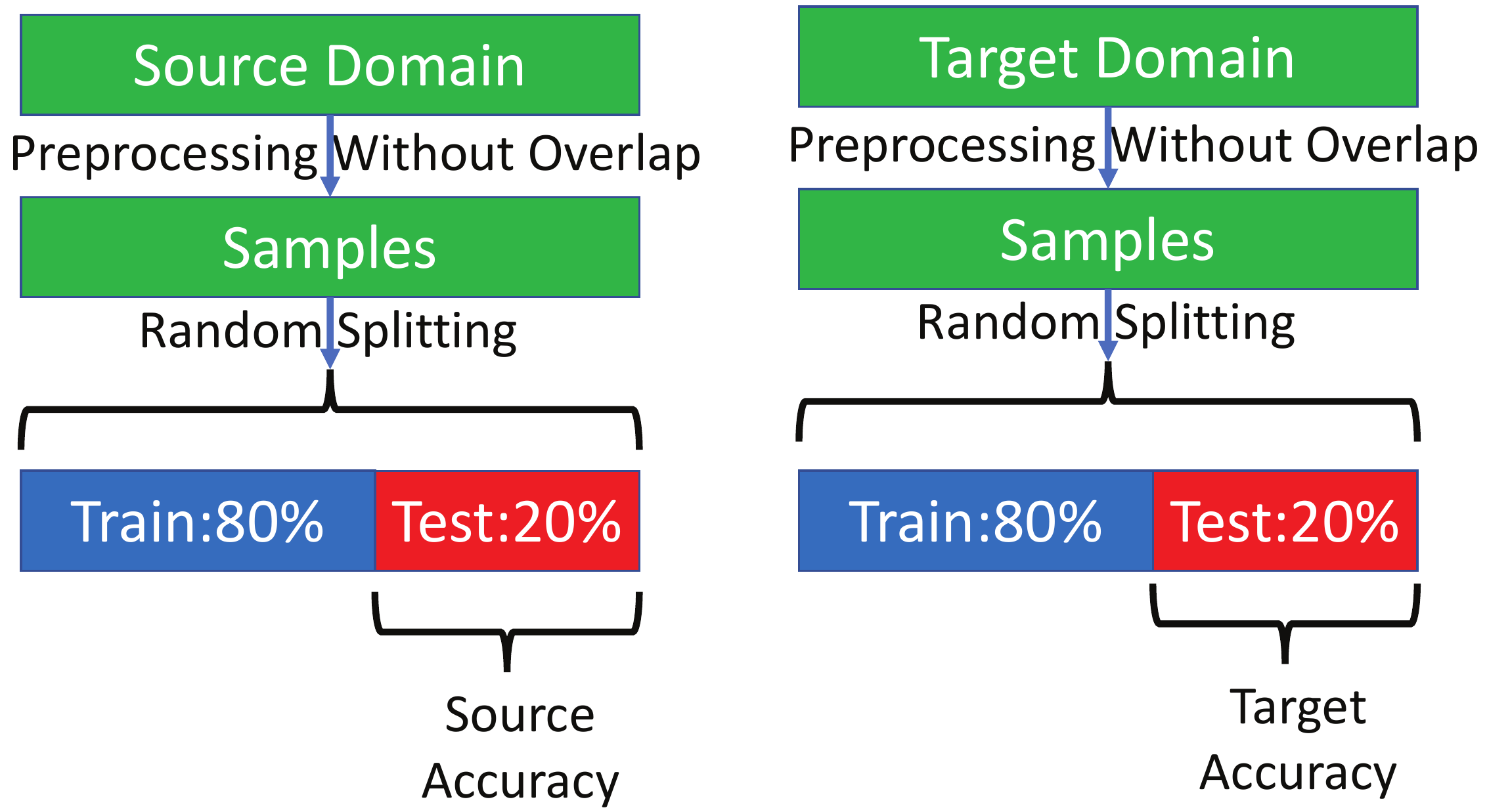}}
	\\ [-5pt]
	\caption{Data splitting for UDTL-based IFD.}
	\label{Fig-data-split}
\end{figure}

Since this paper does not use the validation set to select the best model, the splitting of the validation set is ignored here.
In UDTL-based IFD, data in the target domain are used in the training procedure to realize the domain alignment and are also used as the test sets.
In fact, data in these two situations should not overlap, otherwise there would exist test leakage.
Therefore, as shown in \figref{Fig-data-split}, we take 80\% of total samples as the training set and 20\% of total samples as the test set in source and target domains to avoid this test leakage.

\section{Comparative studies}
\label{S:7}
We will discuss evaluation results, which are shown in \textbf{Appendix A}.
To make the accuracy more readable, we use some visualization methods to present the results.
\subsection{Training details}
We implement all UDTL-based IFD methods in \textbf{Pytorch} and put them into a unified code framework.
Each model is trained for 300 epochs and model training and test processes are alternated during the training procedure.
We adapt mini-batch Adam optimizer and the batch size is equal to 64.
The ``step'' strategy in \textbf{Pytorch} is used as the learning rate annealing method and the initial learning rate is 0.001 with a decay (multiplied by 0.1) in the epoch 150 and 250, respectively.
We use a progressive training method increasing the trade-off parameter from 0 to 1 via multiplying by $ \frac{1-\exp(-\zeta \kappa)}{1+\exp(-\zeta \kappa)} $ \cite{long2018conditional}, where $ \zeta = 10 $ and $ \kappa $ means the training progress changing from 0 to 1 after transfer learning strategies are activated.

All experiments are executed under Window 10 and \textbf{Pytorch 1.3} running on a computer with an Intel Core i7-9700K, GeForce RTX 2080Ti, and 16G RAM.

\subsection{Label-consistent UDTL}
For MK-MMD, JMMD, CORAL, DANN, and CDAN, we train models with source samples in the former 50 epochs to get a so-called pre-trained model, and then transfer learning strategies are activated.
For AdaBN, we update the statistics of BN layers via each batch for 3 extra epochs.
\subsubsection{Evaluation metrics}
For simplicity, we use the overall accuracy, which is the number of correctly classified samples divided by the total number of samples in test data, to verify the performance of different models.
To avoid the randomness, we perform experiments five times, and mean as well as maximum values of the overall accuracy are used to evaluate the final performance because variance of five experiments is not statistically useful.
In this paper, we use mean and maximum accuracy in the last epoch denoted as Last-Mean and Last-Max to represent the test accuracy without any test leakage.
Meanwhile, we also list mean and maximum accuracy denoted as Best-Mean and Best-Max in the epoch where models achieve the best performance.

\subsubsection{Results of datasets}
\begin{figure}[!t]
	\centering
	\subfigure{\includegraphics[scale = 0.6]{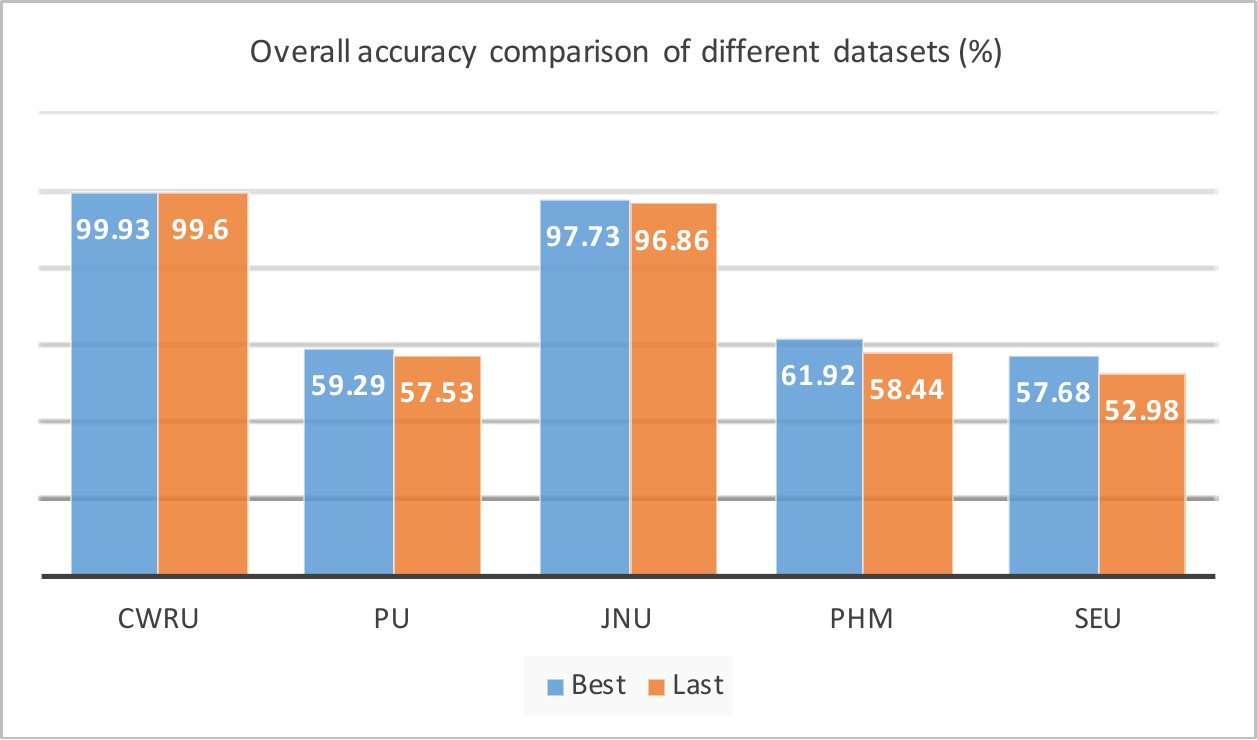}}
	\\ [-10pt]
	\caption{The highest average accuracy of different datasets among all methods.}
	\label{Datasets}
\end{figure}

To make comparisons clearer, we summarize the highest average accuracy of different datasets among all methods, and results are shown in \figref{Datasets}.
We can observe that CWRU and JNU can achieve the accuracy over 95\% and other datasets can only achieve an accuracy of around 60\%.
It is also worth mentioning that the accuracy is just a lower bound due to the fact that it is very hard to fine-tune every parameter in detail.

\subsubsection{Results of models}
\begin{figure*}[!t]
	\centering
	\subfigure{\includegraphics[scale = 0.7]{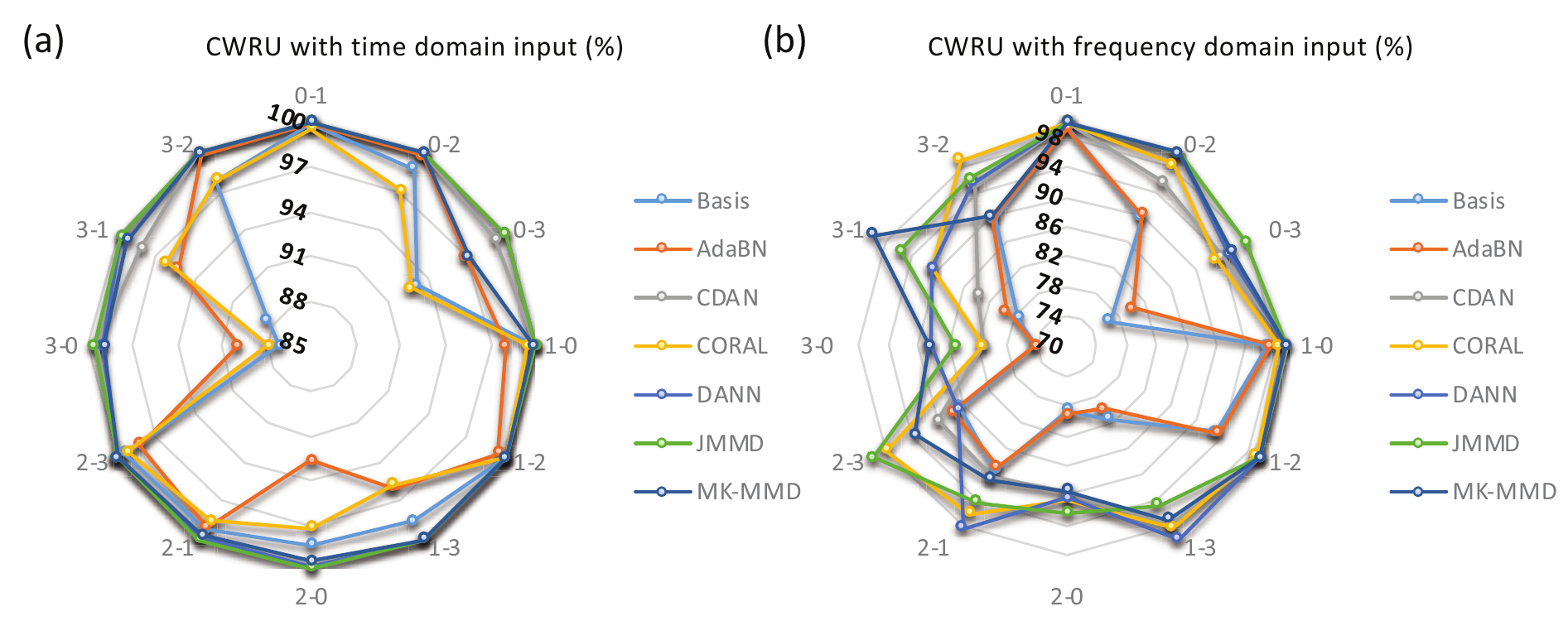}}
	\\ [-10pt]
	\caption{The accuracy comparisons of different methods in CWRU.}
	\label{Models-CWRU}
\end{figure*}

\begin{figure*}[!t]
	\centering
	\subfigure{\includegraphics[scale = 0.7]{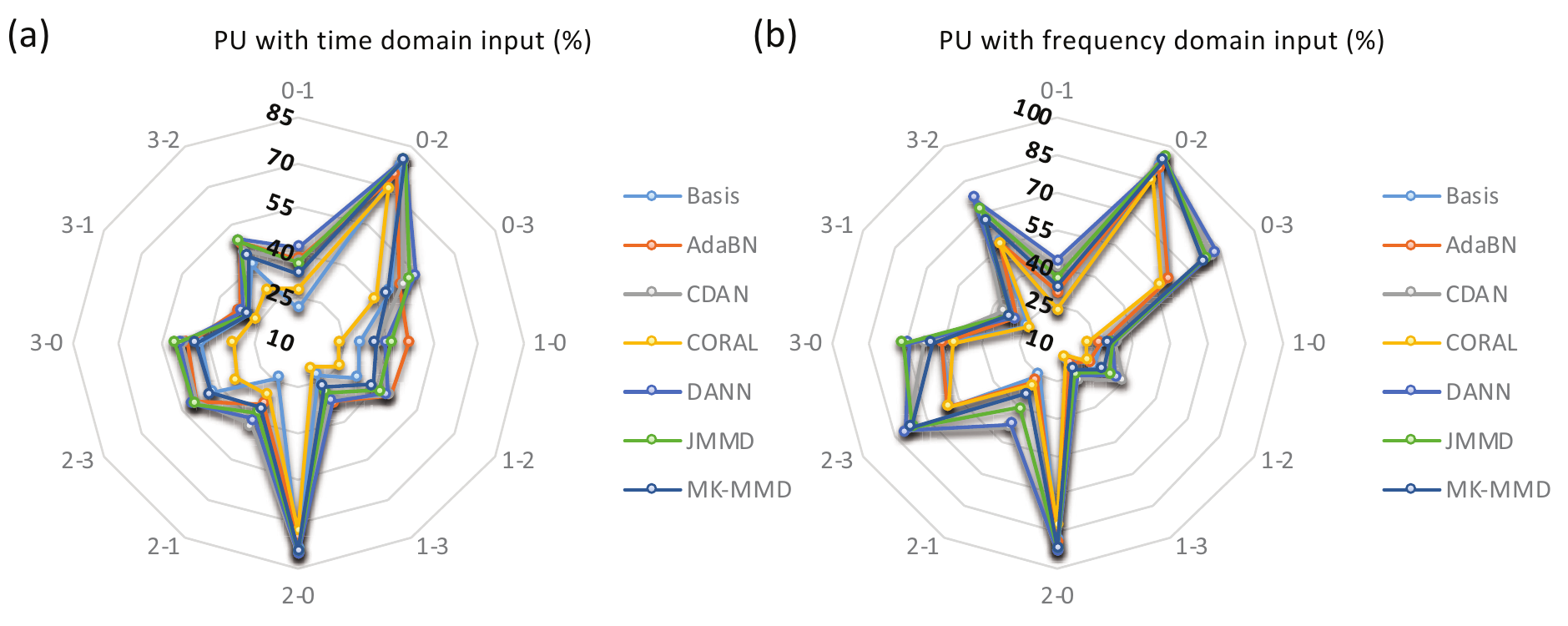}}
	\\ [-10pt]
	\caption{The accuracy comparisons of different methods in PU.}
	\label{Models-PU}
\end{figure*}

\begin{figure*}[!t]
	\centering
	\subfigure{\includegraphics[scale = 0.7]{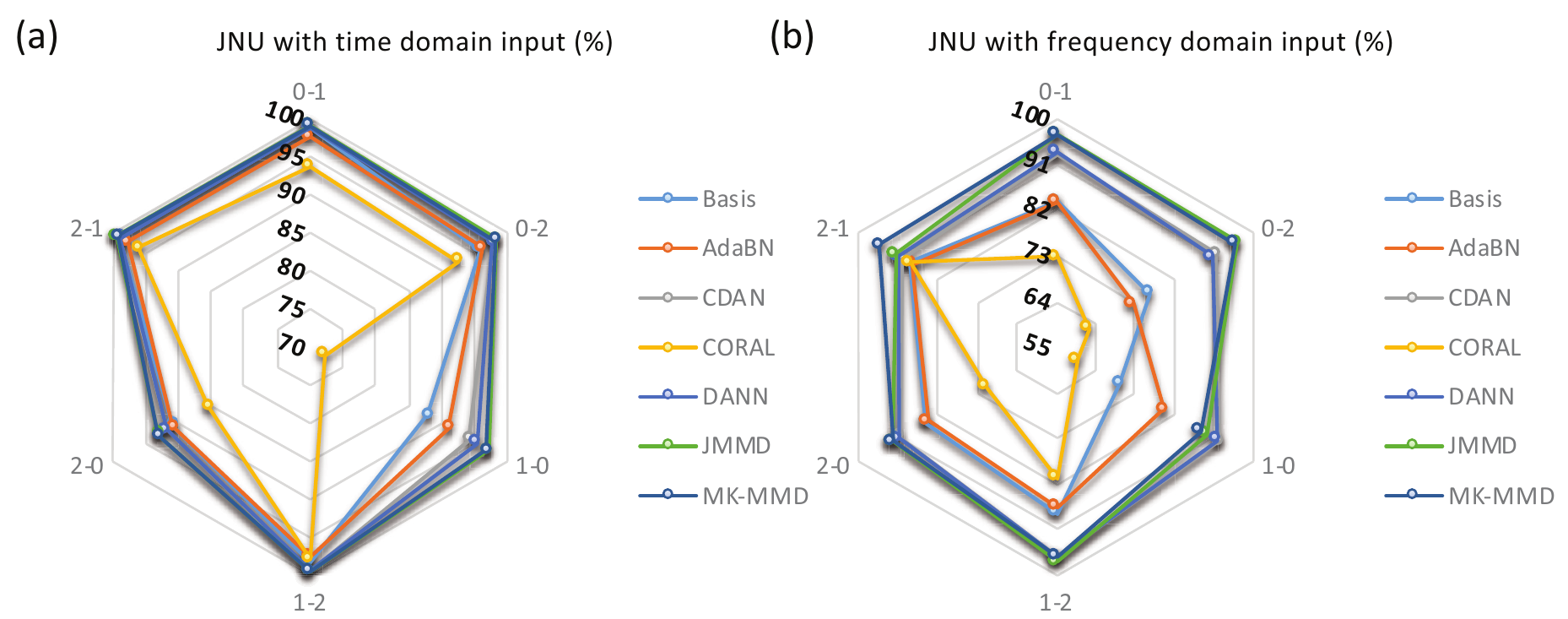}}
	\\ [-10pt]
	\caption{The accuracy comparisons of different methods in JNU.}
	\label{Models-JNU}
\end{figure*}

\begin{figure*}[!t]
	\centering
	\subfigure{\includegraphics[scale = 0.7]{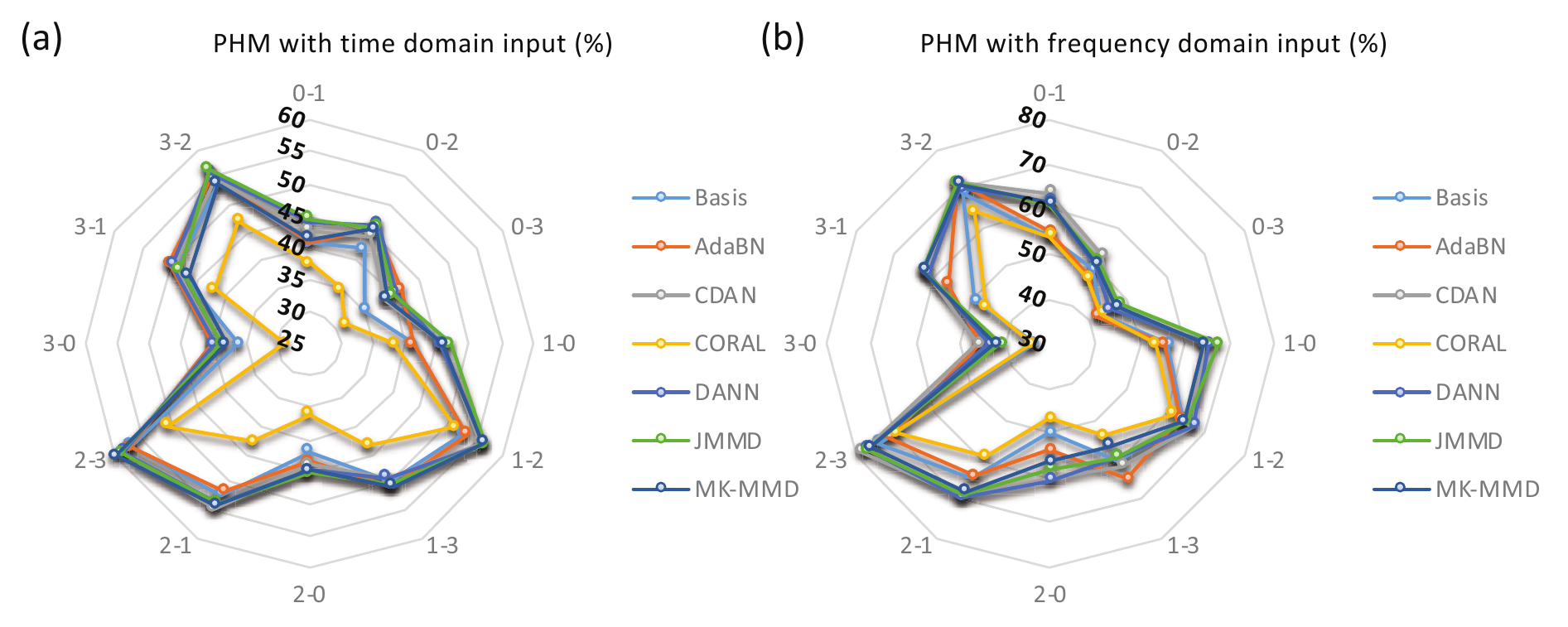}}
	\\ [-10pt]
	\caption{The accuracy comparisons of different methods in PHM2009.}
	\label{Models-PHM}
\end{figure*}

\begin{figure*}[!t]
	\centering
	\subfigure{\includegraphics[scale = 0.8]{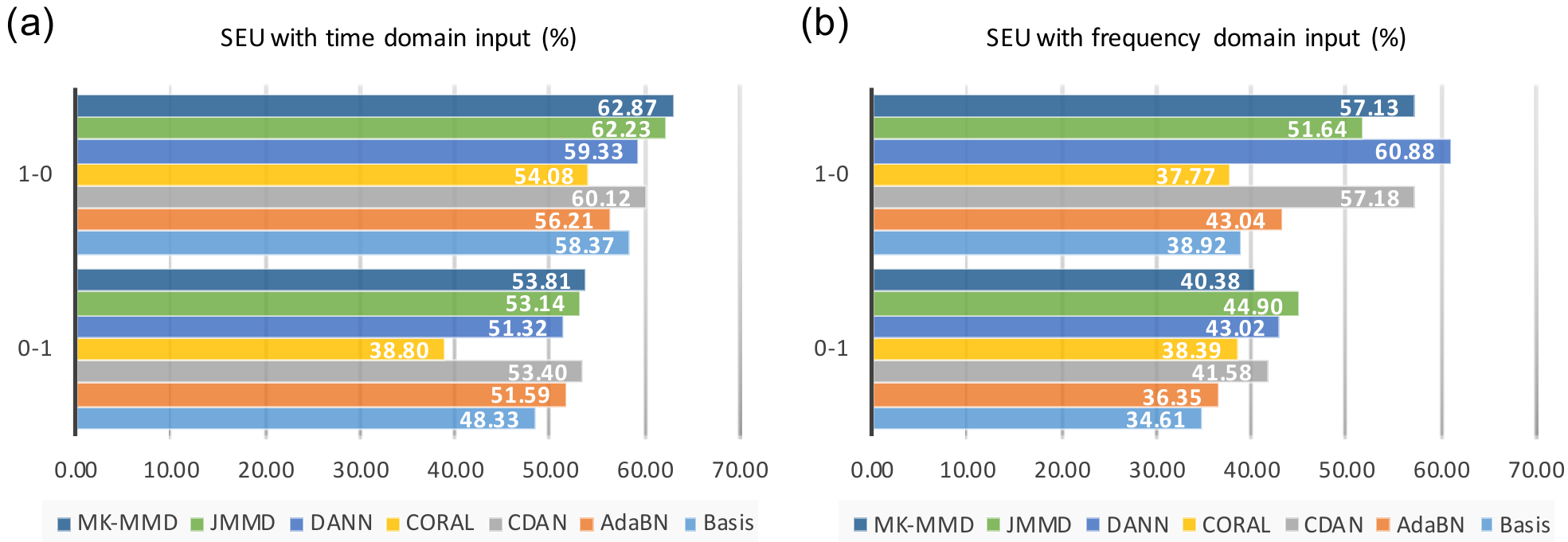}}
	\\ [-10pt]
	\caption{The accuracy comparisons of different methods in SEU.}
	\label{Models-SEU}
\end{figure*}

Results of different methods are shown in \figref{Models-CWRU} to \figref{Models-SEU}, and \figref{Models-SEU} is not set as the radar chart because two transfer tasks are not suitable for this visualization.
For all datasets, methods discussed in this paper can improve the accuracy of Basis, except CORAL.
For CORAL, it can only improve the accuracy in CWRU with the frequency domain input or in some transfer tasks.
For AdaBN, the improvement is much smaller than other methods.

In general, results of JMMD are better than those of MK-MMD, which indicates that the assumption of joint distribution in source and target domains is useful for improving the performance.
Results of DANN and CDAN are generally better than those of MK-MMD, which indicates that adversarial training is helpful for aligning the domain shift.


\subsubsection{Results of input types}
\begin{figure*}[!t]
	\centering
	\subfigure{\includegraphics[scale = 0.7]{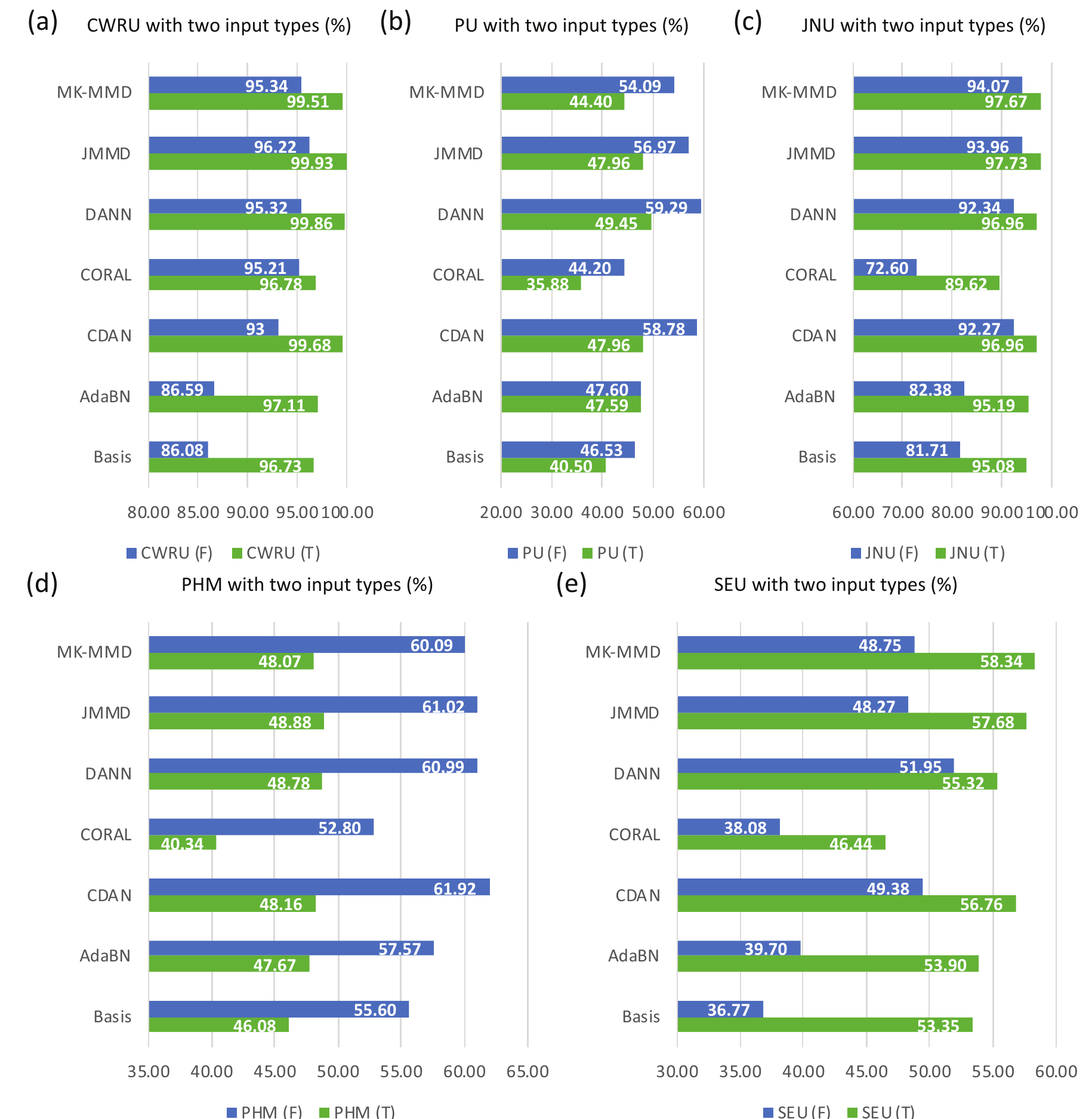}}
	\\ [-10pt]
	\caption{The accuracy comparisons of two input types with different datasets. (F) means the frequency domain input, and (T) means the time domain input.}
	\label{Input-types}
\end{figure*}

Accuracy comparisons of two input types are shown in \figref{Input-types}, and it can be concluded that the time domain input achieves better accuracy in CWRU, JNU, and SEU, while the frequency domain input gets better accuracy in PU and PHM2009.
Besides, the accuracy gap between these two input types is relatively large, and we cannot simply infer which one is better due to the influence of backbones.

Thus, for a new dataset, we should test results of different input types instead of just using the more advanced techniques to improve the performance of one input type, because using a different input type might improve the accuracy more efficient than using advanced techniques.

\subsubsection{Results of accuracy types}
\begin{figure*}[!t]
	\centering
	\subfigure{\includegraphics[scale = 0.8]{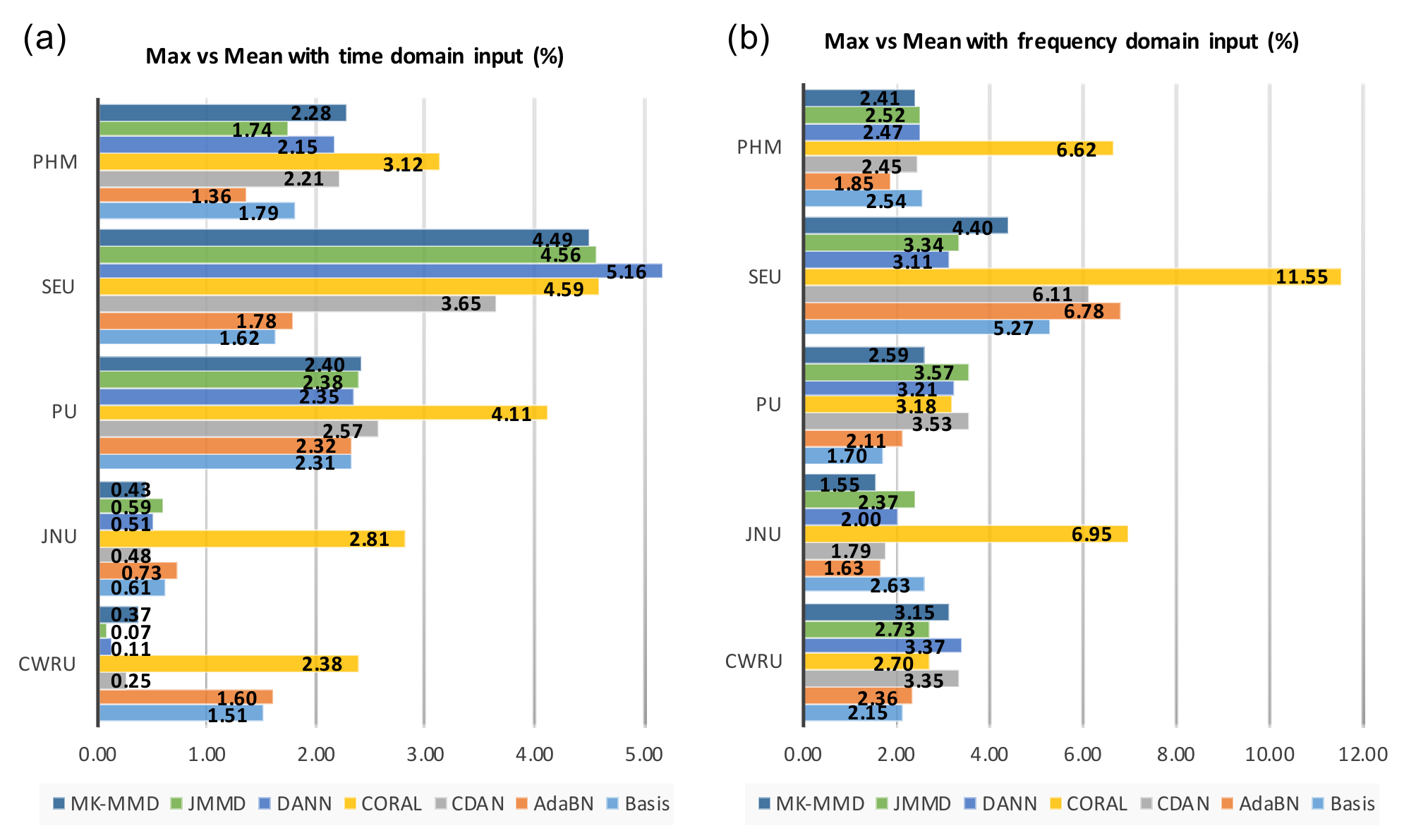}}
	\\ [-10pt]
	\caption{The difference between Max and Mean according to Best average.}
	\label{Max-Mean}
\end{figure*}

As mentioned in Section \ref{S:7}, we use four kinds of accuracy including Best-Mean, Best-Max, Last-Mean, and Last-Max to evaluate the performance.
As shown in \figref{Max-Mean}, the fluctuation of different experiments is sometimes large, especially for those datasets whose overall accuracy is not very high, which indicates that the used algorithms are not very stable and robust.
Besides, it seems that the fluctuation of the time domain input is smaller than that of the frequency domain input, and the reason might be that the backbone used in this paper is more suitable for the time domain input.

\begin{figure*}[!t]
	\centering
	\subfigure{\includegraphics[scale = 0.8]{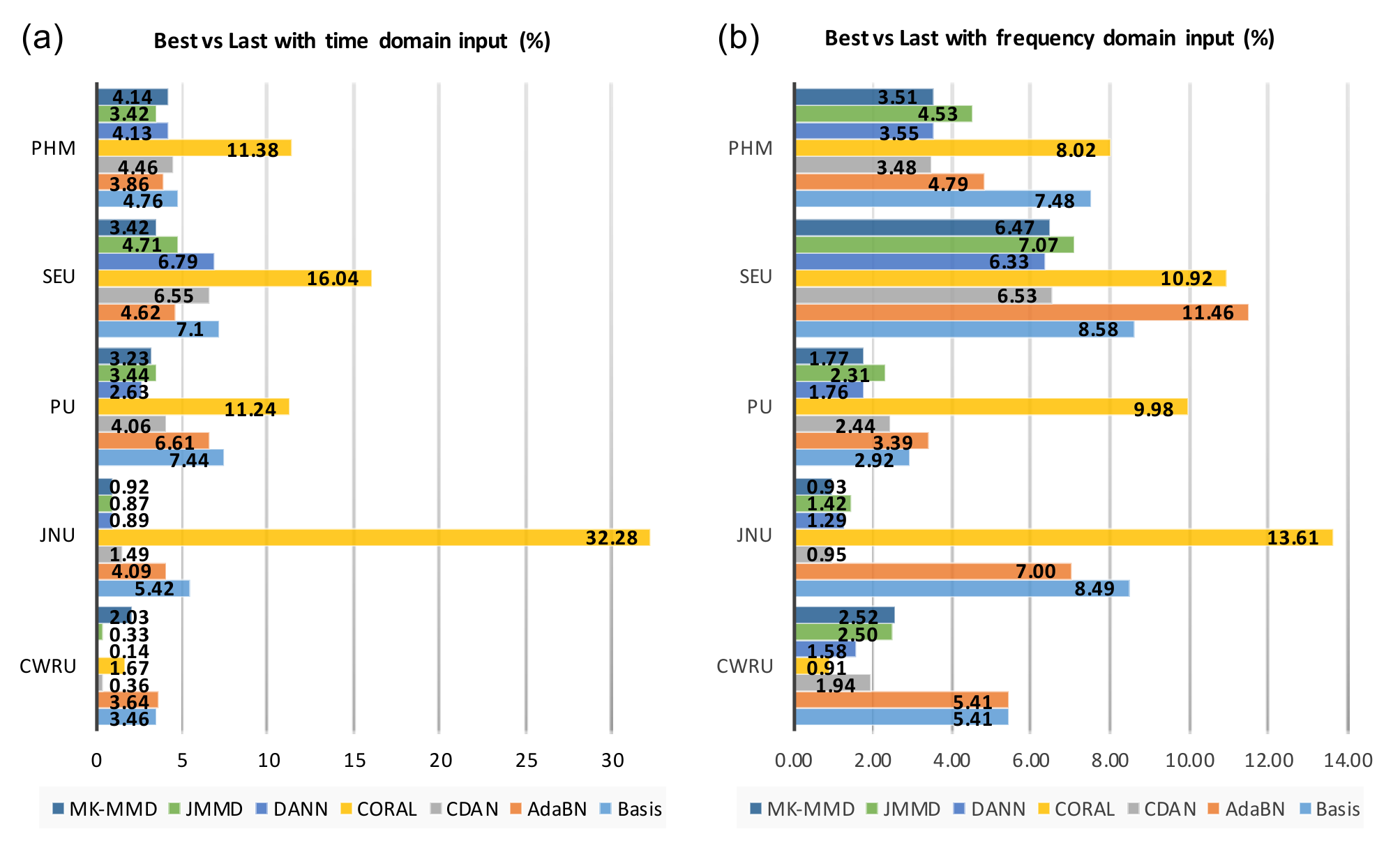}}
	\\ [-10pt]
	\caption{The difference between Best average and Last average according to Mean.}
	\label{Best-Last}
\end{figure*}

As shown in \figref{Best-Last}, the fluctuation of different experiments is also large, which is dangerous for evaluating the true performance. 
Since Best uses the test set to choose the best model (it is a kind of test leakage), Last may be more suitable for representing the generalization accuracy.

Thus, on the one hand, the stability and robustness of UDTL-based IFD need more attention instead of just improving the accuracy.
On the other hand, as we analyze above, the accuracy of the last epoch (Last) is more suitable for representing the generalization ability of algorithms when the fluctuation between Best and Last is large.

\subsection{Label-inconsistent UDTL}
In these methods, the transfer learning strategies are activated from the beginning.
For UAN, the trade-off parameter of the loss of non-adversarial domain discriminator is fixed as 1.
The value $\tau$ of OSBP and the threshold $\omega_0$ of UAN are both set to 0.5 for all tasks.
\subsubsection{Evaluation metrics}
For partial-based transfer learning, the evaluation metrics are the same as that of label-consistent UDTL, including Last-Mean, Last-Max, Best-Mean, and Best-Max.
For open set and universal transfer learning, due to the existence of unknown classes, only the overall accuracy is not sufficient for evaluating the model performance.
To clearly explain evaluation metrics, several mathematical notations are defined.
$M_S$ and $M_U$ are the number of correctly classified shared-class and successfully detected unknown-class test samples, respectively.
$N_S$ and $N_U$ are the number of all shared-class and unknown-class test samples, respectively.
$Acc_c$ is the accuracy of test samples from the $c$-th class.

Following previous work in \cite{9394793,liu2019separate,saito2018open,fu2020learning}, five evaluation metrics are employed:
\begin{enumerate}[1)]
	\item \textit{Accuracy of shared classes:} $\text{ALL}^* = \frac{M_S}{N_S}$
	\item \textit{Accuracy of unknown classes:} $\text{UNK} = \frac{M_U}{N_U}$
	\item \textit{Averaged accuracy of all classes:} $\text{OS} = \frac{1}{C + 1}\sum_{c = 0}^C {Acc_c} $
	\item \textit{Accuracy of all test samples:} $\text{ALL} = \frac{ M_S + M_U }{ N_S + N_U}$
	\item \textit{Harmonic mean:} $\text{H-score} = \frac{2 \text{ALL}^* \; \text{UNK}}{\text{ALL}^* + \text{UNK}}$
\end{enumerate}

Similar to the label-consistent UDTL, mean and maximum values of the overall accuracy are used to evaluate the final performance.
We use the mean accuracy of all five evaluation metrics in the last epoch denoted as Last-Mean-ALL*, Last-Mean-UNK, Last-Mean-OS, Last-Mean-ALL, and Last-Mean-H-score.
We use the accuracy when models perform the best H-score among five tests in the last epoch, denoted as Last-Max-ALL*, Last-Max-UNK, Last-Max-OS, Last-Max-ALL, and Last-Max-H-score.
Meanwhile, we also list the mean accuracy denoted as Best-Mean-ALL*, Best-Mean-UNK, Best-Mean-OS, Best-Mean-ALL, and Best-Mean-H-score in the epoch where models achieve the best performance on H-score.
We also list the accuracy denoted as Best-Max-ALL*, Best-Max-UNK, Best-Max-OS, Best-Max-ALL, and Best-Max-H-score where models achieve the best performance on H-score among five tests.

\subsubsection{Dataset settings}
CWRU is selected for testing the performance.
Following recent works in \cite{9394793}, different classes are randomly selected to form transfer learning tasks to validate the effectiveness of models on different label sets.
The fault diagnosis tasks for partial, open set, and universal transfer learning are presented in Table \ref{CWRU_Tab} respectively. 

\begin{table*}[!t]
  \caption{The fault diagnosis tasks of CWRU.}
  \centering
  \label{CWRU_Tab}
  \begin{tabular}{ccccccc}
  \hline
    \multirow{3}[0]{*}{Task} & \multicolumn{2}{c}{Partial UDTL} & \multicolumn{2}{c}{Open Set UDTL} & \multicolumn{2}{c}{ Universal UDTL} \\
              & \multirow{2}[0]{*}{Source Label Set} & \multirow{2}[0]{*}{Target Label Set} & \multirow{2}[0]{*}{Source Label Set} & \multirow{2}[0]{*}{Target Label Set} & \multirow{2}[0]{*}{Source Label Set} & \multirow{2}[0]{*}{Target Label Set} \\
              &       &       &       &       &       &  \\

  \hline
    0-1   & \multirow{2}[0]{*}{0$\sim$9} & \multirow{2}[0]{*}{0,1,2,4,5,7,8,9} & \multirow{2}[0]{*}{0,2,3,5,6,7,8,9} & \multirow{2}[0]{*}{0$\sim$9} & \multirow{2}[0]{*}{0,1,2,4,5,6,7,8,9} & \multirow{2}[0]{*}{1,2,3,4,5,7,8,9} \\
    0-2   &       &       &       &       &       &  \\
    0-3   & \multirow{2}[0]{*}{0$\sim$9} & \multirow{2}[0]{*}{1,2,4,6,7,8,9} & \multirow{2}[0]{*}{0,1,2,3,4,5,6} & \multirow{2}[0]{*}{0$\sim$9} & \multirow{2}[0]{*}{0,1,2,3,4,5,7,8} & \multirow{2}[0]{*}{0,2,3,4,5,6,7,8,9} \\
    1-0   &       &       &       &       &       &  \\
    1-2   & \multirow{2}[0]{*}{0$\sim$9} & \multirow{2}[0]{*}{0,1,3,7,9} & \multirow{2}[0]{*}{1,3,4,5,7,8} & \multirow{2}[0]{*}{0$\sim$9} & \multirow{2}[0]{*}{1,2,4,5,7,8,9} & \multirow{2}[0]{*}{1,3,6,8} \\
    1-3   &       &       &       &       &       &  \\
    2-0   & \multirow{2}[0]{*}{0$\sim$9} & \multirow{2}[0]{*}{1,2,4,7} & \multirow{2}[0]{*}{0,2,4,5,6,9} & \multirow{2}[0]{*}{0$\sim$9} & \multirow{2}[0]{*}{1,3,4,6,7,8} & \multirow{2}[0]{*}{0,1,2,3,5,6,8,9} \\
    2-1   &       &       &       &       &       &  \\
    2-3   & \multirow{2}[0]{*}{0$\sim$9} & \multirow{2}[0]{*}{0,2,6} & \multirow{2}[0]{*}{0,1,2,5,7} & \multirow{2}[0]{*}{0$\sim$9} & \multirow{2}[0]{*}{0,1,2,7,8} & \multirow{2}[0]{*}{1,2,6,9} \\
    3-0   &       &       &       &       &       &  \\
    3-1   & \multirow{2}[0]{*}{0$\sim$9} & \multirow{2}[0]{*}{5,8} & \multirow{2}[0]{*}{4,8,9} & \multirow{2}[0]{*}{0$\sim$9} & \multirow{2}[0]{*}{0,1,5,7,8} & \multirow{2}[0]{*}{0,4,8} \\
    3-2   &       &       &       &       &       &  \\ \hline
    \end{tabular}
\end{table*}

\begin{figure}[!t]
	\centering
	\subfigure{\includegraphics[scale = 0.8]{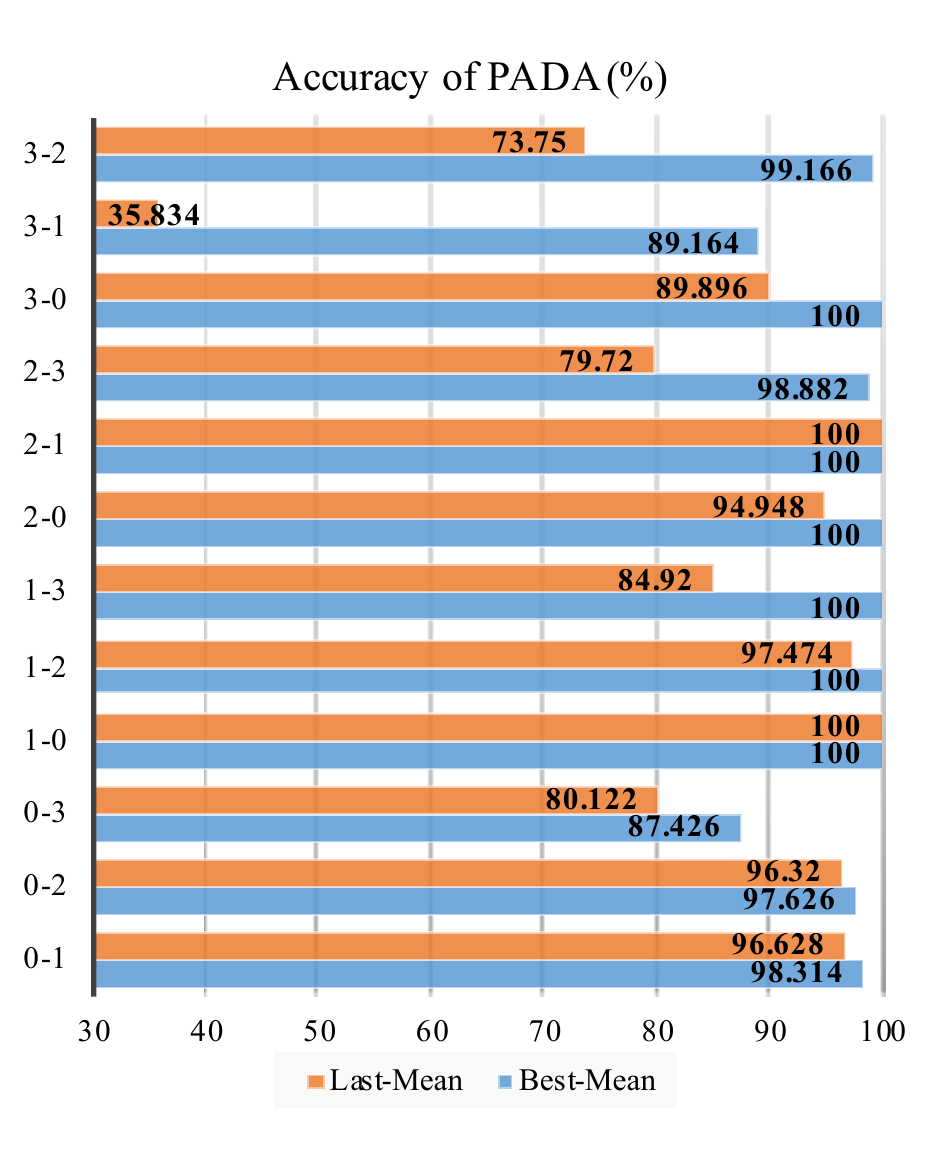}}
	\\ [-10pt]
	\caption{The overall accuracy of PADA with the time domain input.}
	\label{accuracy_pada}
\end{figure}

\begin{figure*}[!t]
	\centering
	\subfigure{\includegraphics[scale = 0.7]{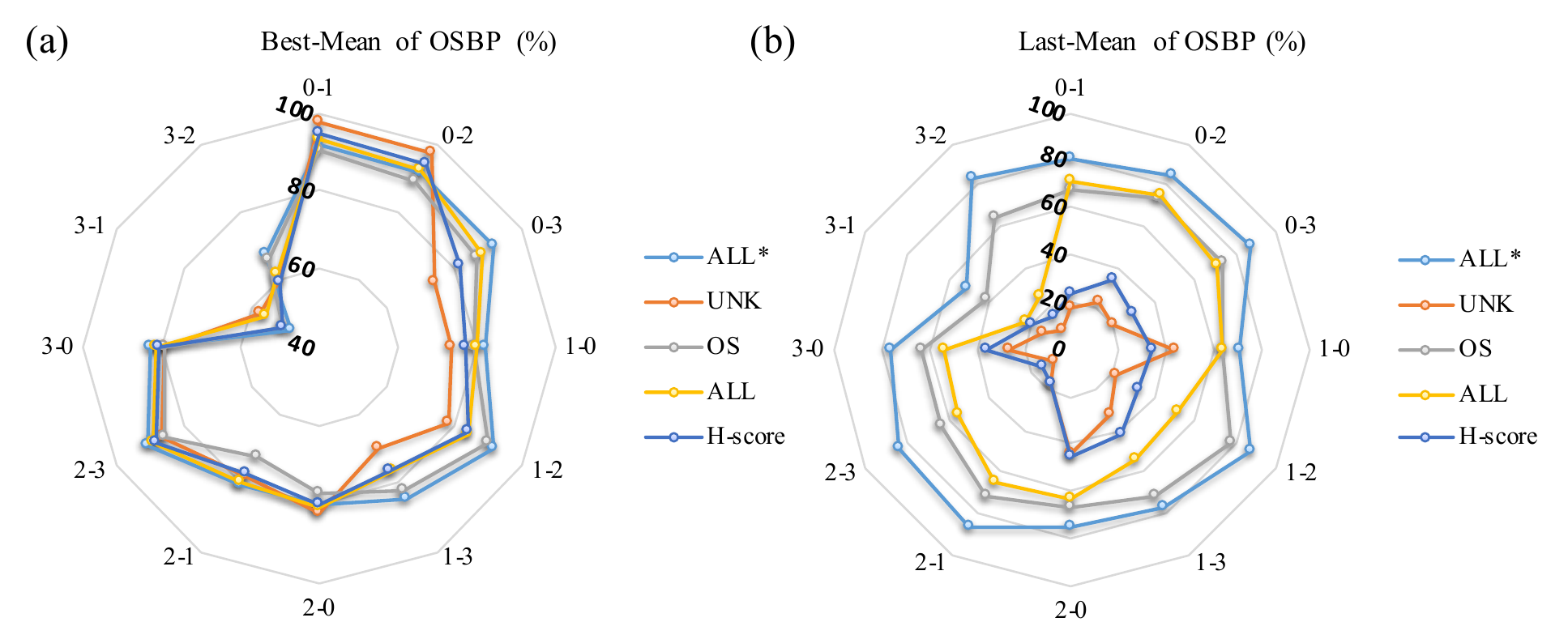}}
	\\ [-10pt]
	\caption{The overall accuracy of OSBP with the time domain input: (a) Best-Mean and (b) Last-Mean.}
	\label{accuracy_OSBP}
\end{figure*}

\begin{figure*}[!t]
	\centering
	\subfigure{\includegraphics[scale = 0.7]{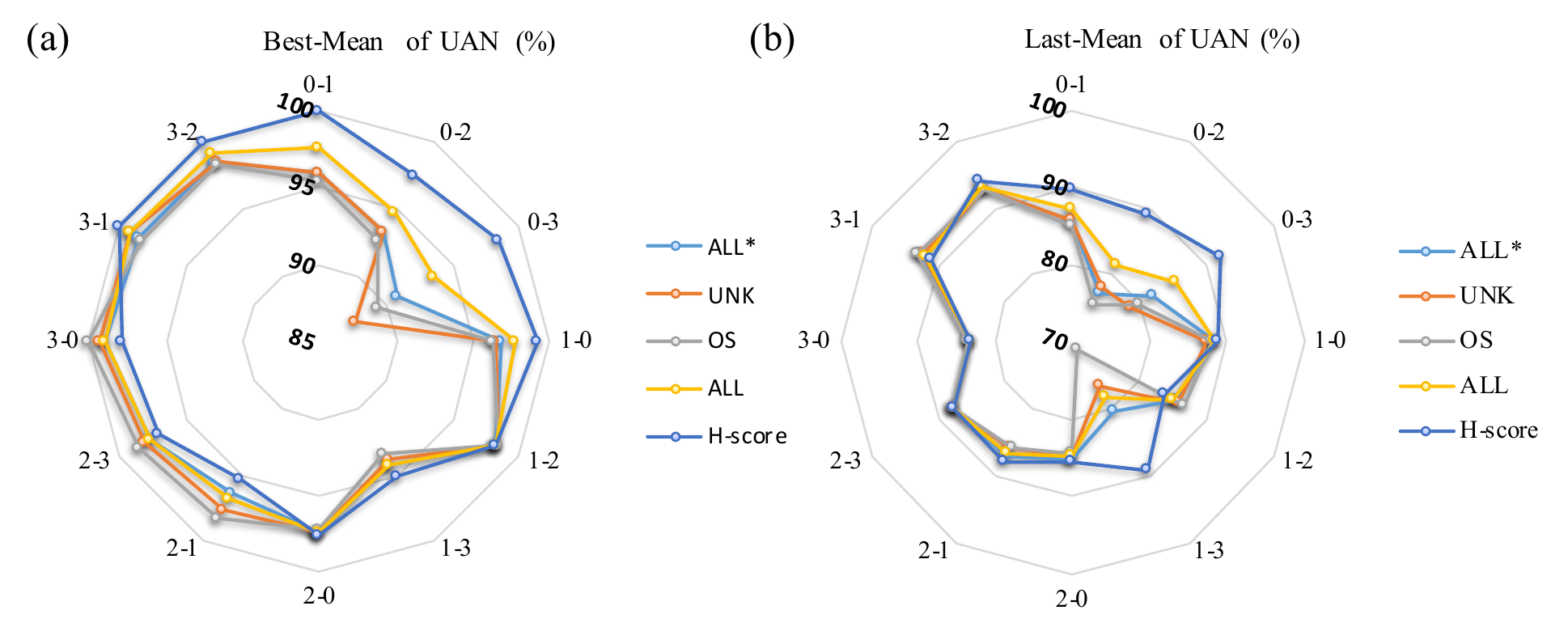}}
	\\ [-10pt]
	\caption{The overall accuracy of UAN with the time domain input: (a) Best-Mean and (b) Last-Mean.}
	\label{accuracy_UAN}
\end{figure*}

\subsubsection{Results of partial UDTL}
For simplicity, as shown in \figref{accuracy_pada}, we only list Best-Mean and Last-Mean of PADA with the time domain input due to the similarity between time and frequency domain inputs.
We can observe that PADA can achieve good performance on most tasks according to the overall training phase.
But for tasks 3-1, 2-3, and 3-2, Last-Mean is obviously lower than Best-Mean, indicating that negative transfer resulting from extra source labels cannot be addressed totally by PADA and there exist the overfitting problem during the training procedure.

\subsubsection{Results of open set UDTL}
Best-Mean and Last-Mean accuracy of OSBP with the time domain input are shown in \figref{accuracy_OSBP}.
From \figref{accuracy_OSBP} (a), it can be seen that OSBP can achieve relatively good performance on most transfer tasks.
However as shown in \figref{accuracy_OSBP} (b), performance obviously degrades on the later stage, especially for $\text{UNK}$, which reveals that the model overfits on the source samples, and thus unknown-class samples cannot be effectively recognized.
Moreover, the lowest $\text{ALL}^*$ is only about 50 \%, which means that only half of shared-class samples can be correctly classified.
Therefore, more effective models, which not only have the ability to detect unknown-class samples but also ensure accurate shared-class classification, are required.

\subsubsection{Results of universal UDTL}
Best-Mean and Last-Mean accuracy of UAN with the time domain input are shown in \figref{accuracy_UAN}.
Generally speaking, UAN can achieve excellent performance on the CWRU dataset according to \figref{accuracy_UAN} (a).
Similar to the results of OSBP, the performance of UAN also degrades on the later stage because of the overfitting problem and wrong feature alignment.
In addition, the shared-class classification accuracy still need be improved.
It is still difficult for the model to separate extra source classes and detect unknown classes from the target domain.

\subsubsection{Results of multi-criterion evaluation metric}
\begin{figure}[!t]
	\centering
	\subfigure{\includegraphics[scale = 0.65]{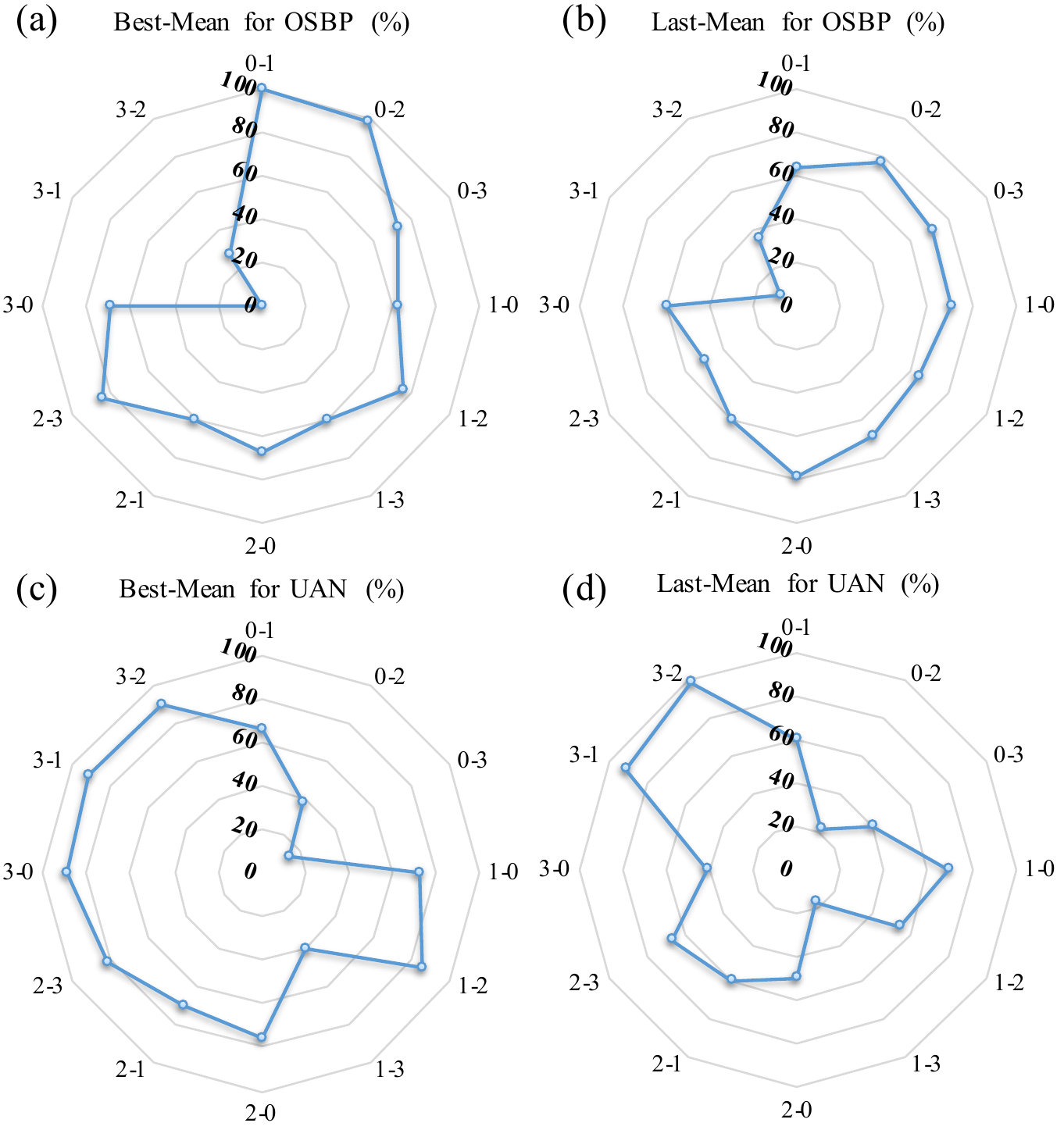}}
	\\ [-10pt]
	\caption{The TOPSIS of OSBP and UAN: (a) Best-Mean of OSBP, (b) Last-Mean of OSBP, (c) Best-Mean of UAN, and (d) Last-Mean of UAN.}
	\label{accuracy_TOPSIS}
\end{figure}

Due to the fact that we have five evaluation metrics for open set UDTL and universal UDTL, it is better to have a final score concerning different metrics for a better understanding of the result.
Thus, we use Technique for Order Preference by Similarity to an
Ideal Solution (TOPSIS), which is a famous method in the multi-criterion
evluation metric, as the final score.
Meanwhile, TOPSIS was also widely applied to the field of fault diagnosis\cite{wang2010integration,he2016fuzzy,jiang2019fault}.
In this paper, we use $\text{ALL}^*$, UNK, OS, ALL, and H-score to calculate TOPSIS for the multi-criterion evaluation.
For the sake of simplicity, the weight of every index is set to 0.25 in TOPSIS.
As shown in \figref{accuracy_TOPSIS}, we can observe the TOPSIS comparsions of OSBP and UAN for different transfer learning tasks.
It is not unexpected that the evaluation using TOPSIS is similar to these metrics in \figref{accuracy_OSBP} and \figref{accuracy_UAN}.
The overall performance also degrades at the later stage due to the overfitting problem and wrong feature alignment.
The shared-class classification accuracy has a relatively large space to be promoted.
In addition, separating extra source classes and detecting unknown classes in the target domain are still not well solved.

\subsection{Multi-domain UDTL}
For multi-domain UDTL, the evaluation metrics are the same as that of label-consistent UDTL, including Last-Mean, Last-Max, Best-mean, and Best-Max.

\subsubsection{Dataset settings}
Similarity to label-inconsistent UDTL, CWRU is selected to test the performance of multi-domain UDTL, including MS-UADA and IAN.
The types of inputs consist of the time and frequency domain inputs.
The fault diagnosis tasks for multi-domain UDTL are listed in Table \ref{CWRU_multi_Tab}.
For example, 123-0\_T means that task 1, 2, and 3 (shown in Table \ref{Tab_2}) are used as multiple source domains; task 0 is used as the target domain; the time domain input is used as the model input.
It should be mentioned that we do not use the target data in the training phase when testing the performance of DG.
\begin{table}
\center
\caption{The task of multi-domain UDTL.}
\label{CWRU_multi_Tab}
\begin{tabular}{ccccc}
\hline
Task&Source 1&Source 2&Source 3&Target \\ \hline
012-3 & 0 & 1 & 2 & 3 \\ \hline
123-0 & 1 & 2 & 3 & 0 \\ \hline
013-2 & 0 & 1 & 3 & 2\\ \hline
230-1 & 2 & 3 & 0 & 1\\ \hline
\end{tabular}
\end{table}
\begin{figure*}[!t]
	\centering
	\subfigure{\includegraphics[scale = 0.7]{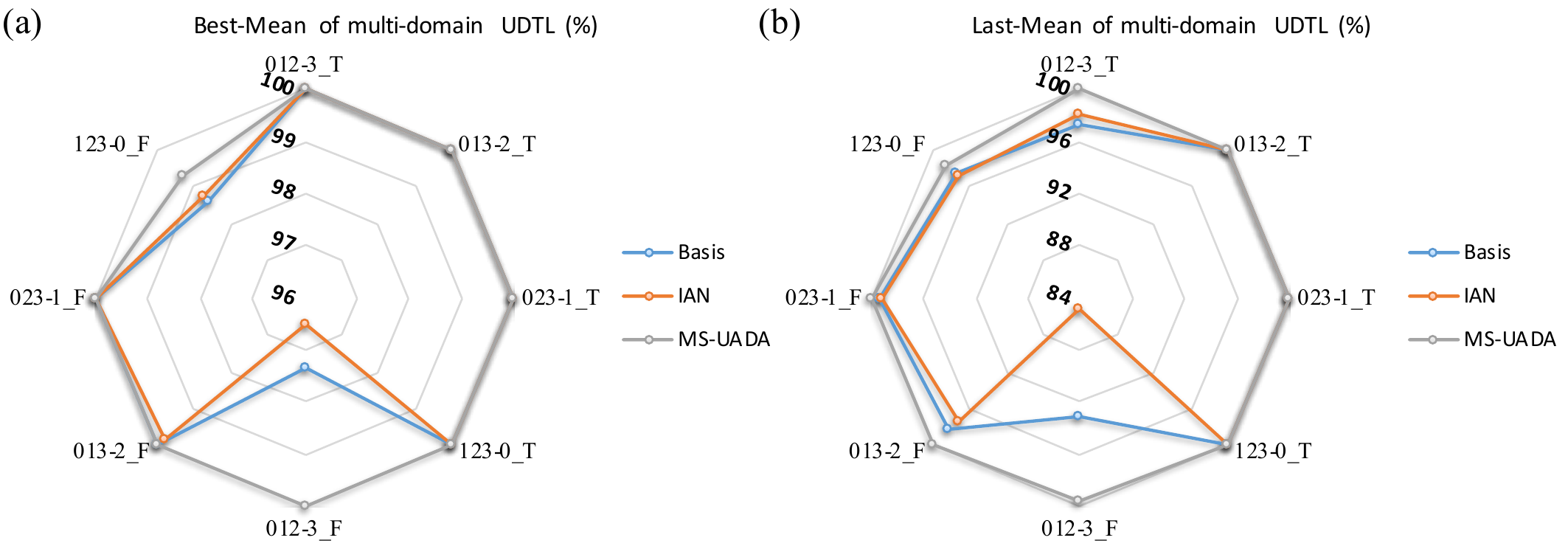}}
	\\ [-10pt]
	\caption{The overall accuracy of multi-domain UDTL (F and T mean the time domain and the frequency domain inputs, respectively): (a) Best-Mean, (b) Last-Mean.}
	\label{accuracy_MD}
\end{figure*}

\subsubsection{Results of multi-domain adaptation}
As shown in \figref{accuracy_MD} (a) and (b), we can observe that MS-UADA can always improve the accuracy of CWRU compared with Basis which directly transfers the trained model using multiple source domains to the target domain.
Performance of the time domain input is slightly better than that of the frequency domain input, but the overall difference is very tiny.

\subsubsection{Results of DG}
As shown in \figref{accuracy_MD} (a) and (b),
we can observe that the performance of IAN for CWRU is similar to that of Basis in most tasks.
However, for the task 012-3\_F, the accuracy of IAN decreases greatly.
The main reason might be that IAN only uses multiple sources to find the domain-invariant features, which is not suitable for the unseen target domain.
Thus, more complex DG methods should be further designed to dig the discriminative and domain-invariant features.

\section{Further discussions}
\label{S:8}
\subsection{Transferability of features}
The reason why DL models embedded transfer learning methods can achieve breakthrough performance in computer vision is that many studies have shown and proved that DL models can learn more transferable features for these tasks than traditional hand-crafted features \cite{glorot2011domain,yosinski2014transferable}.
In spite of the ability to learn general and transferable features, DL models also exist transition from general features to specific features and their transferability drops significantly in the last layers \cite{yosinski2014transferable}.
Therefore, fine-tuning DL models or adding various transfer learning strategies into the training process need to be investigated for realizing the valid transfer.

However, for IFD, there is no research about how transferable are features in DL models, and actually, answering this problem is the most important cornerstone in UDTL-based IFD.
Since the aim of this paper is to give a comparative accuracy and release a code library, we just assume that the bottleneck layer is the task-specific layer and its output features are restrained with various transfer learning strategies.
Thus, it is imperative and vital for scholars to study transferability of features and answer the question about how transferable features are learned.
In order to make transferability of features more reasonable, we suggest that scholars might need to visualize neurons to analyze learned features by existing visualization algorithms \cite{zeiler2014visualizing,selvaraju2017grad}.

\subsection{Influence of backbones and bottleneck}
In the field of computer vision, many strong CNN models (also called backbones), such as VGG \cite{simonyan2014very} and ResNet \cite{he2016deep} can be extended without caring about the model selection.
Scholars often use the same backbones to test the performance of proposed algorithms and can pay more attention to construct specific algorithms to align source and target domains.

However, backbones of published UDTL-based IFD are often different, which makes results hard to compare directly, and influence of different backbones has never been studied thoroughly.
Whereas, backbones of UDTL-based algorithms do have a huge impact on results from comparisons between CWRU with the frequency domain input and ``Table II'' in \cite{wang2019domain} (the main difference is the backbone used in this paper and \cite{wang2019domain}).
We can observe that the accuracy related to the task 3 in CWRU with the frequency domain input is much worse than that in ``Table II'' \cite{wang2019domain}.
However, the backbone used in this paper can achieve excellent results with the time domain input and some accuracies are even higher than those in \cite{wang2019domain}.

\begin{figure*}[!t]
	\centering
	\subfigure{\includegraphics[scale = 0.7]{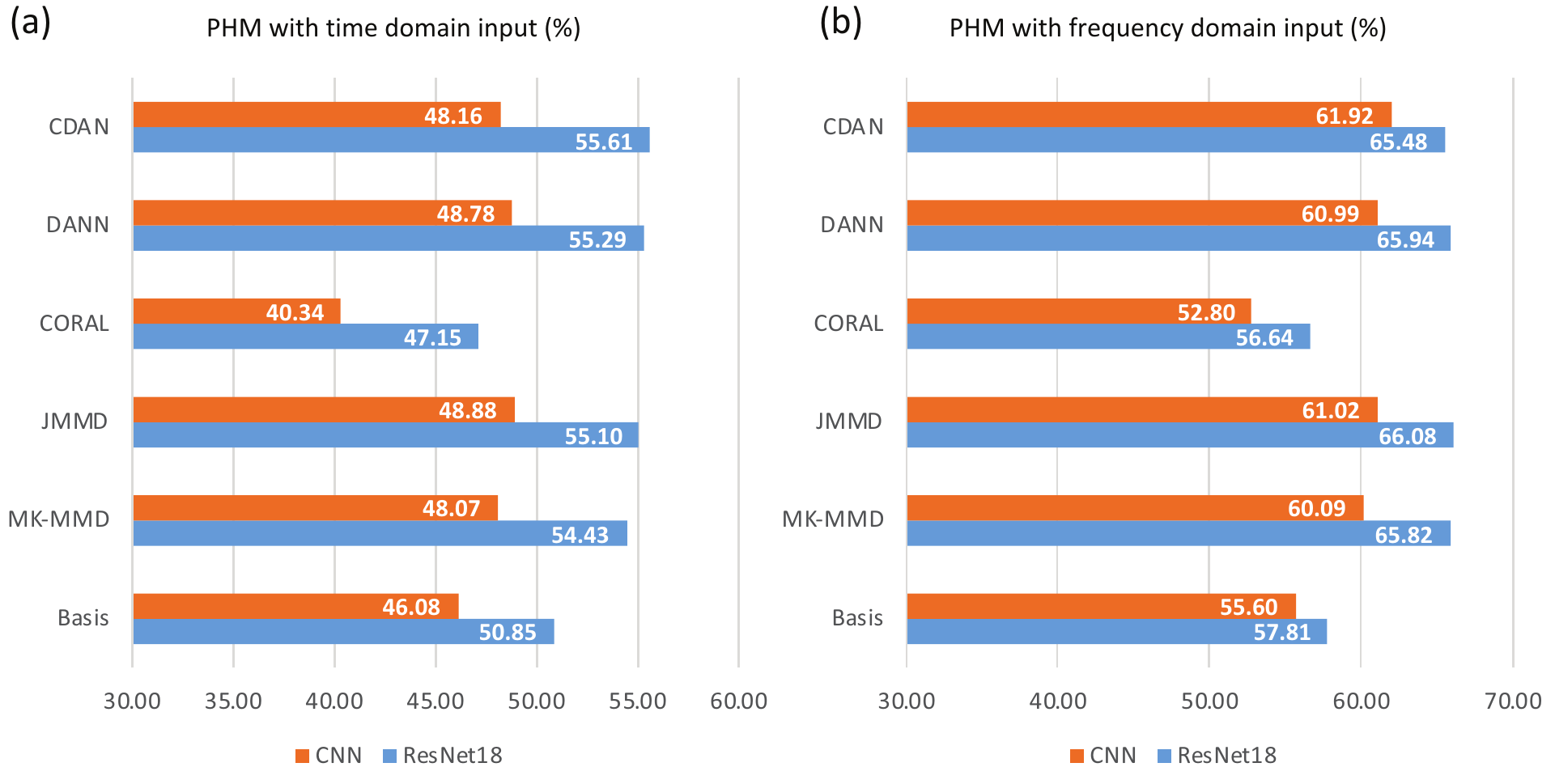}}
	\\ [-10pt]
	\caption{Comparisons of PHM2009.}
	\label{Others-PHM}
\end{figure*}

\begin{figure*}[!t]
	\centering
	\subfigure{\includegraphics[scale = 0.7]{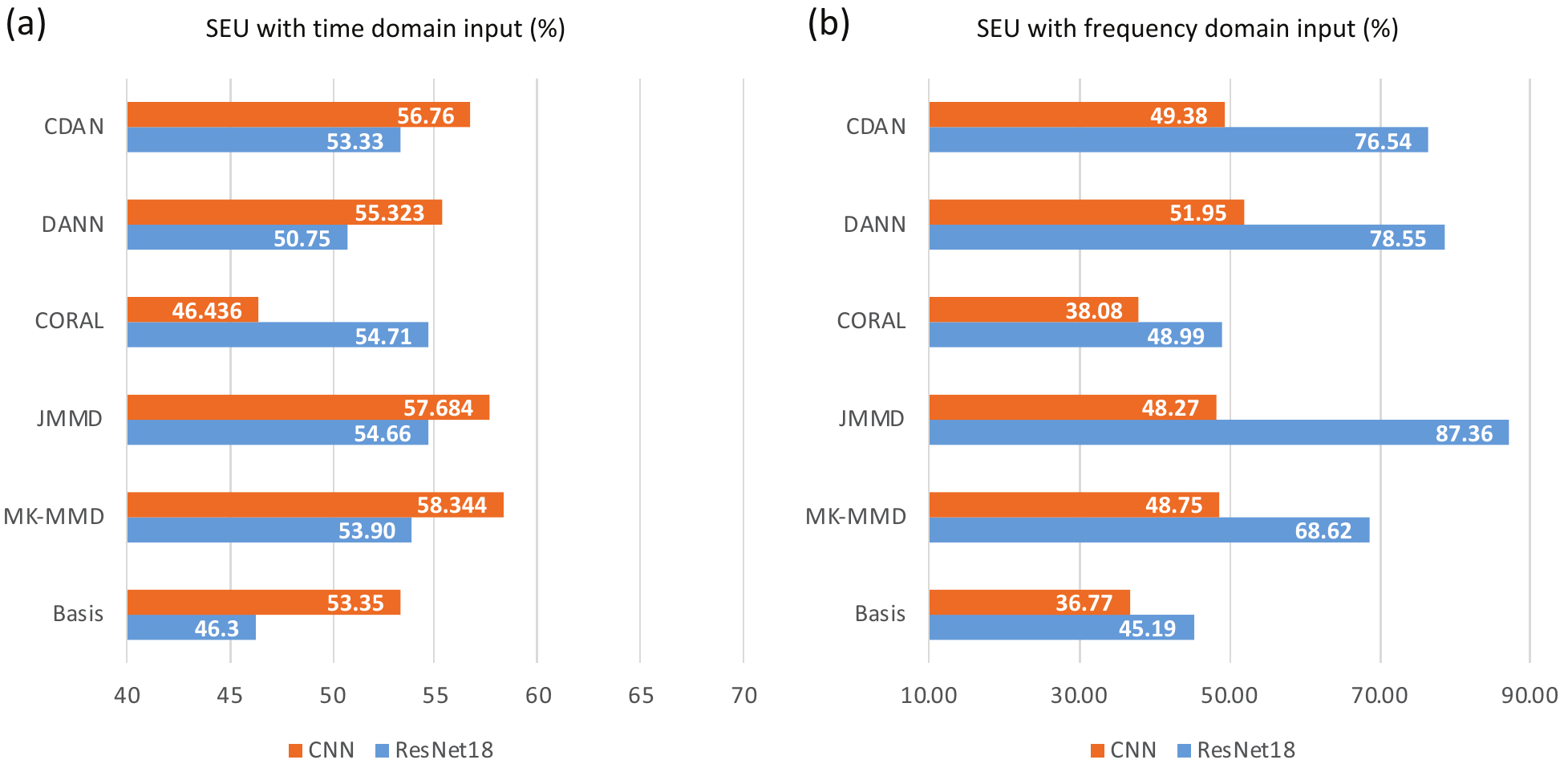}}
	\\ [-10pt]
	\caption{Comparisons of SEU.}
	\label{Others-SEU}
\end{figure*}

To make a stronger statement, we also use the well-known backbone called ResNet18 (we modify the structure of ResNet18 to adapt one dimensional input) to test SEU and PHM2009 datasets for explaining the huge impact of backbones. 
From comparisons of PHM2009 shown in \figref{Others-PHM}, ResNet18 can improve the accuracy of each algorithm significantly. 
Besides, from comparisons of SEU shown in \figref{Others-SEU}, ResNet18 with the time domain input actually reduces the accuracy, and on the contrary, ResNet18 with the frequency domain input improve the accuracy significantly.
In summary, different backbones behave differently with variant datasets and input types.

Therefore, finding a strong and suitable backbone, which can learn more transferable features for IFD, is also very important for UDTL-based methods (sometimes choosing a more effective backbone is even more important than using a more advanced algorithm)
We suggest that scholars should first find a strong backbone and then use the same backbone to compare results for avoiding unfair comparisons.

In the top comparsion, we discuss the influence of backbones.
However, in our designed structure, the bottleneck layer in the source domain also shares parameters with that in the target domain.
Thus, it is necessary to discuss the influence of the bottleneck layer during the transfer learning procedure.
For the sake of simplicity, we only use CWRU with two different inputs to test two representative UDTL methods, including MK-MMD and DANN.

\begin{figure}[!t]
	\centering
	\subfigure{\includegraphics[scale = 0.65]{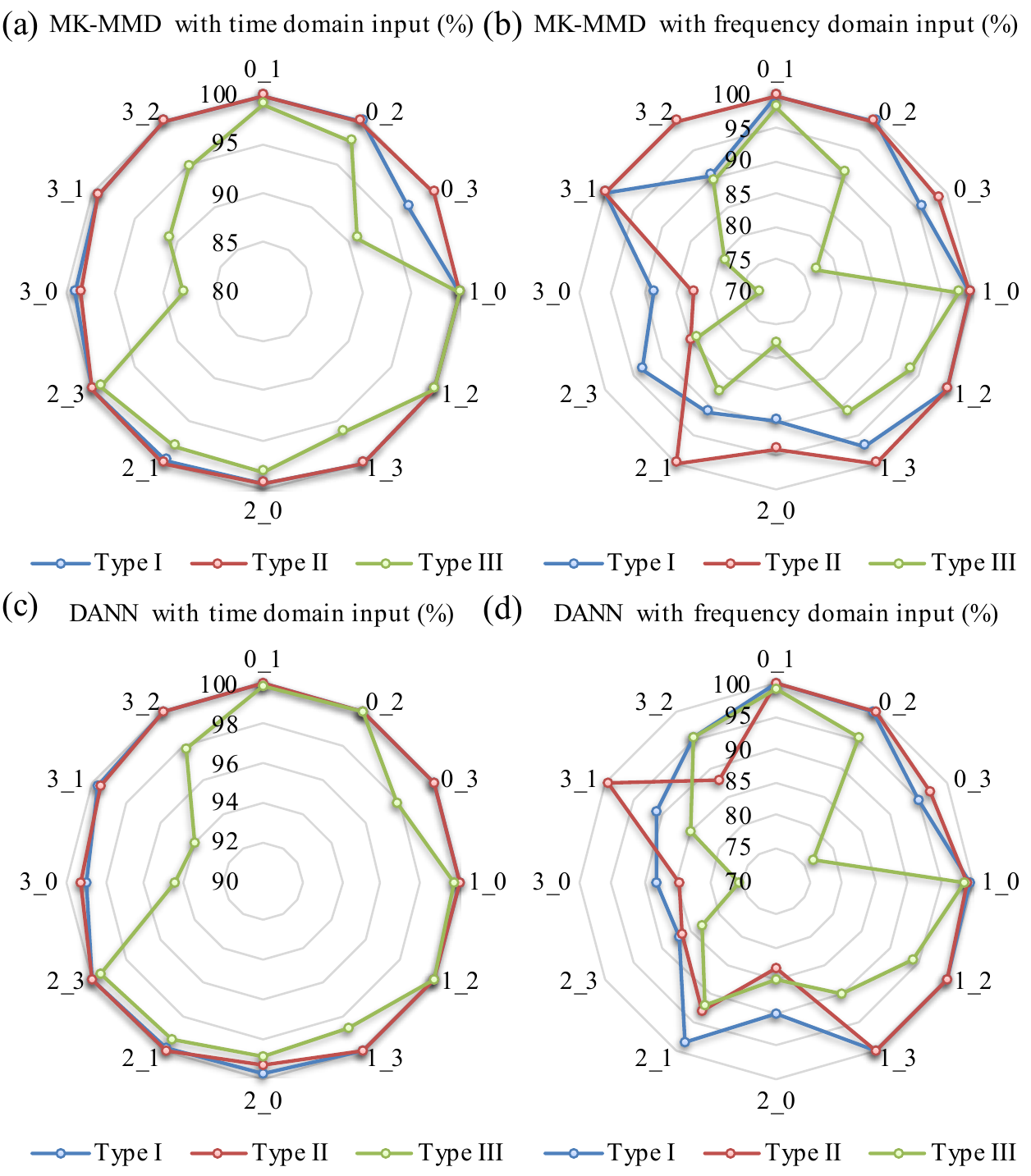}}
	\\ [-10pt]
	\caption{Comparisons of three conditions related to the bottleneck layer.}
	\label{BottleneckResults}
\end{figure}

We use Type \uppercase\expandafter{\romannumeral1} to represent original models in this paper, Type \uppercase\expandafter{\romannumeral2} to represent models without the bottleneck layer, and Type \uppercase\expandafter{\romannumeral3} to represent models with fixed parameters of backbones (their parameters are pretrained by the source data) when starting transfer learning (only updating parameters of the bottleneck layer during the transfer learning procedure).
The comparison results are shown in \figref{BottleneckResults}.
We can observe that for the time domain input, it is almost the same with and without the bottleneck layer.
Likewise, for the frequency domain input, it is also difficult to judge which one is better.
Thus, choosing a suitable network (according to datasets, transfer learning methods, input types, etc.), which can learn more transferable features, is very important for UDTL-based methods.
In addition, it is clear that when parameters of backbones are fixed during the transfer learning procedure, the accuracy in the target domain decreases dramatically, which means that backbones trained using the source data cannot be transferred directly to the target domain. 

\subsection{Negative transfer}
As we discussed in Section \ref{S:4}, there are mainly four kinds of scenarios of UDTL-based IFD, but all experiments with five datasets are about transfer between different working conditions.
To state that these scenarios are not always suitable for generating the positive transfer, we use the PU dataset to design another transfer task considering the transfer between different methods of generating damages. 
Each task consists of three health conditions, and detailed information is listed in Table \ref{PU_Tab_3}.
There are two transfer learning settings in total.
\begin{table}[!t]
	\caption{The information of bearings with artificial damages.} 
	\centering
	\label{PU_Tab_3}  
	\begin{tabular}{cp{1.2cm}p{1cm}p{1cm}p{1.2cm}c}
		\hline
		Task               & Precast Method    & Damage Location & Damage Extent & Bearing Code &Label \\ \hline
		\multirow{3}{*}{0} & Electric Engraver & OR              & 1             & KA05 & 0         \\ \cline{2-6} 
		& Electric Engraver & OR              & 2             & KA03 & 1         \\ \cline{2-6} 
		& Electric Engraver & IR              & 1            & KI03    & 2     \\ \hline
		\multirow{3}{*}{1} & EDM and Drilling               & OR              & 1             & KA01 and KA07 & 0         \\ \cline{2-6} 
		& Drilling      & OR              & 2            &  KA08 & 1   \\ \cline{2-6} 
		& EDM          & IR              & 1             & KI01 & 2         \\ \hline
	\end{tabular}
\end{table}

\begin{figure}[!t]
	\centering
	\subfigure{\includegraphics[scale = 0.6]{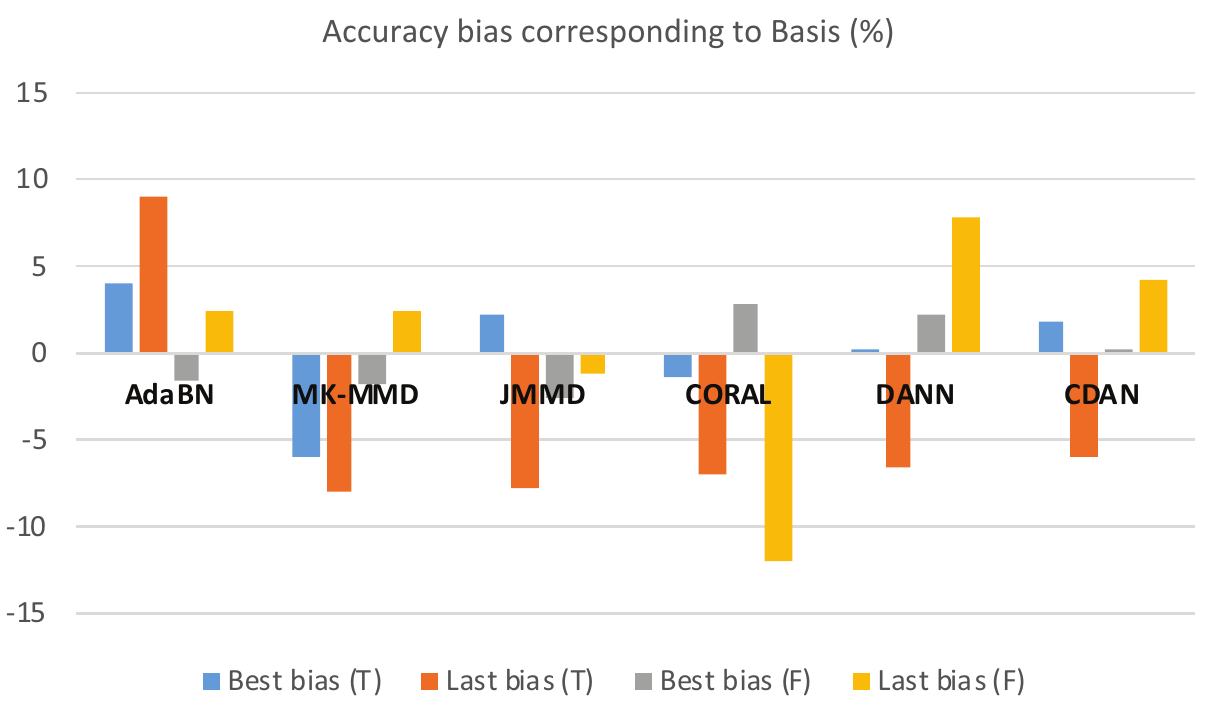}}
	\\ [-10pt]
	\caption{The accuracy biases of these five methods corresponding to Basis. (F) means the frequency domain input, and (T) means the time domain input.}
	\label{Transfer-PU}
\end{figure}

The transfer results are shown in \figref{Transfer-PU} and \textbf{Appendix A} called PU-Types.
We can observe that each method has a negative transfer with the time or frequency domain inputs, and this phenomenon indicates that this constructed task may not be suitable for the transfer learning task.
Actually, there are also some published papers designing transfer learning tasks which tackle transferring the gear samples to the bearing samples (it may not be a reliable transfer task) or transferring the experimental data to the real data (if structures of two machines are different, it also may not be a reliable transfer task).
Thus, it is very important to first figure out whether this task is suitable for transfer learning and whether two domains do have shared features.

\subsection{Physical priors}
In the field of computer vision and natural language processing, new transfer learning methods often use the existing knowledge or laws to provide a meaningful explanation, such as attention mechanism \cite{vaswani2017attention} and multi-modal structures \cite{long2018conditional}.

However, for UDTL-based IFD, many scholars only introduce methods, which have already existed in other fields, to perform IFD tasks and pay less attention to the prior knowledge behind the data (lack of using special phenomena or rules in physical systems).
Therefore, we suggest that scholars can learn from core ideas in the field of transfer learning (not just use the existing methods) and introduce prior knowledge of physical systems into the proposed method to construct more targeted and suitable diagnostic models with higher recognition rates in industrial applications.

\subsection{Label-inconsistent transfer}
Recently, some scholars have considered the label-inconsistent scenario and proposed some specific methods to allow the model to adapt to this situation (detailed references can be found in the above review).
However, as discussed in comparative results of label-inconsistent transfer, selected methods often face the risk of overfitting.
That is, although the best average accuracy is acceptable, the last average accuracy often has a big drop.
The main reason might be that models cannot focus on shared classes effectively, leading to poor domain alignment.

Hence, more attention should be paid to the label-inconsistent scenario to realize effective extra source classes separation and unknown classes detection from the target domain.
A possible solution is to combine other valid open set recognition algorithms for better unknown class detection \cite{jing2021balanced,9394793}.
For example, an EVT model using deep features of source samples, was applied to detect unknown-class samples \cite{9394793}.

\subsection{Multi-domain transfer}
Most of published papers are based on a single source domain, but in real applications, the labeled data might be from multiple source domains.
These domains often follow different distributions, but shared or related features exist among multiple source domains.
A common step is to align the shared features via multi-domain adaptation or DG.
However, how to balance contributions of multiple source domains is still not well solved.
For example, in comparative analysis, we simply assume that each domain contributes equally to transfer learning.
Thus, suitable weights should be carefully designed and added into the process of multi-domain transfer.

Additionally, to make better use of some data in unlabeled source domains, semi-supervised multi-domain learning \cite{liao2020deep} might also be worth focusing on.
To further improve the accuracy, minimizing the gap of conditional distributions might be an effective way to align shared features \cite{zhang2021ConditionalAdversarialDG,DG2018ECCV}.

\subsection{Other aspects}
Although a large amount of data in different conditions can be collected, fault data in some conditions are still scarce.
Due to the fact that most machines operate in a normal condition, the class-imbalanced problem often naturally exists in real applications.
Thus, imbalanced learning or few shot learning combined with transfer learning methods \cite{wu2020few} might also be an important direction for better getting constructed algorithms off the ground.

Federated transfer learning (FTL) \cite{liu2020secure} provides a safer and more reliable approach for specific industries.
At the same time, based on characteristics of transfer learning, FTL participants can own their own feature space without requiring all participants to own or use the same feature data, which makes FTL suitable for more application scenarios.
FTL was initially used in IFD \cite{zhang2021federated} and more in-depth research is required.

Uncertainty quantification plays a critical role in assessing the safety of DL models during construction, optimization, and decision making procedures.
Bayesian networks \cite{gal2016dropout} and ensemble learning techniques \cite{lakshminarayanan2017simple} are two widely-used uncertainty quantification methods, and their effectiveness has been verified by different kinds of applications, such as bioinformatics, self-driving car, etc.
Thus, uncertainty as an auxiliary term can be used to further correct some inappropriate predictions or results during the transfer learning.
For example, the prediction uncertainty is explicitly estimated  during training to rectify the pseudo label learning for UDTL of semantic segmentation \cite{zheng2021rectifying}.
\section{Conclusion}
\label{S:9}
In this paper, we construct a new taxonomy and perform a comprehensive review of UDTL-based IFD according to different tasks of UDTL.
Five publicly available datasets are gathered to perform a comparative analysis of different UDTL-based IFD methods from several perspectives.
Based on the systematically comparative study, we conclude that some useful results might be helpful for further research.
Firstly, the accuracy of CWRU and JNU is larger than 95\%.
Secondly, results of different methods indicate that the assumption of joint distributions and adversarial training are two helpful techniques for promoting the accuracy.
Thirdly, different input types often behave differently on each dataset, and choosing a suitable input type might also be important to improve the accuracy.
Finally, the stability and robustness of UDTL-based IFD need to be taken seriously.
To sum up, it might be useful for scholars to think ahead of these results before developing new models.

Also, we release the code library at \url{https://github.com/ZhaoZhibin/UDTL} and try to give a basic performance of current algorithms to find the core that determines the transfer performance of algorithms to guide future research.

\bibliographystyle{References/IEEEbib}
\bibliography{References/Reference}\ 

\begin{thebibliography}{100}

\bibitem{kankar2011fault}
Pavan~Kumar Kankar, Satish~C Sharma, and Suraj~Prakash Harsha,
\newblock ``Fault diagnosis of ball bearings using machine learning methods,''
\newblock {\em Expert Systems with applications}, vol. 38, no. 3, pp.
  1876--1886, 2011.

\bibitem{cerrada2016fault}
Mariela Cerrada, Grover Zurita, Diego Cabrera, Ren{\'e}-Vinicio S{\'a}nchez,
  Mariano Art{\'e}s, and Chuan Li,
\newblock ``Fault diagnosis in spur gears based on genetic algorithm and random
  forest,''
\newblock {\em Mechanical Systems and Signal Processing}, vol. 70, pp. 87--103,
  2016.

\bibitem{widodo2007support}
Achmad Widodo and Bo-Suk Yang,
\newblock ``Support vector machine in machine condition monitoring and fault
  diagnosis,''
\newblock {\em Mechanical systems and signal processing}, vol. 21, no. 6, pp.
  2560--2574, 2007.

\bibitem{tong2020ridge}
Chaowei Tong, Shibin Wang, Ivan Selesnick, Ruqiang Yan, and Xuefeng Chen,
\newblock ``Ridge-aware weighted sparse time-frequency representation,''
\newblock {\em IEEE Transactions on Signal Processing}, vol. 69, pp. 136--149,
  2020.

\bibitem{zhao2018enhanced}
Zhibin Zhao, Shuming Wu, Baijie Qiao, Shibin Wang, and Xuefeng Chen,
\newblock ``Enhanced sparse period-group lasso for bearing fault diagnosis,''
\newblock {\em IEEE Transactions on Industrial Electronics}, vol. 66, no. 3,
  pp. 2143--2153, 2018.

\bibitem{zhao2021fast}
Zhibin Zhao, Shibin Wang, David Wong, Wendong Wang, Ruqiang Yan, and Xuefeng
  Chen,
\newblock ``Fast sparsity-assisted signal decomposition with non-convex
  enhancement for bearing fault diagnosis,''
\newblock {\em IEEE/ASME Transactions on Mechatronics}, 2021.

\bibitem{lecun2015deep}
Yann LeCun, Yoshua Bengio, and Geoffrey Hinton,
\newblock ``Deep learning,''
\newblock {\em nature}, vol. 521, no. 7553, pp. 436, 2015.

\bibitem{ravi2016deep}
Daniele Rav{\`\i}, Charence Wong, Fani Deligianni, Melissa Berthelot, Javier
  Andreu-Perez, Benny Lo, and Guang-Zhong Yang,
\newblock ``Deep learning for health informatics,''
\newblock {\em IEEE journal of biomedical and health informatics}, vol. 21, no.
  1, pp. 4--21, 2016.

\bibitem{min2017deep}
Seonwoo Min, Byunghan Lee, and Sungroh Yoon,
\newblock ``Deep learning in bioinformatics,''
\newblock {\em Briefings in bioinformatics}, vol. 18, no. 5, pp. 851--869,
  2017.

\bibitem{yang2019robust}
Lei Yang, Yunfei Wang, Yanjie Guo, Weiqiang Zhang, and Zhibin Zhao,
\newblock ``Robust working mechanism of water droplet-driven triboelectric
  nanogenerator: Triboelectric output versus dynamic motion of water droplet,''
\newblock {\em Advanced Materials Interfaces}, vol. 6, no. 24, pp. 1901547,
  2019.

\bibitem{yang2020particle}
Lei Yang, Yunfei Wang, Zhibin Zhao, Yanjie Guo, Sicheng Chen, Weiqiang Zhang,
  and Xiao Guo,
\newblock ``Particle-laden droplet-driven triboelectric nanogenerator for
  real-time sediment monitoring using a deep learning method,''
\newblock {\em ACS Applied Materials \& Interfaces}, vol. 12, no. 34, pp.
  38192--38201, 2020.

\bibitem{zhu2020stacked}
Haiping Zhu, Jiaxin Cheng, Cong Zhang, Jun Wu, and Xinyu Shao,
\newblock ``Stacked pruning sparse denoising autoencoder based intelligent
  fault diagnosis of rolling bearings,''
\newblock {\em Applied Soft Computing}, p. 106060, 2020.

\bibitem{zhang2020knowledge}
Wenfeng Zhang, Gautam Biswas, Qi~Zhao, Hongbo Zhao, and Wenquan Feng,
\newblock ``Knowledge distilling based model compression and feature learning
  in fault diagnosis,''
\newblock {\em Applied Soft Computing}, vol. 88, pp. 105958, 2020.

\bibitem{xu2018roller}
Fan Xu, Yiu~Lun Tse, et~al.,
\newblock ``Roller bearing fault diagnosis using stacked denoising autoencoder
  in deep learning and gath--geva clustering algorithm without principal
  component analysis and data label,''
\newblock {\em Applied Soft Computing}, vol. 73, pp. 898--913, 2018.

\bibitem{krizhevsky2012imagenet}
Alex Krizhevsky, Ilya Sutskever, and Geoffrey~E Hinton,
\newblock ``Imagenet classification with deep convolutional neural networks,''
\newblock in {\em Advances in neural information processing systems}, 2012, pp.
  1097--1105.

\bibitem{ng2011sparse}
Andrew Ng et~al.,
\newblock ``Sparse autoencoder,''
\newblock {\em CS294A Lecture notes}, vol. 72, no. 2011, pp. 1--19, 2011.

\bibitem{zhao2020deep}
Zhibin Zhao, Tianfu Li, Jingyao Wu, Chuang Sun, Shibin Wang, Ruqiang Yan, and
  Xuefeng Chen,
\newblock ``Deep learning algorithms for rotating machinery intelligent
  diagnosis: An open source benchmark study,''
\newblock {\em ISA transactions}, vol. 107, pp. 224--255, 2020.

\bibitem{zheng2019cross}
Huailiang Zheng, Rixin Wang, Yuantao Yang, Jiancheng Yin, Yongbo Li, Yuqing Li,
  and Minqiang Xu,
\newblock ``Cross-domain fault diagnosis using knowledge transfer strategy: a
  review,''
\newblock {\em IEEE Access}, vol. 7, pp. 129260--129290, 2019.

\bibitem{yan2019knowledge}
Ruqiang Yan, Fei Shen, Chuang Sun, and Xuefeng Chen,
\newblock ``Knowledge transfer for rotary machine fault diagnosis,''
\newblock {\em IEEE Sensors Journal}, vol. 20, no. 15, pp. 8374--8393, 2019.

\bibitem{lei2020applications}
Yaguo Lei, Bin Yang, Xinwei Jiang, Feng Jia, Naipeng Li, and Asoke~K Nandi,
\newblock ``Applications of machine learning to machine fault diagnosis: A
  review and roadmap,''
\newblock {\em Mechanical Systems and Signal Processing}, vol. 138, pp. 106587,
  2020.

\bibitem{long2015learning}
Mingsheng Long, Yue Cao, Jianmin Wang, and Michael Jordan,
\newblock ``Learning transferable features with deep adaptation networks,''
\newblock in {\em International conference on machine learning}. PMLR, 2015,
  pp. 97--105.

\bibitem{tan2018survey}
Chuanqi Tan, Fuchun Sun, Tao Kong, Wenchang Zhang, Chao Yang, and Chunfang Liu,
\newblock ``A survey on deep transfer learning,''
\newblock in {\em International Conference on Artificial Neural Networks}.
  Springer, 2018, pp. 270--279.

\bibitem{li2019multi}
Xiang Li, Wei Zhang, Qian Ding, and Jian-Qiao Sun,
\newblock ``Multi-layer domain adaptation method for rolling bearing fault
  diagnosis,''
\newblock {\em Signal Processing}, vol. 157, pp. 180--197, 2019.

\bibitem{simonyan2014very}
Karen Simonyan and Andrew Zisserman,
\newblock ``Very deep convolutional networks for large-scale image
  recognition,''
\newblock {\em arXiv preprint arXiv:1409.1556}, 2014.

\bibitem{he2016deep}
Kaiming He, Xiangyu Zhang, Shaoqing Ren, and Jian Sun,
\newblock ``Deep residual learning for image recognition,''
\newblock in {\em Proceedings of the IEEE conference on computer vision and
  pattern recognition}, 2016, pp. 770--778.

\bibitem{zhang2017transfer}
Ran Zhang, Hongyang Tao, Lifeng Wu, and Yong Guan,
\newblock ``Transfer learning with neural networks for bearing fault diagnosis
  in changing working conditions,''
\newblock {\em IEEE Access}, vol. 5, pp. 14347--14357, 2017.

\bibitem{zhang2018method}
Cheng Zhang, Liqing Xu, Xingwang Li, and Huiyun Wang,
\newblock ``A method of fault diagnosis for rotary equipment based on deep
  learning,''
\newblock in {\em 2018 Prognostics and System Health Management Conference
  (PHM-Chongqing)}. IEEE, 2018, pp. 958--962.

\bibitem{chen2018incipient}
Danmin Chen, Shuai Yang, and Funa Zhou,
\newblock ``Incipient fault diagnosis based on dnn with transfer learning,''
\newblock in {\em 2018 International Conference on Control, Automation and
  Information Sciences (ICCAIS)}. IEEE, 2018, pp. 303--308.

\bibitem{hasan20181d}
Md~Junayed Hasan, Muhammad Sohaib, and Jong-Myon Kim,
\newblock ``1d cnn-based transfer learning model for bearing fault diagnosis
  under variable working conditions,''
\newblock in {\em International Conference on Computational Intelligence in
  Information System}. Springer, 2018, pp. 13--23.

\bibitem{kim2019new}
Hyunjae Kim and Byeng~D Youn,
\newblock ``A new parameter repurposing method for parameter transfer with
  small dataset and its application in fault diagnosis of rolling element
  bearings,''
\newblock {\em IEEE Access}, vol. 7, pp. 46917--46930, 2019.

\bibitem{hasan2019acoustic}
Md~Junayed Hasan, MM~Manjurul Islam, and Jong-Myon Kim,
\newblock ``Acoustic spectral imaging and transfer learning for reliable
  bearing fault diagnosis under variable speed conditions,''
\newblock {\em Measurement}, vol. 138, pp. 620--631, 2019.

\bibitem{sun2018deep}
Chuang Sun, Meng Ma, Zhibin Zhao, Shaohua Tian, Ruqiang Yan, and Xuefeng Chen,
\newblock ``Deep transfer learning based on sparse autoencoder for remaining
  useful life prediction of tool in manufacturing,''
\newblock {\em IEEE Transactions on Industrial Informatics}, vol. 15, no. 4,
  pp. 2416--2425, 2018.

\bibitem{shao2018highly}
Siyu Shao, Stephen McAleer, Ruqiang Yan, and Pierre Baldi,
\newblock ``Highly accurate machine fault diagnosis using deep transfer
  learning,''
\newblock {\em IEEE Transactions on Industrial Informatics}, vol. 15, no. 4,
  pp. 2446--2455, 2018.

\bibitem{chen2019transfer}
Danmin Chen, Shuai Yang, and Funa Zhou,
\newblock ``Transfer learning based fault diagnosis with missing data due to
  multi-rate sampling,''
\newblock {\em Sensors}, vol. 19, no. 8, pp. 1826, 2019.

\bibitem{mao2019online}
Wentao Mao, Ling Ding, Siyu Tian, and Xihui Liang,
\newblock ``Online detection for bearing incipient fault based on deep transfer
  learning,''
\newblock {\em Measurement}, p. 107278, 2019.

\bibitem{he2020deep}
Zhiyi He, Haidong Shao, Ping Wang, Janet~Jing Lin, Junsheng Cheng, and Yu~Yang,
\newblock ``Deep transfer multi-wavelet auto-encoder for intelligent fault
  diagnosis of gearbox with few target training samples,''
\newblock {\em Knowledge-Based Systems}, vol. 191, pp. 105313, 2020.

\bibitem{li2020cross}
Fudong Li, Jinglong Chen, Jun Pan, and Tongyang Pan,
\newblock ``Cross-domain learning in rotating machinery fault diagnosis under
  various operating conditions based on parameter transfer,''
\newblock {\em Measurement Science and Technology}, vol. 31, no. 8, pp. 085104,
  2020.

\bibitem{sharaf2018beam}
Sayed~Ali Sharaf,
\newblock ``Beam pump dynamometer card prediction using artificial neural
  networks,''
\newblock {\em KnE Engineering}, vol. 3, no. 7, pp. 198--212, 2018.

\bibitem{cao2018preprocessing}
Pei Cao, Shengli Zhang, and Jiong Tang,
\newblock ``Preprocessing-free gear fault diagnosis using small datasets with
  deep convolutional neural network-based transfer learning,''
\newblock {\em IEEE Access}, vol. 6, pp. 26241--26253, 2018.

\bibitem{chen2019intelligent}
Zhuyun Chen, Konstantinos Gryllias, and Weihua Li,
\newblock ``Intelligent fault diagnosis for rotary machinery using transferable
  convolutional neural network,''
\newblock {\em IEEE Transactions on Industrial Informatics}, vol. 16, no. 1,
  pp. 339--349, 2019.

\bibitem{iba2019vibration}
D~Iba, Y~Ishii, Y~Tsutsui, N~Miura, T~Iizuka, A~Masuda, A~Sone, and I~Moriwaki,
\newblock ``Vibration analysis of a meshing gear pair by neural network
  (visualization of meshing vibration and detection of a crack at tooth root by
  vgg16 with transfer learning),''
\newblock in {\em Smart Structures and NDE for Energy Systems and Industry
  4.0}. International Society for Optics and Photonics, 2019, vol. 10973, p.
  109730Y.

\bibitem{ma2019novel}
Ping Ma, Hongli Zhang, Wenhui Fan, Cong Wang, Guangrui Wen, and Xining Zhang,
\newblock ``A novel bearing fault diagnosis method based on 2d image
  representation and transfer learning-convolutional neural network,''
\newblock {\em Measurement Science and Technology}, vol. 30, no. 5, pp. 055402,
  2019.

\bibitem{he2020ensemble}
Zhiyi He, Haidong Shao, Xiang Zhong, and Xianzhu Zhao,
\newblock ``Ensemble transfer cnns driven by multi-channel signals for fault
  diagnosis of rotating machinery cross working conditions,''
\newblock {\em Knowledge-Based Systems}, vol. 207, pp. 106396, 2020.

\bibitem{di2021ensemble}
ZiYang Di, HaiDong Shao, and JiaWei Xiang,
\newblock ``Ensemble deep transfer learning driven by multisensor signals for
  the fault diagnosis of bevel-gear cross-operation conditions,''
\newblock {\em Science China Technological Sciences}, vol. 64, no. 3, pp.
  481--492, 2021.

\bibitem{wang2020deformable}
Zheng Wang, Qingxiu Liu, Hansi Chen, and Xuening Chu,
\newblock ``A deformable cnn-dlstm based transfer learning method for fault
  diagnosis of rolling bearing under multiple working conditions,''
\newblock {\em International Journal of Production Research}, pp. 1--15, 2020.

\bibitem{wang2019deep}
Jianyu Wang, Zhenling Mo, Heng Zhang, and Qiang Miao,
\newblock ``A deep learning method for bearing fault diagnosis based on
  time-frequency image,''
\newblock {\em IEEE Access}, vol. 7, pp. 42373--42383, 2019.

\bibitem{shao2020intelligent}
Haidong Shao, Min Xia, Guangjie Han, Yu~Zhang, and Jiafu Wan,
\newblock ``Intelligent fault diagnosis of rotor-bearing system under varying
  working conditions with modified transfer convolutional neural network and
  thermal images,''
\newblock {\em IEEE Transactions on Industrial Informatics}, vol. 17, no. 5,
  pp. 3488--3496, 2020.

\bibitem{qureshi2017wind}
Aqsa~Saeed Qureshi, Asifullah Khan, Aneela Zameer, and Anila Usman,
\newblock ``Wind power prediction using deep neural network based meta
  regression and transfer learning,''
\newblock {\em Applied Soft Computing}, vol. 58, pp. 742--755, 2017.

\bibitem{zhong2019novel}
Shi-sheng Zhong, Song Fu, and Lin Lin,
\newblock ``A novel gas turbine fault diagnosis method based on transfer
  learning with cnn,''
\newblock {\em Measurement}, vol. 137, pp. 435--453, 2019.

\bibitem{han2019learning}
Te~Han, Chao Liu, Wenguang Yang, and Dongxiang Jiang,
\newblock ``Learning transferable features in deep convolutional neural
  networks for diagnosing unseen machine conditions,''
\newblock {\em ISA transactions}, vol. 93, pp. 341--353, 2019.

\bibitem{xu2019online}
Gaowei Xu, Min Liu, Zhuofu Jiang, Weiming Shen, and Chenxi Huang,
\newblock ``Online fault diagnosis method based on transfer convolutional
  neural networks,''
\newblock {\em IEEE Transactions on Instrumentation and Measurement}, vol. 69,
  no. 2, pp. 509--520, 2019.

\bibitem{zhao2020deepmultiscale}
Bo~Zhao, Xianmin Zhang, Zhenhui Zhan, and Shuiquan Pang,
\newblock ``Deep multi-scale convolutional transfer learning network: A novel
  method for intelligent fault diagnosis of rolling bearings under variable
  working conditions and domains,''
\newblock {\em Neurocomputing}, vol. 407, pp. 24--38, 2020.

\bibitem{dai2007boosting}
Wenyuan Dai, Qiang Yang, Gui-Rong Xue, and Yong Yu,
\newblock ``Boosting for transfer learning,''
\newblock in {\em Proceedings of the 24th international conference on Machine
  learning}. ACM, 2007, pp. 193--200.

\bibitem{li2016revisiting}
Yanghao Li, Naiyan Wang, Jianping Shi, Jiaying Liu, and Xiaodi Hou,
\newblock ``Revisiting batch normalization for practical domain adaptation,''
\newblock {\em arXiv preprint arXiv:1603.04779}, 2016.

\bibitem{ioffe2015batch}
Sergey Ioffe and Christian Szegedy,
\newblock ``Batch normalization: Accelerating deep network training by reducing
  internal covariate shift,''
\newblock {\em arXiv preprint arXiv:1502.03167}, 2015.

\bibitem{xiao2019transfer}
Dengyu Xiao, Yixiang Huang, Chengjin Qin, Zhiyu Liu, Yanming Li, and Chengliang
  Liu,
\newblock ``Transfer learning with convolutional neural networks for small
  sample size problem in machinery fault diagnosis,''
\newblock {\em Proceedings of the Institution of Mechanical Engineers, Part C:
  Journal of Mechanical Engineering Science}, p. 0954406219840381, 2019.

\bibitem{zhang2017new}
Wei Zhang, Gaoliang Peng, Chuanhao Li, Yuanhang Chen, and Zhujun Zhang,
\newblock ``A new deep learning model for fault diagnosis with good anti-noise
  and domain adaptation ability on raw vibration signals,''
\newblock {\em Sensors}, vol. 17, no. 2, pp. 425, 2017.

\bibitem{qian2018new}
Weiwei Qian, Shunming Li, and Jinrui Wang,
\newblock ``A new transfer learning method and its application on rotating
  machine fault diagnosis under variant working conditions,''
\newblock {\em IEEE Access}, vol. 6, pp. 69907--69917, 2018.

\bibitem{sun2016deep}
Baochen Sun and Kate Saenko,
\newblock ``Deep coral: Correlation alignment for deep domain adaptation,''
\newblock in {\em European Conference on Computer Vision}. Springer, 2016, pp.
  443--450.

\bibitem{borgwardt2006integrating}
Karsten~M Borgwardt, Arthur Gretton, Malte~J Rasch, Hans-Peter Kriegel,
  Bernhard Sch{\"o}lkopf, and Alex~J Smola,
\newblock ``Integrating structured biological data by kernel maximum mean
  discrepancy,''
\newblock {\em Bioinformatics}, vol. 22, no. 14, pp. e49--e57, 2006.

\bibitem{sejdinovic2013equivalence}
Dino Sejdinovic, Bharath Sriperumbudur, Arthur Gretton, Kenji Fukumizu, et~al.,
\newblock ``Equivalence of distance-based and rkhs-based statistics in
  hypothesis testing,''
\newblock {\em The Annals of Statistics}, vol. 41, no. 5, pp. 2263--2291, 2013.

\bibitem{gretton2012optimal}
Arthur Gretton, Dino Sejdinovic, Heiko Strathmann, Sivaraman Balakrishnan,
  Massimiliano Pontil, Kenji Fukumizu, and Bharath~K Sriperumbudur,
\newblock ``Optimal kernel choice for large-scale two-sample tests,''
\newblock in {\em Advances in neural information processing systems}, 2012, pp.
  1205--1213.

\bibitem{long2013transfer}
Mingsheng Long, Jianmin Wang, Guiguang Ding, Jiaguang Sun, and Philip~S Yu,
\newblock ``Transfer feature learning with joint distribution adaptation,''
\newblock in {\em Proceedings of the IEEE international conference on computer
  vision}, 2013, pp. 2200--2207.

\bibitem{wang2017balanced}
Jindong Wang, Yiqiang Chen, Shuji Hao, Wenjie Feng, and Zhiqi Shen,
\newblock ``Balanced distribution adaptation for transfer learning,''
\newblock in {\em 2017 IEEE International Conference on Data Mining (ICDM)}.
  IEEE, 2017, pp. 1129--1134.

\bibitem{long2017deep}
Mingsheng Long, Han Zhu, Jianmin Wang, and Michael~I Jordan,
\newblock ``Deep transfer learning with joint adaptation networks,''
\newblock in {\em Proceedings of the 34th International Conference on Machine
  Learning-Volume 70}. JMLR. org, 2017, pp. 2208--2217.

\bibitem{pan2010domain}
Sinno~Jialin Pan, Ivor~W Tsang, James~T Kwok, and Qiang Yang,
\newblock ``Domain adaptation via transfer component analysis,''
\newblock {\em IEEE Transactions on Neural Networks}, vol. 22, no. 2, pp.
  199--210, 2010.

\bibitem{tzeng2014deep}
Eric Tzeng, Judy Hoffman, Ning Zhang, Kate Saenko, and Trevor Darrell,
\newblock ``Deep domain confusion: Maximizing for domain invariance,''
\newblock {\em arXiv preprint arXiv:1412.3474}, 2014.

\bibitem{zynda2020weights}
T{\L}~{\.Z}ynda,
\newblock ``On weights which admit reproducing kernel of szeg{\H{o}} type,''
\newblock {\em Journal of Contemporary Mathematical Analysis (Armenian Academy
  of Sciences)}, vol. 55, no. 5, pp. 320--327, 2020.

\bibitem{sun2016return}
Baochen Sun, Jiashi Feng, and Kate Saenko,
\newblock ``Return of frustratingly easy domain adaptation,''
\newblock in {\em Thirtieth AAAI Conference on Artificial Intelligence}, 2016.

\bibitem{wang2018power}
Kaijie Wang and Bin Wu,
\newblock ``Power equipment fault diagnosis model based on deep transfer
  learning with balanced distribution adaptation,''
\newblock in {\em International Conference on Advanced Data Mining and
  Applications}. Springer, 2018, pp. 178--188.

\bibitem{wang2020network}
Yujing Wang, Chao Wang, Shouqiang Kang, Jinbao Xie, Qingyan Wang, and
  VI~Mikulovich,
\newblock ``Network-combined broad learning and transfer learning: a new
  intelligent fault diagnosis method for rolling bearings,''
\newblock {\em Measurement Science and Technology}, vol. 31, no. 11, pp.
  115013, 2020.

\bibitem{wang2019hierarchical}
Xiaoxia Wang, Haibo He, and Lusi Li,
\newblock ``A hierarchical deep domain adaptation approach for fault diagnosis
  of power plant thermal system,''
\newblock {\em IEEE Transactions on Industrial Informatics}, vol. 15, no. 9,
  pp. 5139--5148, 2019.

\bibitem{an2020deep}
Jing An, Ping Ai, and Dakun Liu,
\newblock ``Deep domain adaptation model for bearing fault diagnosis with
  domain alignment and discriminative feature learning,''
\newblock {\em Shock and Vibration}, vol. 2020, 2020.

\bibitem{zhang2020unsupervised}
Zhongwei Zhang, Huaihai Chen, Shunming Li, and Zenghui An,
\newblock ``Unsupervised domain adaptation via enhanced transfer joint matching
  for bearing fault diagnosis,''
\newblock {\em Measurement}, vol. 165, pp. 108071, 2020.

\bibitem{qian2019deep}
Weiwei Qian, Shunming Li, and Xingxing Jiang,
\newblock ``Deep transfer network for rotating machine fault analysis,''
\newblock {\em Pattern Recognition}, vol. 96, pp. 106993, 2019.

\bibitem{hu2020cross}
Chaofan Hu, Yanxue Wang, and Jiawei Gu,
\newblock ``Cross-domain intelligent fault classification of bearings based on
  tensor-aligned invariant subspace learning and two-dimensional convolutional
  neural networks,''
\newblock {\em Knowledge-Based Systems}, vol. 209, pp. 106214, 2020.

\bibitem{lu2016deep}
Weining Lu, Bin Liang, Yu~Cheng, Deshan Meng, Jun Yang, and Tao Zhang,
\newblock ``Deep model based domain adaptation for fault diagnosis,''
\newblock {\em IEEE Transactions on Industrial Electronics}, vol. 64, no. 3,
  pp. 2296--2305, 2016.

\bibitem{zhang2018intelligent}
Bo~Zhang, Wei Li, Xiao-Li Li, and See-Kiong Ng,
\newblock ``Intelligent fault diagnosis under varying working conditions based
  on domain adaptive convolutional neural networks,''
\newblock {\em IEEE Access}, vol. 6, pp. 66367--66384, 2018.

\bibitem{wen2017new}
Long Wen, Liang Gao, and Xinyu Li,
\newblock ``A new deep transfer learning based on sparse auto-encoder for fault
  diagnosis,''
\newblock {\em IEEE Transactions on Systems, Man, and Cybernetics: Systems},
  vol. 49, no. 1, pp. 136--144, 2017.

\bibitem{yang2019intelligent}
Bin Yang, Yaguo Lei, Feng Jia, and Saibo Xing,
\newblock ``An intelligent fault diagnosis approach based on transfer learning
  from laboratory bearings to locomotive bearings,''
\newblock {\em Mechanical Systems and Signal Processing}, vol. 122, pp.
  692--706, 2019.

\bibitem{tang2019transfer}
Shanxuan Tang, Hailong Tang, and Min Chen,
\newblock ``Transfer-learning based gas path analysis method for gas
  turbines,''
\newblock {\em Applied Thermal Engineering}, vol. 155, pp. 1--13, 2019.

\bibitem{li2018cross}
Xiang Li, Wei Zhang, and Qian Ding,
\newblock ``Cross-domain fault diagnosis of rolling element bearings using deep
  generative neural networks,''
\newblock {\em IEEE Transactions on Industrial Electronics}, vol. 66, no. 7,
  pp. 5525--5534, 2018.

\bibitem{xu2019digital}
Yan Xu, Yanming Sun, Xiaolong Liu, and Yonghua Zheng,
\newblock ``A digital-twin-assisted fault diagnosis using deep transfer
  learning,''
\newblock {\em IEEE Access}, vol. 7, pp. 19990--19999, 2019.

\bibitem{an2019intelligent}
Zenghui An, Shunming Li, Yu~Xin, Kun Xu, and Huijie Ma,
\newblock ``An intelligent fault diagnosis framework dealing with arbitrary
  length inputs under different working conditions,''
\newblock {\em Measurement Science and Technology}, vol. 30, no. 12, pp.
  125107, 2019.

\bibitem{li2020intelligent}
Xiang Li, Xiao-Dong Jia, Wei Zhang, Hui Ma, Zhong Luo, and Xu~Li,
\newblock ``Intelligent cross-machine fault diagnosis approach with deep
  auto-encoder and domain adaptation,''
\newblock {\em Neurocomputing}, vol. 383, pp. 235--247, 2020.

\bibitem{tong2018bearing1}
Zhe Tong, Wei Li, Bo~Zhang, and Meng Zhang,
\newblock ``Bearing fault diagnosis based on domain adaptation using
  transferable features under different working conditions,''
\newblock {\em Shock and Vibration}, vol. 2018, 2018.

\bibitem{tong2018bearing2}
Zhe Tong, Wei Li, Bo~Zhang, Fan Jiang, and Gongbo Zhou,
\newblock ``Bearing fault diagnosis under variable working conditions based on
  domain adaptation using feature transfer learning,''
\newblock {\em IEEE Access}, vol. 6, pp. 76187--76197, 2018.

\bibitem{wang2020multi}
Xu~Wang, Changqing Shen, Min Xia, Dong Wang, Jun Zhu, and Zhongkui Zhu,
\newblock ``Multi-scale deep intra-class transfer learning for bearing fault
  diagnosis,''
\newblock {\em Reliability Engineering \& System Safety}, vol. 202, pp. 107050,
  2020.

\bibitem{li2020deepbalanced}
Qikang Li, Baoping Tang, Lei Deng, Yanling Wu, and Yi~Wang,
\newblock ``Deep balanced domain adaptation neural networks for fault diagnosis
  of planetary gearboxes with limited labeled data,''
\newblock {\em Measurement}, vol. 156, pp. 107570, 2020.

\bibitem{lu2021new}
Nannan Lu, Hanhan Xiao, Yanjing Sun, Min Han, and Yanfen Wang,
\newblock ``A new method for intelligent fault diagnosis of machines based on
  unsupervised domain adaptation,''
\newblock {\em Neurocomputing}, vol. 427, pp. 96--109, 2021.

\bibitem{yang2020polynomial}
Bin Yang, Yaguo Lei, Feng Jia, Naipeng Li, and Zhaojun Du,
\newblock ``A polynomial kernel induced distance metric to improve deep
  transfer learning for fault diagnosis of machines,''
\newblock {\em IEEE Transactions on Industrial Electronics}, vol. 67, no. 11,
  pp. 9747--9757, 2020.

\bibitem{cao2021domain}
Xincheng Cao, Yu~Wang, Binqiang Chen, and Nianyin Zeng,
\newblock ``Domain-adaptive intelligence for fault diagnosis based on deep
  transfer learning from scientific test rigs to industrial applications,''
\newblock {\em Neural Computing and Applications}, vol. 33, no. 9, pp.
  4483--4499, 2021.

\bibitem{zhao2020novel}
Ke~Zhao, Hongkai Jiang, Zhenghong Wu, and Tengfei Lu,
\newblock ``A novel transfer learning fault diagnosis method based on manifold
  embedded distribution alignment with a little labeled data,''
\newblock {\em Journal of Intelligent Manufacturing}, pp. 1--15, 2020.

\bibitem{zheng2020new}
Huailiang Zheng, Rixin Wang, Jiancheng Yin, Yuqing Li, Haiqing Lu, and Minqiang
  Xu,
\newblock ``A new intelligent fault identification method based on transfer
  locality preserving projection for actual diagnosis scenario of rotating
  machinery,''
\newblock {\em Mechanical Systems and Signal Processing}, vol. 135, pp. 106344,
  2020.

\bibitem{zhang2020novel}
Zhongwei Zhang, Huaihai Chen, Shunming Li, and Zenghui An,
\newblock ``A novel unsupervised domain adaptation based on deep neural network
  and manifold regularization for mechanical fault diagnosis,''
\newblock {\em Measurement Science and Technology}, vol. 31, no. 8, pp. 085101,
  2020.

\bibitem{qian2021discriminative}
Weiwei Qian, Shunming Li, Tong Yao, and Kun Xu,
\newblock ``Discriminative feature-based adaptive distribution alignment
  (dfada) for rotating machine fault diagnosis under variable working
  conditions,''
\newblock {\em Applied Soft Computing}, vol. 99, pp. 106886, 2021.

\bibitem{li2018robust}
Xiang Li, Wei Zhang, and Qian Ding,
\newblock ``A robust intelligent fault diagnosis method for rolling element
  bearings based on deep distance metric learning,''
\newblock {\em Neurocomputing}, vol. 310, pp. 77--95, 2018.

\bibitem{yang2018transfer}
Bin Yang, Yaguo Lei, Feng Jia, and Saibo Xing,
\newblock ``A transfer learning method for intelligent fault diagnosis from
  laboratory machines to real-case machines,''
\newblock in {\em 2018 International Conference on Sensing, Diagnostics,
  Prognostics, and Control (SDPC)}. IEEE, 2018, pp. 35--40.

\bibitem{an2019generalization}
Zenghui An, Shunming Li, Jinrui Wang, Yu~Xin, and Kun Xu,
\newblock ``Generalization of deep neural network for bearing fault diagnosis
  under different working conditions using multiple kernel method,''
\newblock {\em Neurocomputing}, vol. 352, pp. 42--53, 2019.

\bibitem{zhu2019new}
Jun Zhu, Nan Chen, and Changqing Shen,
\newblock ``A new deep transfer learning method for bearing fault diagnosis
  under different working conditions,''
\newblock {\em IEEE Sensors Journal}, vol. 20, no. 15, pp. 8394--8402, 2019.

\bibitem{che2020domain}
Changchang Che, Huawei Wang, Xiaomei Ni, and Qiang Fu,
\newblock ``Domain adaptive deep belief network for rolling bearing fault
  diagnosis,''
\newblock {\em Computers \& Industrial Engineering}, vol. 143, pp. 106427,
  2020.

\bibitem{han2019deep}
Te~Han, Chao Liu, Wenguang Yang, and Dongxiang Jiang,
\newblock ``Deep transfer network with joint distribution adaptation: A new
  intelligent fault diagnosis framework for industry application,''
\newblock {\em ISA transactions}, vol. 97, pp. 269--281, 2020.

\bibitem{qian2019novel}
Weiwei Qian, Shunming Li, Pengxing Yi, and Kaicheng Zhang,
\newblock ``A novel transfer learning method for robust fault diagnosis of
  rotating machines under variable working conditions,''
\newblock {\em Measurement}, vol. 138, pp. 514--525, 2019.

\bibitem{wu2020adaptive}
Zhenghong Wu, Hongkai Jiang, Ke~Zhao, and Xingqiu Li,
\newblock ``An adaptive deep transfer learning method for bearing fault
  diagnosis,''
\newblock {\em Measurement}, vol. 151, pp. 107227, 2020.

\bibitem{cao2020deep}
Xincheng Cao, Binqiang Chen, and Nianyin Zeng,
\newblock ``A deep domain adaption model with multi-task networks for planetary
  gearbox fault diagnosis,''
\newblock {\em Neurocomputing}, vol. 409, pp. 173--190, 2020.

\bibitem{ganin2016domain}
Yaroslav Ganin, Evgeniya Ustinova, Hana Ajakan, Pascal Germain, Hugo
  Larochelle, Fran{\c{c}}ois Laviolette, Mario Marchand, and Victor Lempitsky,
\newblock ``Domain-adversarial training of neural networks,''
\newblock {\em The Journal of Machine Learning Research}, vol. 17, no. 1, pp.
  2096--2030, 2016.

\bibitem{long2018conditional}
Mingsheng Long, Zhangjie Cao, Jianmin Wang, and Michael~I Jordan,
\newblock ``Conditional adversarial domain adaptation,''
\newblock in {\em Advances in Neural Information Processing Systems}, 2018, pp.
  1640--1650.

\bibitem{zhang2018adversarial}
Bo~Zhang, Wei Li, Jie Hao, Xiao-Li Li, and Meng Zhang,
\newblock ``Adversarial adaptive 1-d convolutional neural networks for bearing
  fault diagnosis under varying working condition,''
\newblock {\em arXiv preprint arXiv:1805.00778}, 2018.

\bibitem{han2019novel}
Te~Han, Chao Liu, Wenguang Yang, and Dongxiang Jiang,
\newblock ``A novel adversarial learning framework in deep convolutional neural
  network for intelligent diagnosis of mechanical faults,''
\newblock {\em Knowledge-Based Systems}, vol. 165, pp. 474--487, 2019.

\bibitem{guo2018deep}
Liang Guo, Yaguo Lei, Saibo Xing, Tao Yan, and Naipeng Li,
\newblock ``Deep convolutional transfer learning network: A new method for
  intelligent fault diagnosis of machines with unlabeled data,''
\newblock {\em IEEE Transactions on Industrial Electronics}, vol. 66, no. 9,
  pp. 7316--7325, 2018.

\bibitem{wang2019domain}
Qin Wang, Gabriel Michau, and Olga Fink,
\newblock ``Domain adaptive transfer learning for fault diagnosis,''
\newblock in {\em 2019 Prognostics and System Health Management Conference
  (PHM-Paris)}. IEEE, 2019, pp. 279--285.

\bibitem{chen2020domain}
Zhuyun Chen, Guolin He, Jipu Li, Yixiao Liao, Konstantinos Gryllias, and Weihua
  Li,
\newblock ``Domain adversarial transfer network for cross-domain fault
  diagnosis of rotary machinery,''
\newblock {\em IEEE Transactions on Instrumentation and Measurement}, vol. 69,
  no. 11, pp. 8702--8712, 2020.

\bibitem{zou2020adversarial}
Lei Zou, Yang Li, and Feiyun Xu,
\newblock ``An adversarial denoising convolutional neural network for fault
  diagnosis of rotating machinery under noisy environment and limited sample
  size case,''
\newblock {\em Neurocomputing}, vol. 407, pp. 105--120, 2020.

\bibitem{li2021knowledge}
Qi~Li, Changqing Shen, Liang Chen, and Zhongkui Zhu,
\newblock ``Knowledge mapping-based adversarial domain adaptation: A novel
  fault diagnosis method with high generalizability under variable working
  conditions,''
\newblock {\em Mechanical Systems and Signal Processing}, vol. 147, pp. 107095,
  2021.

\bibitem{li2021domain}
Tianfu Li, Zhibin Zhao, Chuang Sun, Ruqiang Yan, and Xuefeng Chen,
\newblock ``Domain adversarial graph convolutional network for fault diagnosis
  under variable working conditions,''
\newblock {\em IEEE Transactions on Instrumentation and Measurement}, vol. 70,
  pp. 1--10, 2021.

\bibitem{jiao2019unsupervised}
Jinyang Jiao, Ming Zhao, and Jing Lin,
\newblock ``Unsupervised adversarial adaptation network for intelligent fault
  diagnosis,''
\newblock {\em IEEE Transactions on Industrial Electronics}, vol. 67, no. 11,
  pp. 9904--9913, 2019.

\bibitem{yu2020symmetric}
Kun Yu, Hongzheng Han, Qiang Fu, Hui Ma, and Jin Zeng,
\newblock ``Symmetric co-training based unsupervised domain adaptation approach
  for intelligent fault diagnosis of rolling bearing,''
\newblock {\em Measurement Science and Technology}, vol. 31, no. 11, pp.
  115008, 2020.

\bibitem{jiao2020double}
Jinyang Jiao, Jing Lin, Ming Zhao, and Kaixuan Liang,
\newblock ``Double-level adversarial domain adaptation network for intelligent
  fault diagnosis,''
\newblock {\em Knowledge-Based Systems}, vol. 205, pp. 106236, 2020.

\bibitem{li2019deep}
Xiang Li, Wei Zhang, Nan-Xi Xu, and Qian Ding,
\newblock ``Deep learning-based machinery fault diagnostics with domain
  adaptation across sensors at different places,''
\newblock {\em IEEE Transactions on Industrial Electronics}, vol. 67, no. 8,
  pp. 6785--6794, 2019.

\bibitem{jia2020novel}
Sixiang Jia, Jinrui Wang, Baokun Han, Guowei Zhang, Xiaoyu Wang, and Jingtao
  He,
\newblock ``A novel transfer learning method for fault diagnosis using maximum
  classifier discrepancy with marginal probability distribution adaptation,''
\newblock {\em IEEE Access}, vol. 8, pp. 71475--71485, 2020.

\bibitem{zhang2020new}
Yongchao Zhang, Zhaohui Ren, and Shihua Zhou,
\newblock ``A new deep convolutional domain adaptation network for bearing
  fault diagnosis under different working conditions,''
\newblock {\em Shock and Vibration}, vol. 2020, 2020.

\bibitem{jiao2020residual}
Jinyang Jiao, Ming Zhao, Jing Lin, and Kaixuan Liang,
\newblock ``Residual joint adaptation adversarial network for intelligent
  transfer fault diagnosis,''
\newblock {\em Mechanical Systems and Signal Processing}, vol. 145, pp. 106962,
  2020.

\bibitem{li2021intelligent}
Yibin Li, Yan Song, Lei Jia, Shengyao Gao, Qiqiang Li, and Meikang Qiu,
\newblock ``Intelligent fault diagnosis by fusing domain adversarial training
  and maximum mean discrepancy via ensemble learning,''
\newblock {\em IEEE Transactions on Industrial Informatics}, vol. 17, no. 4,
  pp. 2833--2841, 2021.

\bibitem{qin2021multi}
Yi~Qin, Xin Wang, Quan Qian, Huayan Pu, and Jun Luo,
\newblock ``Multiscale transfer voting mechanism: A new strategy for domain
  adaption,''
\newblock {\em IEEE Transactions on Industrial Informatics}, vol. 17, no. 10,
  pp. 7103--7113, 2021.

\bibitem{qin2021parameter}
Yi~Qin, Qunwang Yao, Yi~Wang, and Yongfang Mao,
\newblock ``Parameter sharing adversarial domain adaptation networks for fault
  transfer diagnosis of planetary gearboxes,''
\newblock {\em Mechanical Systems and Signal Processing}, vol. 160, pp. 107936,
  2021.

\bibitem{yao2021multiscale}
Qunwang Yao, Yi~Qin, Xin Wang, and Quan Qian,
\newblock ``Multiscale domain adaption models and their application in fault
  transfer diagnosis of planetary gearboxes,''
\newblock {\em Engineering Applications of Artificial Intelligence}, vol. 104,
  pp. 104383, 2021.

\bibitem{cheng2020wasserstein}
Cheng Cheng, Beitong Zhou, Guijun Ma, Dongrui Wu, and Ye~Yuan,
\newblock ``Wasserstein distance based deep adversarial transfer learning for
  intelligent fault diagnosis with unlabeled or insufficient labeled data,''
\newblock {\em Neurocomputing}, vol. 409, pp. 35--45, 2020.

\bibitem{zhang2019deep}
Ming Zhang, Duo Wang, Weining Lu, Jun Yang, Zhiheng Li, and Bin Liang,
\newblock ``A deep transfer model with wasserstein distance guided
  multi-adversarial networks for bearing fault diagnosis under different
  working conditions,''
\newblock {\em IEEE Access}, vol. 7, pp. 65303--65318, 2019.

\bibitem{wang2020triplet}
Xiaodong Wang and Feng Liu,
\newblock ``Triplet loss guided adversarial domain adaptation for bearing fault
  diagnosis,''
\newblock {\em Sensors}, vol. 20, no. 1, pp. 320, 2020.

\bibitem{she2020wasserstein}
D~She, N~Peng, M~Jia, and MG~Pecht,
\newblock ``Wasserstein distance based deep multi-feature adversarial transfer
  diagnosis approach under variable working conditions,''
\newblock {\em Journal of Instrumentation}, vol. 15, no. 06, pp. P06002, 2020.

\bibitem{yu2020conditional}
Xiaolei Yu, Zhibin Zhao, Xingwu Zhang, Chuang Sun, Baogui Gong, Ruqiang Yan,
  and Xuefeng Chen,
\newblock ``Conditional adversarial domain adaptation with discrimination
  embedding for locomotive fault diagnosis,''
\newblock {\em IEEE Transactions on Instrumentation and Measurement}, vol. 70,
  pp. 1--12, 2020.

\bibitem{li2021deep}
Feng Li, Tuojiang Tang, Baoping Tang, and Qiyuan He,
\newblock ``Deep convolution domain-adversarial transfer learning for fault
  diagnosis of rolling bearings,''
\newblock {\em Measurement}, vol. 169, pp. 108339, 2021.

\bibitem{xie2018transfer}
Yuan Xie and Tao Zhang,
\newblock ``A transfer learning strategy for rotation machinery fault diagnosis
  based on cycle-consistent generative adversarial networks,''
\newblock in {\em 2018 Chinese Automation Congress (CAC)}. IEEE, 2018, pp.
  1309--1313.

\bibitem{dixit2021intelligent}
Sonal Dixit, Nishchal~K Verma, and AK~Ghosh,
\newblock ``Intelligent fault diagnosis of rotary machines: Conditional
  auxiliary classifier gan coupled with meta learning using limited data,''
\newblock {\em IEEE Transactions on Instrumentation and Measurement}, vol. 70,
  pp. 1--11, 2021.

\bibitem{cao2018partial1}
Zhangjie Cao, Mingsheng Long, Jianmin Wang, and Michael~I Jordan,
\newblock ``Partial transfer learning with selective adversarial networks,''
\newblock in {\em Proceedings of the IEEE conference on computer vision and
  pattern recognition}, 2018, pp. 2724--2732.

\bibitem{cao2018partial2}
Zhangjie Cao, Lijia Ma, Mingsheng Long, and Jianmin Wang,
\newblock ``Partial adversarial domain adaptation,''
\newblock in {\em Proceedings of the European Conference on Computer Vision
  (ECCV)}, 2018, pp. 135--150.

\bibitem{jiao2019classifier}
Jinyang Jiao, Ming Zhao, Jing Lin, and Chuancang Ding,
\newblock ``Classifier inconsistency-based domain adaptation network for
  partial transfer intelligent diagnosis,''
\newblock {\em IEEE Transactions on Industrial Informatics}, vol. 16, no. 9,
  pp. 5965--5974, 2019.

\bibitem{li2020novel}
Weihua Li, Zhuyun Chen, and Guolin He,
\newblock ``A novel weighted adversarial transfer network for partial domain
  fault diagnosis of machinery,''
\newblock {\em IEEE Transactions on Industrial Informatics}, vol. 17, no. 3,
  pp. 1753--1762, 2020.

\bibitem{li2020deep}
Xiang Li and Wei Zhang,
\newblock ``Deep learning-based partial domain adaptation method on intelligent
  machinery fault diagnostics,''
\newblock {\em IEEE Transactions on Industrial Electronics}, vol. 68, no. 5,
  pp. 4351--4361, 2020.

\bibitem{li2020partial}
Xiang Li, Wei Zhang, Hui Ma, Zhong Luo, and Xu~Li,
\newblock ``Partial transfer learning in machinery cross-domain fault
  diagnostics using class-weighted adversarial networks,''
\newblock {\em Neural Networks}, vol. 129, pp. 313--322, 2020.

\bibitem{wang2020missing}
Qin Wang, Gabriel Michau, and Olga Fink,
\newblock ``Missing-class-robust domain adaptation by unilateral alignment,''
\newblock {\em IEEE Transactions on Industrial Electronics}, vol. 68, no. 1,
  pp. 663--671, 2020.

\bibitem{deng2021double}
Yafei Deng, Delin Huang, Shichang Du, Guilong Li, Chen Zhao, and Jun Lv,
\newblock ``A double-layer attention based adversarial network for partial
  transfer learning in machinery fault diagnosis,''
\newblock {\em Computers in Industry}, vol. 127, pp. 103399, 2021.

\bibitem{yang2021deep}
Bin Yang, Chi-Guhn Lee, Yaguo Lei, Naipeng Li, and Na~Lu,
\newblock ``Deep partial transfer learning network: A method to selectively
  transfer diagnostic knowledge across related machines,''
\newblock {\em Mechanical Systems and Signal Processing}, vol. 156, pp. 107618,
  2021.

\bibitem{saito2018open}
Kuniaki Saito, Shohei Yamamoto, Yoshitaka Ushiku, and Tatsuya Harada,
\newblock ``Open set domain adaptation by backpropagation,''
\newblock in {\em Proceedings of the European Conference on Computer Vision
  (ECCV)}, 2018, pp. 153--168.

\bibitem{li2020two}
Jipu Li, Ruyi Huang, Guolin He, Yixiao Liao, Zhen Wang, and Weihua Li,
\newblock ``A two-stage transfer adversarial network for intelligent fault
  diagnosis of rotating machinery with multiple new faults,''
\newblock {\em IEEE/ASME Transactions on Mechatronics}, vol. 26, no. 3, pp.
  1591--1601, 2020.

\bibitem{li2020deep2}
Jipu Li, Ruyi Huang, Guolin He, Shuhua Wang, Guanghui Li, and Weihua Li,
\newblock ``A deep adversarial transfer learning network for machinery emerging
  fault detection,''
\newblock {\em IEEE Sensors Journal}, vol. 20, no. 15, pp. 8413--8422, 2020.

\bibitem{zhang2021open}
Wei Zhang, Xiang Li, Hui Ma, Zhong Luo, and Xu~Li,
\newblock ``Open set domain adaptation in machinery fault diagnostics using
  instance-level weighted adversarial learning,''
\newblock {\em IEEE Transactions on Industrial Informatics}, vol. 17, no. 11,
  pp. 7445--7455, 2021.

\bibitem{you2019universal}
Kaichao You, Mingsheng Long, Zhangjie Cao, Jianmin Wang, and Michael~I Jordan,
\newblock ``Universal domain adaptation,''
\newblock in {\em Proceedings of the IEEE/CVF Conference on Computer Vision and
  Pattern Recognition}, 2019, pp. 2720--2729.

\bibitem{zhang2021universal}
Wei Zhang, Xiang Li, Hui Ma, Zhong Luo, and Xu~Li,
\newblock ``Universal domain adaptation in fault diagnostics with hybrid
  weighted deep adversarial learning,''
\newblock {\em IEEE Transactions on Industrial Informatics}, vol. 17, no. 12,
  pp. 7957--7967, 2021.

\bibitem{9394793}
Xiaolei Yu, Zhibin Zhao, Xingwu Zhang, Qiyang Zhang, Yilong Liu, Chuang Sun,
  and Xuefeng Chen,
\newblock ``Deep learning-based open set fault diagnosis by extreme value
  theory,''
\newblock {\em IEEE Transactions on Industrial Informatics}, pp. 1--1, 2021.

\bibitem{dai2020adversarial}
Yong Dai, Jian Liu, Xiancong Ren, and Zenglin Xu,
\newblock ``Adversarial training based multi-source unsupervised domain
  adaptation for sentiment analysis,''
\newblock in {\em Proceedings of the AAAI Conference on Artificial
  Intelligence}, 2020, vol.~34, pp. 7618--7625.

\bibitem{zhao2020multi}
Sicheng Zhao, Guangzhi Wang, Shanghang Zhang, Yang Gu, Yaxian Li, Zhichao Song,
  Pengfei Xu, Runbo Hu, Hua Chai, and Kurt Keutzer,
\newblock ``Multi-source distilling domain adaptation,''
\newblock in {\em Proceedings of the AAAI Conference on Artificial
  Intelligence}, 2020, vol.~34, pp. 12975--12983.

\bibitem{ANewPenaltyYAN}
Fei Shen, Yun Hui, Ruqiang Yan, Chuang Sun, and Jiawen Xu,
\newblock ``A new penalty domain selection machine enabled transfer learning
  for gearbox fault recognition,''
\newblock {\em IEEE Transactions on Industrial Electronics}, vol. 67, no. 10,
  pp. 8743--8754, 2020.

\bibitem{ANewMultipleSourceDomainAdaptation}
Jun Zhu, Nan Chen, and Changqing Shen,
\newblock ``A new multiple source domain adaptation fault diagnosis method
  between different rotating machines,''
\newblock {\em IEEE Transactions on Industrial Informatics}, vol. 17, no. 7,
  pp. 4788--4797, 2021.

\bibitem{REZAEIANJOUYBARI2021109359}
Behnoush Rezaeianjouybari and Yi~Shang,
\newblock ``A novel deep multi-source domain adaptation framework for bearing
  fault diagnosis based on feature-level and task-specific distribution
  alignment,''
\newblock {\em Measurement}, vol. 178, pp. 109359, 2021.

\bibitem{Zhang_2020}
Yongchao Zhang, Zhaohui Ren, Shihua Zhou, and Tianzhuang Yu,
\newblock ``Adversarial domain adaptation with classifier alignment for
  cross-domain intelligent fault diagnosis of multiple source domains,''
\newblock {\em Measurement Science and Technology}, vol. 32, no. 3, pp. 035102,
  dec 2020.

\bibitem{He2020access}
Ya~He, Minghui Hu, Kun Feng, and Zhinong Jiang,
\newblock ``An intelligent fault diagnosis scheme using transferred samples for
  intershaft bearings under variable working conditions,''
\newblock {\em IEEE Access}, vol. 8, pp. 203058--203069, 2020.

\bibitem{WEI2021107744}
Dongdong Wei, Te~Han, Fulei Chu, and Ming~Jian Zuo,
\newblock ``Weighted domain adaptation networks for machinery fault
  diagnosis,''
\newblock {\em Mechanical Systems and Signal Processing}, vol. 158, pp. 107744,
  2021.

\bibitem{DiagnosingRotating2020}
Xiang Li, Wei Zhang, Qian Ding, and Xu~Li,
\newblock ``Diagnosing rotating machines with weakly supervised data using deep
  transfer learning,''
\newblock {\em IEEE Transactions on Industrial Informatics}, vol. 16, no. 3,
  pp. 1688--1697, 2020.

\bibitem{huang2021multi}
Ziling Huang, Zihao Lei, Guangrui Wen, Xin Huang, Haoxuan Zhou, Ruqiang Yan,
  and Xuefeng Chen,
\newblock ``A multi-source dense adaptation adversarial network for fault
  diagnosis of machinery,''
\newblock {\em IEEE Transactions on Industrial Electronics}, 2021.

\bibitem{DG2018CVPR}
Haoliang Li, Sinno~Jialin Pan, Shiqi Wang, and Alex~C. Kot,
\newblock ``Domain generalization with adversarial feature learning,''
\newblock in {\em 2018 IEEE/CVF Conference on Computer Vision and Pattern
  Recognition}, 2018, pp. 5400--5409.

\bibitem{LI2020409}
Xiang Li, Wei Zhang, Hui Ma, Zhong Luo, and Xu~Li,
\newblock ``Domain generalization in rotating machinery fault diagnostics using
  deep neural networks,''
\newblock {\em Neurocomputing}, vol. 403, pp. 409--420, 2020.

\bibitem{DG2018ECCV}
Ya~Li, Xinmei Tian, Mingming Gong, Yajing Liu, Tongliang Liu, Kun Zhang, and
  Dacheng Tao,
\newblock ``Deep domain generalization via conditional invariant adversarial
  networks,''
\newblock in {\em Proceedings of the European Conference on Computer Vision
  (ECCV)}, 2018, pp. 624--639.

\bibitem{DGPriori}
Huailiang Zheng, Yuantao Yang, Jiancheng Yin, Yuqing Li, Rixin Wang, and
  Minqiang Xu,
\newblock ``Deep domain generalization combining a priori diagnosis knowledge
  toward cross-domain fault diagnosis of rolling bearing,''
\newblock {\em IEEE Transactions on Instrumentation and Measurement}, vol. 70,
  pp. 1--11, 2021.

\bibitem{liao2020deep}
Yixiao Liao, Ruyi Huang, Jipu Li, Zhuyun Chen, and Weihua Li,
\newblock ``Deep semisupervised domain generalization network for rotary
  machinery fault diagnosis under variable speed,''
\newblock {\em IEEE Transactions on Instrumentation and Measurement}, vol. 69,
  no. 10, pp. 8064--8075, 2020.

\bibitem{centerlossDG2020}
Yuantao Yang, Jiancheng Yin, Huailiang Zheng, Yuqing Li, Minqiang Xu, and Yushu
  Chen,
\newblock ``Learn generalization feature via convolutional neural network: A
  fault diagnosis scheme toward unseen operating conditions,''
\newblock {\em IEEE Access}, vol. 8, pp. 91103--91115, 2020.

\bibitem{zhang2021ConditionalAdversarialDG}
Qiyang Zhang, Zhibin Zhao, Xingwu Zhang, Yilong Liu, Chuang Sun, Ming Li,
  Shibin Wang, and Xuefeng Chen,
\newblock ``Conditional adversarial domain generalization with a single
  discriminator for bearing fault diagnosis,''
\newblock {\em IEEE Transactions on Instrumentation and Measurement}, vol. 70,
  pp. 1--15, 2021.

\bibitem{han2021hybrid}
Te~Han, Yan-Fu Li, and Min Qian,
\newblock ``A hybrid generalization network for intelligent fault diagnosis of
  rotating machinery under unseen working conditions,''
\newblock {\em IEEE Transactions on Instrumentation and Measurement}, 2021.

\bibitem{CWRU}
{Case Western Reserve University},
\newblock ``{Case Western Reserve University (CWRU) Bearing Data Center,
  [Online]},'' Available:
  \url{https://csegroups.case.edu/bearingdatacenter/pages/download-data-file/},
  accessed on August 2019.

\bibitem{lessmeier2016condition}
Christian Lessmeier, James~Kuria Kimotho, Detmar Zimmer, and Walter Sextro,
\newblock ``Condition monitoring of bearing damage in electromechanical drive
  systems by using motor current signals of electric motors: A benchmark data
  set for data-driven classification,''
\newblock in {\em Proceedings of the European conference of the prognostics and
  health management society}, 2016, pp. 05--08.

\bibitem{PU}
Christian Lessmeier,
\newblock ``et al. {KAt-DataCenter, Chair of Design and Drive Technology,
  Paderborn University},'' Available:
  \url{https://mb.uni-paderborn.de/kat/forschung/datacenter/bearing-datacenter/},
  accessed on August 2019.

\bibitem{JNU}
Ke~Li,
\newblock ``{School of Mechanical Engineering, Jiangnan University},''
  Available: \url{http://mad-net.org:8765/explore.html?t=0.5831516555847212.},
  accessed on August 2019.

\bibitem{li2013sequential}
Ke~Li, Xueliang Ping, Huaqing Wang, Peng Chen, and Yi~Cao,
\newblock ``Sequential fuzzy diagnosis method for motor roller bearing in
  variable operating conditions based on vibration analysis,''
\newblock {\em Sensors}, vol. 13, no. 6, pp. 8013--8041, 2013.

\bibitem{PHM}
PHMSociety,
\newblock ``{PHM09 Data Challenge},'' Available:
  \url{https://www.phmsociety.org/competition/PHM/09/apparatus}, accessed on
  August 2019.

\bibitem{SEU}
Siyu Shao, Stephen McAleer, Ruqiang Yan, and Pierre Baldi,
\newblock ``{Mechanical dataset},'' Available:
  \url{http://mlmechanics.ics.uci.edu./}, accessed on August 2019.

\bibitem{liu2019separate}
Hong Liu, Zhangjie Cao, Mingsheng Long, Jianmin Wang, and Qiang Yang,
\newblock ``Separate to adapt: Open set domain adaptation via progressive
  separation,''
\newblock in {\em Proceedings of the IEEE/CVF Conference on Computer Vision and
  Pattern Recognition}, 2019, pp. 2927--2936.

\bibitem{fu2020learning}
Bo~Fu, Zhangjie Cao, Mingsheng Long, and Jianmin Wang,
\newblock ``Learning to detect open classes for universal domain adaptation,''
\newblock in {\em European Conference on Computer Vision}. Springer, 2020, pp.
  567--583.

\bibitem{wang2010integration}
Jianrong Wang, Kai Fan, and Wanshan Wang,
\newblock ``Integration of fuzzy ahp and fpp with topsis methodology for
  aeroengine health assessment,''
\newblock {\em Expert Systems with Applications}, vol. 37, no. 12, pp.
  8516--8526, 2010.

\bibitem{he2016fuzzy}
Yi-Hai He, Lin-Bo Wang, Zhen-Zhen He, and Min Xie,
\newblock ``A fuzzy topsis and rough set based approach for mechanism analysis
  of product infant failure,''
\newblock {\em Engineering Applications of Artificial Intelligence}, vol. 47,
  pp. 25--37, 2016.

\bibitem{jiang2019fault}
Wen Jiang, Meijuan Wang, Xinyang Deng, and Linfeng Gou,
\newblock ``Fault diagnosis based on topsis method with manhattan distance,''
\newblock {\em Advances in Mechanical Engineering}, vol. 11, no. 3, pp.
  1687814019833279, 2019.

\bibitem{glorot2011domain}
Xavier Glorot, Antoine Bordes, and Yoshua Bengio,
\newblock ``Domain adaptation for large-scale sentiment classification: A deep
  learning approach,''
\newblock in {\em Proceedings of the 28th international conference on machine
  learning (ICML-11)}, 2011, pp. 513--520.

\bibitem{yosinski2014transferable}
Jason Yosinski, Jeff Clune, Yoshua Bengio, and Hod Lipson,
\newblock ``How transferable are features in deep neural networks?,''
\newblock in {\em Advances in neural information processing systems}, 2014, pp.
  3320--3328.

\bibitem{zeiler2014visualizing}
Matthew~D Zeiler and Rob Fergus,
\newblock ``Visualizing and understanding convolutional networks,''
\newblock in {\em European conference on computer vision}. Springer, 2014, pp.
  818--833.

\bibitem{selvaraju2017grad}
Ramprasaath~R Selvaraju, Michael Cogswell, Abhishek Das, Ramakrishna Vedantam,
  Devi Parikh, and Dhruv Batra,
\newblock ``Grad-cam: Visual explanations from deep networks via gradient-based
  localization,''
\newblock in {\em Proceedings of the IEEE International Conference on Computer
  Vision}, 2017, pp. 618--626.

\bibitem{vaswani2017attention}
Ashish Vaswani, Noam Shazeer, Niki Parmar, Jakob Uszkoreit, Llion Jones,
  Aidan~N Gomez, {\L}ukasz Kaiser, and Illia Polosukhin,
\newblock ``Attention is all you need,''
\newblock in {\em Advances in neural information processing systems}, 2017, pp.
  5998--6008.

\bibitem{jing2021balanced}
Mengmeng Jing, Jingjing Li, Lei Zhu, Zhengming Ding, Ke~Lu, and Yang Yang,
\newblock ``Balanced open set domain adaptation via centroid alignment,''
\newblock in {\em Proceedings of the AAAI Conference on Artificial
  Intelligence}, 2021, vol.~35, pp. 8013--8020.

\bibitem{wu2020few}
Jingyao Wu, Zhibin Zhao, Chuang Sun, Ruqiang Yan, and Xuefeng Chen,
\newblock ``Few-shot transfer learning for intelligent fault diagnosis of
  machine,''
\newblock {\em Measurement}, vol. 166, pp. 108202, 2020.

\bibitem{liu2020secure}
Yang Liu, Yan Kang, Chaoping Xing, Tianjian Chen, and Qiang Yang,
\newblock ``A secure federated transfer learning framework,''
\newblock {\em IEEE Intelligent Systems}, vol. 35, no. 4, pp. 70--82, 2020.

\bibitem{zhang2021federated}
Wei Zhang and Xiang Li,
\newblock ``Federated transfer learning for intelligent fault diagnostics using
  deep adversarial networks with data privacy,''
\newblock {\em IEEE/ASME Transactions on Mechatronics}, 2021.

\bibitem{gal2016dropout}
Yarin Gal and Zoubin Ghahramani,
\newblock ``Dropout as a bayesian approximation: Representing model uncertainty
  in deep learning,''
\newblock in {\em international conference on machine learning}. PMLR, 2016,
  pp. 1050--1059.

\bibitem{lakshminarayanan2017simple}
Balaji Lakshminarayanan, Alexander Pritzel, and Charles Blundell,
\newblock ``Simple and scalable predictive uncertainty estimation using deep
  ensembles,''
\newblock in {\em Proceedings of the 31st International Conference on Neural
  Information Processing Systems}, 2017, pp. 6405--6416.

\bibitem{zheng2021rectifying}
Zhedong Zheng and Yi~Yang,
\newblock ``Rectifying pseudo label learning via uncertainty estimation for
  domain adaptive semantic segmentation,''
\newblock {\em International Journal of Computer Vision}, vol. 129, no. 4, pp.
  1106--1120, 2021.

\end{thebibliography}

\clearpage



\section*{Supplementary material}

## Appendix A. Evaluation Results

### CWRU with the time domain input

| Task | Loc | Basis | | AdaBN | | MK-MMD | | JMMD | | CORAL | | DANN | | CDAN | |
|---|---|---|---|---|---|---|---|---|---|---|---|---|---|---|---|
| | | Max | Mean | Max | Mean | Max | Mean | Max | Mean | Max | Mean | Max | Mean | Max | Mean |
| 0-1 | Best | 100.00 | 99.91 | 100.00 | 99.91 | 100.00 | **100.00** | 100.00 | **100.00** | 100.00 | 99.55 | 100.00 | **100.00** | 100.00 | **100.00** |
| | Last | 99.48 | 99.12 | 100.00 | 99.77 | 100.00 | 98.51 | 100.00 | **100.00** | 100.00 | 96.75 | 100.00 | 99.94 | 100.00 | **100.00** |
| 0-2 | Best | 99.94 | 98.73 | 100.00 | 99.77 | 100.00 | **100.00** | 100.00 | **100.00** | 100.00 | 97.01 | 100.00 | **100.00** | 100.00 | **100.00** |
| | Last | 96.30 | 94.07 | 99.74 | 99.26 | 100.00 | 99.74 | 100.00 | 99.61 | 100.00 | 94.80 | 100.00 | **100.00** | 100.00 | **100.00** |
| 0-3 | Best | 98.32 | 93.15 | 98.77 | 96.96 | 99.68 | 97.09 | 100.00 | **100.00** | 99.03 | 92.69 | 100.00 | **100.00** | 100.00 | 99.29 |
| | Last | 92.42 | 86.92 | 94.69 | 92.20 | 99.35 | 93.72 | 100.00 | 99.68 | 96.76 | 90.87 | 100.00 | **100.00** | 100.00 | 96.89 |
| 1-0 | Best | 99.92 | 99.86 | 99.92 | 98.04 | 100.00 | 99.85 | 100.00 | 99.68 | 100.00 | 99.46 | 100.00 | **100.00** | 100.00 | **100.00** |
| | Last | 99.77 | 99.48 | 98.08 | 94.07 | 93.87 | 93.33 | 100.00 | 98.54 | 99.62 | 95.33 | 100.00 | 99.92 | 100.00 | **100.00** |
| 1-2 | Best | 100.00 | **100.00** | 99.94 | 99.48 | 100.00 | **100.00** | 100.00 | **100.00** | 100.00 | **100.00** | 100.00 | **100.00** | 100.00 | **100.00** |
| | Last | 99.94 | 99.87 | 99.87 | 96.60 | 100.00 | **100.00** | 100.00 | 99.74 | 100.00 | 99.87 | 100.00 | **100.00** | 100.00 | **100.00** |
| 1-3 | Best | 99.68 | 98.59 | 97.93 | 96.02 | 100.00 | **100.00** | 100.00 | **100.00** | 100.00 | 95.79 | 100.00 | **100.00** | 100.00 | **100.00** |
| | Last | 98.32 | 94.33 | 96.18 | 90.93 | 100.00 | 99.94 | 100.00 | 98.90 | 100.00 | 94.60 | 100.00 | **100.00** | 100.00 | 99.87 |
| 2-0 | Best | 99.23 | 98.36 | 97.16 | 92.69 | 99.62 | 99.39 | 100.00 | **99.92** | 99.23 | 97.24 | 100.00 | 99.77 | 99.62 | 99.54 |
| | Last | 98.70 | 96.25 | 96.70 | 85.13 | 98.47 | 93.72 | 100.00 | **99.62** | 97.32 | 95.02 | 99.62 | 99.31 | 99.62 | 99.31 |
| 2-1 | Best | 99.42 | 99.25 | 99.03 | 99.05 | 100.00 | 99.68 | 100.00 | 99.68 | 99.68 | 98.51 | 100.00 | 99.74 | 100.00 | **99.87** |
| | Last | 98.31 | 97.35 | 98.25 | 97.04 | 100.00 | 99.61 | 100.00 | **99.74** | 98.70 | 97.92 | 100.00 | 99.55 | 100.00 | 99.55 |
| 2-3 | Best | 99.94 | 99.45 | 99.03 | 98.30 | 100.00 | **100.00** | 100.00 | **100.00** | 100.00 | 99.22 | 100.00 | **100.00** | 100.00 | **100.00** |
| | Last | 99.87 | 97.78 | 98.90 | 96.43 | 100.00 | **100.00** | 100.00 | 98.90 | 100.00 | 98.90 | 100.00 | **100.00** | 100.00 | **100.00** |
| 3-0 | Best | 91.26 | 87.11 | 95.40 | 88.91 | 99.23 | 98.93 | 100.00 | **99.54** | 94.25 | 87.89 | 99.62 | 99.00 | 99.62 | 99.39 |
| | Last | 87.36 | 81.10 | 95.33 | 84.12 | 95.02 | 92.18 | 100.00 | **98.85** | 93.49 | 85.44 | 99.23 | 98.54 | 99.23 | 98.77 |
| 3-1 | Best | 91.55 | 88.51 | 96.82 | 95.34 | 100.00 | 99.22 | 100.00 | 99.68 | 97.73 | 96.23 | 100.00 | **99.74** | 100.00 | 98.12 |
| | Last | 88.04 | 82.83 | 94.09 | 89.53 | 100.00 | 99.09 | 100.00 | 99.09 | 97.40 | 95.52 | 100.00 | **99.42** | 100.00 | 97.53 |
| 3-2 | Best | 99.61 | 97.82 | 99.87 | 99.82 | 100.00 | **100.00** | 100.00 | **100.00** | 100.00 | 97.79 | 100.00 | **100.00** | 100.00 | **100.00** |
| Best average | | 98.24 | 96.73 | 98.70 | 97.11 | 99.83 | 99.51 | 100.00 | **99.93** | 99.16 | 96.78 | 99.97 | 99.86 | 99.94 | 99.68 |
| Last average | | 96.23 | 93.27 | 97.57 | 93.47 | 98.89 | 97.49 | 100.00 | 99.60 | 98.55 | 95.11 | 99.90 | **99.71** | 99.90 | 99.32 |
| Best bias | | 0.00 | 0.00 | 0.47 | 0.38 | 1.64 | 2.78 | 1.76 | 3.20 | 0.92 | 0.05 | 1.73 | 3.13 | 1.70 | 2.96 |
| Last bias | | 0.00 | 0.00 | 1.34 | 0.20 | 2.66 | 4.22 | 3.77 | 6.33 | 2.33 | 1.84 | 3.68 | 6.44 | 3.68 | 6.05 |

\* Best average: the average of Best; Last average: the average of Last; Best bias: the Best bias between Basis and other methods; Last bias: the Last bias between Basis and other methods.

\* The maximum Mean value of each row is bolded.

### CWRU with the frequency domain input

| Task | Loc | Basis | | AdaBN | | MK-MMD | | JMMD | | CORAL | | DANN | | CDAN | |
|---|---|---|---|---|---|---|---|---|---|---|---|---|---|---|---|
| | | Max | Mean | Max | Mean | Max | Mean | Max | Mean | Max | Mean | Max | Mean | Max | Mean |
| 0-1 | Best | 99.61 | 99.46 | 99.48 | 99.13 | 100.00 | 99.94 | 100.00 | 99.87 | 100.00 | 99.94 | 100.00 | **100.00** | 100.00 | **100.00** |
| | Last | 99.48 | 98.60 | 98.64 | 98.25 | 100.00 | 99.81 | 100.00 | 99.48 | 100.00 | 99.81 | 100.00 | **99.87** | 99.68 | 99.61 |
| 0-2 | Best | 90.51 | 89.87 | 91.03 | 90.27 | 100.00 | **100.00** | 100.00 | 99.94 | 100.00 | 98.12 | 100.00 | 99.81 | 100.00 | 95.52 |
| | Last | 89.34 | 86.16 | 89.15 | 87.86 | 100.00 | **99.87** | 100.00 | 98.90 | 100.00 | 97.92 | 100.00 | 99.55 | 100.00 | 92.73 |
| 0-3 | Best | 77.40 | 76.64 | 83.55 | 80.06 | 100.00 | 95.54 | 100.00 | **97.73** | 98.06 | 93.07 | 100.00 | 94.95 | 100.00 | 93.85 |
| | Last | 75.78 | 74.22 | 78.04 | 75.35 | 100.00 | 94.89 | 100.00 | **96.57** | 97.41 | 91.07 | 99.68 | 91.98 | 100.00 | 93.66 |
| 1-0 | Best | 98.01 | 96.89 | 97.78 | 97.35 | 100.00 | 99.31 | 100.00 | 99.39 | 98.85 | 98.39 | 100.00 | 99.31 | 100.00 | **99.54** |
| | Last | 96.17 | 94.53 | 96.93 | 94.37 | 93.49 | 91.42 | 99.62 | 96.32 | 98.08 | 97.16 | 99.23 | **98.85** | 98.08 | 97.78 |
| 1-2 | Best | 97.01 | 93.01 | 95.00 | 93.39 | 100.00 | **100.00** | 100.00 | **100.00** | 100.00 | 99.29 | 100.00 | **100.00** | 100.00 | 99.68 |
| | Last | 96.17 | 91.19 | 90.58 | 89.66 | 100.00 | **100.00** | 100.00 | **100.00** | 100.00 | 99.83 | 100.00 | **100.00** | 100.00 | 98.90 |
| 1-3 | Best | 83.10 | 81.35 | 81.15 | 79.73 | 100.00 | 96.96 | 100.00 | 94.63 | 99.48 | 98.12 | 100.00 | **99.94** | 100.00 | 97.86 |
| | Last | 77.33 | 75.53 | 77.66 | 73.99 | 100.00 | 96.64 | 99.68 | 93.59 | 99.35 | 97.86 | 100.00 | **99.81** | 100.00 | 96.57 |
| 2-0 | Best | 82.61 | 78.55 | 80.54 | 79.11 | 94.64 | 89.43 | 99.62 | **92.41** | 97.70 | 90.65 | 97.70 | 90.35 | 90.80 | 89.50 |
| | Last | 80.61 | 70.25 | 77.24 | 68.69 | 89.27 | 85.75 | 94.25 | 83.14 | 97.32 | **89.89** | 97.32 | 88.35 | 89.27 | 86.21 |
| 2-1 | Best | 91.75 | 89.24 | 90.45 | 88.68 | 92.21 | 90.72 | 100.00 | 94.09 | 99.03 | 95.91 | 100.00 | **98.12** | 92.53 | 90.45 |
| | Last | 87.33 | 83.66 | 86.22 | 83.72 | 86.04 | 85.39 | 100.00 | 90.78 | 98.05 | 94.67 | 99.35 | **96.24** | 92.53 | 86.30 |
| 2-3 | Best | 89.05 | 86.61 | 93.52 | 87.72 | 100.00 | 93.59 | 100.00 | **100.00** | 100.00 | 98.00 | 89.32 | 86.93 | 100.00 | 90.03 |
| | Last | 79.08 | 77.53 | 91.52 | 79.69 | 100.00 | 90.68 | 100.00 | **100.00** | 100.00 | 97.41 | 85.11 | 84.01 | 100.00 | 87.38 |
| 3-0 | Best | 77.78 | 74.15 | 76.48 | 74.16 | 95.02 | **88.43** | 87.74 | 84.90 | 83.52 | 81.23 | 97.32 | 88.28 | 87.74 | 81.07 |
| | Last | 72.95 | 64.84 | 72.03 | 69.40 | 90.04 | **84.98** | 83.52 | 78.39 | 82.76 | 79.69 | 96.93 | 84.60 | 86.97 | 79.00 |
| 3-1 | Best | 80.25 | 77.52 | 83.30 | 79.49 | 100.00 | **99.94** | 100.00 | 95.65 | 98.05 | 90.91 | 100.00 | 90.85 | 85.06 | 83.70 |
| | Last | 72.45 | 70.59 | 80.70 | 73.90 | 100.00 | **99.81** | 100.00 | 93.64 | 97.73 | 93.48 | 99.68 | 87.92 | 84.74 | 82.40 |
| 3-2 | Best | 91.62 | 89.64 | 95.06 | 89.97 | 100.00 | 90.19 | 100.00 | 95.97 | 100.00 | **98.90** | 100.00 | 95.32 | 100.00 | 94.54 |
| | Last | 84.28 | 78.95 | 84.67 | 79.27 | 100.00 | 84.61 | 100.00 | 93.83 | 100.00 | **97.79** | 100.00 | 93.70 | 100.00 | 92.14 |
| Best average | | 88.23 | 86.08 | 88.95 | 86.59 | 98.49 | 95.34 | 98.59 | **96.22** | 97.91 | 95.21 | 98.70 | 95.32 | 96.34 | 93.00 |
| Last average | | 84.04 | 80.67 | 85.28 | 81.18 | 96.57 | 92.82 | 98.09 | 93.72 | 97.56 | **94.30** | 98.11 | 93.74 | 95.94 | 91.06 |
| Best bias | | 0.00 | 0.00 | 0.72 | 0.51 | 10.26 | 9.26 | 10.72 | 10.14 | 9.68 | 9.13 | 10.47 | 9.24 | 8.12 | 6.92 |
| Last bias | | 0.00 | 0.00 | 1.25 | 0.51 | 12.53 | 12.15 | 14.05 | 13.05 | 13.52 | 13.63 | 14.07 | 13.07 | 11.90 | 10.39 |

\* Best average: the average of Best; Last average: the average of Last; Best bias: the Best bias between Basis and other methods; Last bias: the Last bias between Basis and other methods.

\* The maximum Mean value of each row is bolded.

## PU with the time domain input

| Task | Loc | Basis | | AdaBN | | MK-MMD | | JMMD | | CORAL | | DANN | | CDAN | |
|------|-----|-------|------|-------|------|--------|------|------|------|-------|------|------|------|------|------|
| | | Max | Mean | Max | Mean | Max | Mean | Max | Mean | Max | Mean | Max | Mean | Max | Mean |
| 0-1 | Best | 24.16 | 21.44 | 40.04 | 38.26 | 36.81 | 33.10 | 37.88 | 36.50 | 31.13 | 27.97 | 44.17 | **41.72** | 38.04 | 37.02 |
| | Last | 15.54 | 14.02 | 33.59 | 30.00 | 32.36 | 29.14 | 35.43 | 32.79 | 29.75 | 24.69 | 40.64 | **38.19** | 36.35 | 33.68 |
| 0-2 | Best | 79.24 | 77.78 | 76.83 | 74.70 | 82.29 | 80.73 | 82.44 | 80.61 | 76.18 | 69.89 | 83.05 | **80.95** | 81.83 | 79.85 |
| | Last | 78.60 | 76.33 | 75.91 | 72.27 | 81.37 | 79.48 | 81.98 | 79.63 | 57.10 | 40.46 | 83.05 | **79.97** | 80.15 | 78.56 |
| 0-3 | Best | 47.90 | 45.02 | 51.23 | 48.93 | 47.66 | 43.84 | 54.46 | 52.68 | 49.47 | 39.61 | 57.03 | **55.04** | 54.01 | 50.05 |
| | Last | 32.24 | 30.02 | 41.54 | 40.26 | 44.18 | 41.09 | 52.95 | 50.92 | 38.58 | 33.19 | 55.98 | **53.74** | 51.89 | 47.62 |
| 1-0 | Best | 33.43 | 30.55 | 50.02 | **46.56** | 38.71 | 35.85 | 43.16 | 40.77 | 27.80 | 23.50 | 42.24 | 38.99 | 39.02 | 36.31 |
| | Last | 27.99 | 23.57 | 41.41 | **38.27** | 37.48 | 33.24 | 36.71 | 35.94 | 18.13 | 15.61 | 40.25 | 35.42 | 35.33 | 30.75 |
| 1-2 | Best | 34.48 | 33.13 | 49.80 | **44.92** | 41.68 | 38.23 | 47.63 | 41.19 | 32.06 | 25.53 | 45.80 | 43.88 | 46.56 | 44.43 |
| | Last | 26.02 | 24.18 | 43.17 | 37.13 | 40.00 | 35.39 | 43.97 | 35.17 | 28.70 | 21.16 | 41.83 | **39.57** | 43.51 | 38.99 |
| 1-3 | Best | 23.74 | 22.34 | 34.91 | **32.89** | 28.14 | 26.51 | 30.71 | 28.68 | 20.88 | 19.15 | 36.01 | 32.19 | 34.34 | 30.98 |
| | Last | 19.74 | 16.09 | 30.46 | 25.36 | 25.42 | 23.39 | 24.96 | 22.54 | 17.85 | 14.37 | 29.80 | **27.05** | 27.84 | 23.63 |
| 2-0 | Best | 79.08 | 77.70 | 74.75 | 73.99 | 80.80 | 78.86 | 80.65 | 78.93 | 76.19 | 72.63 | 81.41 | 80.27 | 82.95 | **80.40** |
| | Last | 78.43 | 76.73 | 72.84 | 71.35 | 79.57 | **77.79** | 79.88 | 77.30 | 64.82 | 50.41 | 80.18 | 79.20 | 81.87 | 79.11 |
| 2-1 | Best | 26.04 | 23.16 | 35.52 | 33.16 | 36.50 | 35.03 | 38.96 | 36.72 | 33.74 | 29.94 | 41.26 | 39.27 | 45.55 | **41.38** |
| | Last | 16.24 | 14.71 | 27.33 | 25.05 | 34.66 | 31.07 | 33.59 | 32.06 | 31.90 | 26.59 | 39.11 | **36.53** | 38.65 | 34.76 |
| 2-3 | Best | 45.41 | 43.05 | 50.29 | 49.61 | 46.44 | 43.90 | 51.29 | 50.05 | 38.12 | 33.89 | 53.86 | **50.62** | 51.59 | 49.53 |
| | Last | 32.70 | 31.23 | 44.02 | 40.73 | 46.14 | 41.94 | 49.62 | 47.66 | 25.11 | 18.85 | 52.04 | **49.23** | 50.23 | 47.44 |
| 3-0 | Best | 45.75 | 42.35 | 48.73 | 46.88 | 48.39 | 44.02 | 54.69 | **51.12** | 32.87 | 31.52 | 50.38 | 49.62 | 51.77 | 48.69 |
| | Last | 36.01 | 32.16 | 45.16 | 40.94 | 45.42 | 41.66 | 53.92 | **49.31** | 24.73 | 17.23 | 49.62 | 47.93 | 45.62 | 43.90 |
| 3-1 | Best | 31.01 | 29.25 | 34.51 | **32.65** | 30.06 | 29.17 | 30.67 | 29.29 | 28.83 | 26.41 | 33.59 | 31.50 | 33.28 | 31.59 |
| | Last | 26.59 | 25.27 | 32.05 | **28.43** | 28.22 | 22.64 | 27.61 | 23.53 | 17.18 | 14.23 | 30.67 | 27.45 | 30.67 | 27.97 |
| 3-2 | Best | 43.53 | 40.23 | 52.37 | 48.58 | 44.12 | 43.54 | 51.60 | **49.01** | 32.67 | 30.56 | 52.82 | 49.37 | 47.33 | 45.25 |
| | Last | 35.68 | 32.39 | 46.22 | 41.99 | 38.47 | 37.19 | 49.77 | 47.39 | 22.14 | 18.93 | 50.84 | **47.57** | 42.75 | 40.40 |
| Best average | | 42.81 | 40.50 | 49.92 | 47.59 | 46.80 | 44.40 | 50.35 | 47.96 | 40.00 | 35.88 | 51.80 | **49.45** | 50.52 | 47.96 |
| Last average | | 35.48 | 33.06 | 44.48 | 40.98 | 44.46 | 41.17 | 47.53 | 44.52 | 31.33 | 24.64 | 49.50 | **46.82** | 47.07 | 43.90 |
| Best bias | | 0.00 | 0.00 | 7.10 | 7.09 | 3.99 | 3.90 | 7.53 | 7.46 | -2.82 | -4.62 | 8.99 | 8.95 | 7.71 | 7.46 |
| Last bias | | 0.00 | 0.00 | 8.99 | 7.92 | 8.98 | 8.11 | 12.05 | 11.46 | -4.15 | -8.41 | 14.02 | 13.76 | 11.59 | 10.84 |

* Best average: the average of Best; Last average: the average of Last; Best bias: the Best bias between Basis and other methods; Last bias: the Last bias between Basis and other methods.

* The maximum Mean value of each row is bolded.

## PU with the frequency domain input

| Task | Loc | Basis | | AdaBN | | MK-MMD | | JMMD | | CORAL | | DANN | | CDAN | |
|------|-----|-------|------|-------|------|--------|------|------|------|-------|------|------|------|------|------|
| | | Max | Mean | Max | Mean | Max | Mean | Max | Mean | Max | Mean | Max | Mean | Max | Mean |
| 0-1 | Best | 24.87 | 23.28 | 33.77 | 30.21 | 34.97 | 32.79 | 39.72 | 36.19 | 26.84 | 23.44 | 44.48 | **42.91** | 42.48 | 40.06 |
| | Last | 22.72 | 20.96 | 26.80 | 24.15 | 34.36 | 30.43 | 36.96 | 33.28 | 24.54 | 19.45 | 43.56 | **40.34** | 40.18 | 35.46 |
| 0-2 | Best | 93.21 | 92.08 | 92.27 | 90.63 | 94.96 | 94.35 | 96.34 | **95.54** | 88.70 | 85.83 | 95.42 | 94.29 | 97.25 | 95.36 |
| | Last | 91.93 | 90.87 | 91.56 | 89.12 | 94.50 | 93.59 | 95.57 | **94.99** | 82.29 | 66.47 | 94.96 | 93.71 | 96.64 | 94.72 |
| 0-3 | Best | 62.91 | 61.18 | 64.21 | 60.93 | 77.76 | 76.67 | 80.33 | 78.55 | 60.36 | 57.76 | 84.72 | **82.99** | 83.51 | 80.63 |
| | Last | 58.95 | 57.14 | 61.64 | 57.11 | 77.31 | 75.04 | 79.12 | 77.40 | 54.01 | 50.83 | 84.27 | **82.15** | 82.00 | 79.49 |
| 1-0 | Best | 29.31 | 27.65 | 28.08 | 26.91 | 32.57 | 30.51 | 34.41 | 31.58 | 23.66 | 21.87 | 34.56 | 32.29 | 36.56 | **33.12** |
| | Last | 26.76 | 25.13 | 25.59 | 24.61 | 27.50 | 26.27 | 29.19 | 28.02 | 14.59 | 12.81 | 30.88 | 28.48 | 34.56 | **29.37** |
| 1-2 | Best | 27.03 | 26.58 | 25.22 | 24.49 | 32.67 | 30.87 | 40.76 | 33.86 | 25.50 | 23.76 | 38.78 | 36.91 | 45.50 | **40.31** |
| | Last | 25.44 | 24.14 | 23.08 | 21.50 | 30.23 | 27.33 | 39.39 | 30.44 | 18.78 | 15.39 | 38.02 | 35.27 | 44.89 | **38.51** |
| 1-3 | Best | 19.53 | 18.03 | 20.35 | 19.04 | 23.90 | 22.03 | 26.32 | 24.18 | 20.27 | 15.95 | 26.93 | 24.84 | 28.59 | **26.63** |
| | Last | 14.17 | 13.75 | 15.74 | 14.74 | 21.33 | 20.36 | 24.21 | 22.30 | 12.56 | 11.10 | 25.57 | **22.88** | 27.53 | 21.39 |
| 2-0 | Best | 88.69 | 87.79 | 89.59 | 88.98 | 92.32 | 91.46 | 94.32 | 91.89 | 83.72 | 82.34 | 94.16 | 93.09 | 94.01 | **93.27** |
| | Last | 87.47 | 86.40 | 88.91 | 88.37 | 91.40 | 90.48 | 93.86 | 91.37 | 79.88 | 64.82 | 94.01 | 92.50 | 93.09 | **92.57** |
| 2-1 | Best | 26.83 | 24.49 | 28.43 | 27.47 | 36.96 | 33.56 | 44.33 | 40.06 | 32.98 | 29.48 | 53.07 | 47.36 | 50.15 | **48.44** |
| | Last | 22.57 | 20.55 | 23.98 | 22.43 | 35.28 | 32.30 | 43.87 | 37.27 | 29.14 | 24.11 | 51.69 | 46.01 | 49.08 | **46.23** |
| 2-3 | Best | 63.25 | 60.23 | 62.28 | 60.62 | 79.58 | 77.40 | 79.27 | 76.97 | 66.26 | 60.33 | 83.36 | **80.21** | 82.60 | 79.15 |
| | Last | 60.55 | 57.18 | 59.16 | 58.10 | 78.67 | 76.70 | 78.52 | 76.31 | 64.30 | 43.90 | 83.06 | **79.52** | 81.54 | 78.24 |
| 3-0 | Best | 58.00 | 56.25 | 60.18 | 56.28 | 62.52 | 60.06 | 74.81 | **71.58** | 52.53 | 51.33 | 80.03 | 69.40 | 67.74 | 63.29 |
| | Last | 54.78 | 52.58 | 54.75 | 51.63 | 61.14 | 58.77 | 74.04 | **71.03** | 50.69 | 36.80 | 78.96 | 68.76 | 65.90 | 61.50 |
| 3-1 | Best | 25.85 | 24.21 | 30.33 | 29.52 | 40.03 | 32.18 | 34.66 | 31.87 | 24.54 | 22.36 | 30.67 | 29.97 | 39.57 | **35.52** |
| | Last | 21.80 | 20.90 | 26.90 | 24.76 | 39.72 | 30.15 | 30.67 | 23.04 | 19.48 | 16.20 | 28.53 | 24.57 | 37.88 | **32.12** |
| 3-2 | Best | 59.22 | 56.57 | 61.88 | 56.15 | 71.91 | 67.18 | 81.22 | 71.39 | 63.21 | 55.91 | 83.82 | **77.19** | 79.69 | 69.56 |
| | Last | 56.86 | 53.64 | 57.78 | 54.00 | 70.69 | 66.32 | 81.07 | 70.47 | 60.61 | 48.70 | 83.36 | **76.15** | 78.47 | 66.47 |
| Best average | | 48.23 | 46.53 | 49.72 | 47.60 | 56.68 | 54.09 | 60.54 | 56.97 | 47.38 | 44.20 | 62.50 | **59.29** | 62.30 | 58.78 |
| Last average | | 45.33 | 43.60 | 46.32 | 44.21 | 55.18 | 52.31 | 58.87 | 54.66 | 42.57 | 34.22 | 61.41 | **57.53** | 60.98 | 56.34 |
| Best bias | | 0.00 | 0.00 | 1.49 | 1.07 | 8.45 | 7.56 | 12.32 | 10.45 | -0.84 | -2.33 | 14.28 | 12.76 | 14.08 | 12.25 |
| Last bias | | 0.00 | 0.00 | 0.99 | 0.61 | 9.84 | 8.71 | 13.54 | 11.06 | -2.76 | -9.39 | 16.07 | 13.93 | 15.65 | 12.74 |

* Best average: the average of Best; Last average: the average of Last; Best bias: the Best bias between Basis and other methods; Last bias: the Last bias between Basis and other methods.

* The maximum Mean value of each row is bolded.

## JNU with the time domain input

| Task | Loc | Basis | | AdaBN | | MK-MMD | | JMMD | | CORAL | | DANN | | CDAN | |
|------|-----|-------|------|-------|------|--------|------|------|------|-------|------|------|------|------|------|
| | | Max | Mean | Max | Mean | Max | Mean | Max | Mean | Max | Mean | Max | Mean | Max | Mean |
| 0-1 | Best | 99.04 | 98.85 | 98.46 | 97.70 | 99.32 | 99.08 | 99.83 | 99.15 | 95.73 | 93.75 | 98.98 | 98.88 | 99.32 | **99.22** |
| | Last | 97.92 | 97.54 | 95.97 | 95.29 | 98.81 | **98.33** | 98.46 | 97.95 | 72.35 | 61.74 | 98.12 | 97.54 | 98.12 | 97.85 |
| 0-2 | Best | 96.42 | 95.86 | 96.66 | 96.22 | 98.46 | 98.19 | 98.81 | **98.29** | 95.22 | 92.53 | 98.29 | 97.68 | 97.95 | 97.81 |
| | Last | 93.72 | 92.80 | 93.58 | 93.19 | 97.78 | 97.20 | 97.95 | **97.27** | 55.46 | 52.18 | 97.27 | 96.69 | 97.10 | 96.79 |
| 1-0 | Best | 89.49 | 87.87 | 91.60 | 91.16 | 97.78 | 96.96 | 97.95 | **97.17** | 77.65 | 72.15 | 96.25 | 95.26 | 94.30 | 94.30 |
| | Last | 68.87 | 66.31 | 86.89 | 85.45 | 97.27 | 96.21 | 97.61 | **96.59** | 73.38 | 61.74 | 96.08 | 94.85 | 94.20 | 91.50 |
| 1-2 | Best | 98.63 | 98.29 | 98.23 | 97.56 | 99.66 | **99.49** | 99.66 | **99.49** | 99.32 | 97.82 | 99.49 | 99.29 | 99.66 | 99.42 |
| | Last | 98.16 | 97.77 | 96.79 | 95.23 | 99.32 | 99.01 | 99.49 | **99.11** | 72.01 | 58.23 | 99.32 | 98.87 | 98.98 | 98.60 |
| 2-0 | Best | 91.43 | 90.73 | 92.39 | 90.81 | 93.86 | **93.07** | 94.03 | 92.93 | 89.25 | 85.39 | 92.66 | 91.84 | 93.00 | 91.81 |
| | Last | 87.34 | 85.94 | 87.37 | 83.28 | 93.17 | **91.98** | 93.17 | 91.78 | 55.46 | 53.58 | 91.64 | 90.51 | 91.13 | 89.56 |
| 2-1 | Best | 99.15 | 98.89 | 98.16 | 97.69 | 99.49 | 99.22 | 99.66 | **99.35** | 97.44 | 96.11 | 99.15 | 98.84 | 99.32 | 99.18 |
| | Last | 98.57 | 97.62 | 94.71 | 94.16 | 98.29 | 97.78 | 98.81 | 98.46 | 79.69 | 56.59 | 98.63 | 97.98 | 98.81 | **98.50** |
| Best average | | 95.69 | 95.08 | 95.92 | 95.19 | 98.10 | 97.67 | 98.32 | **97.73** | 92.44 | 89.62 | 97.47 | 96.96 | 97.44 | 96.96 |
| Last average | | 90.76 | 89.67 | 92.55 | 91.10 | 97.44 | 96.75 | 97.58 | **96.86** | 68.06 | 57.34 | 96.84 | 96.08 | 96.39 | 95.47 |
| Best bias | | 0.00 | 0.00 | 0.22 | 0.11 | 2.40 | 2.59 | 2.63 | 2.65 | -3.26 | -5.46 | 1.78 | 1.88 | 1.75 | 1.88 |
| Last bias | | 0.00 | 0.00 | 1.79 | 1.44 | 6.68 | 7.09 | 6.82 | 7.20 | -22.71 | -32.32 | 6.08 | 6.41 | 5.63 | 5.80 |

\* Best average: the average of Best; Last average: the average of Last; Best bias: the Best bias between Basis and other methods; Last bias: the Last bias between Basis and other methods.

\* The maximum Mean value of each row is bolded.

## JNU with the frequency domain input

| Task | Loc | Basis | | AdaBN | | MK-MMD | | JMMD | | CORAL | | DANN | | CDAN | |
|------|-----|-------|------|-------|------|--------|------|------|------|-------|------|------|------|------|------|
| | | Max | Mean | Max | Mean | Max | Mean | Max | Mean | Max | Mean | Max | Mean | Max | Mean |
| 0-1 | Best | 89.80 | 83.83 | 84.68 | 83.59 | 97.44 | 96.90 | 97.44 | **97.00** | 82.94 | 72.97 | 96.25 | 93.48 | 97.44 | 93.00 |
| | Last | 76.38 | 73.04 | 82.18 | 77.77 | 97.10 | **96.35** | 96.42 | 96.08 | 60.58 | 54.95 | 95.73 | 92.76 | 96.42 | 91.94 |
| 0-2 | Best | 78.19 | 76.39 | 73.65 | 72.43 | 96.93 | 95.80 | 97.78 | **96.31** | 71.67 | 62.59 | 94.88 | 90.65 | 95.39 | 91.67 |
| | Last | 68.57 | 62.19 | 61.23 | 58.66 | 96.76 | 95.22 | 97.27 | **95.63** | 71.67 | 57.92 | 93.86 | 88.16 | 94.20 | 89.73 |
| 1-0 | Best | 73.75 | 69.38 | 84.10 | 79.58 | 92.15 | 87.92 | 93.34 | 89.31 | 67.41 | 59.97 | 92.83 | 91.78 | 92.83 | **92.22** |
| | Last | 66.79 | 55.33 | 76.25 | 66.58 | 91.30 | 85.87 | 92.83 | 87.06 | 60.24 | 53.62 | 91.47 | 90.82 | 92.15 | **91.43** |
| 1-2 | Best | 89.93 | 87.88 | 87.68 | 86.61 | 96.93 | 96.32 | 97.61 | **97.24** | 83.96 | 80.89 | 97.10 | 96.35 | 96.93 | 96.31 |
| | Last | 87.37 | 84.51 | 84.91 | 83.17 | 96.42 | 95.57 | 96.93 | **96.21** | 77.47 | 59.93 | 96.59 | 95.56 | 96.42 | 95.80 |
| 2-0 | Best | 85.53 | 84.44 | 85.87 | 84.31 | 92.83 | 92.18 | 93.86 | **92.22** | 79.86 | 70.99 | 92.83 | 90.85 | 91.13 | 90.51 |
| | Last | 80.44 | 77.07 | 81.57 | 80.02 | 91.98 | **91.16** | 92.15 | 90.68 | 60.92 | 55.29 | 91.81 | 89.97 | 90.61 | 89.69 |
| 2-1 | Best | 88.84 | 88.35 | 88.12 | 87.79 | 97.44 | **95.32** | 97.95 | 91.67 | 91.47 | 88.19 | 92.15 | 90.92 | 90.61 | 89.90 |
| | Last | 87.92 | 87.17 | 86.48 | 86.10 | 96.93 | **94.71** | 97.61 | 89.59 | 91.30 | 72.22 | 90.44 | 89.05 | 90.10 | 89.32 |
| Best average | | 84.34 | 81.71 | 84.02 | 82.38 | 95.62 | **94.07** | 96.33 | 93.96 | 79.55 | 72.60 | 94.34 | 92.34 | 94.06 | 92.27 |
| Last average | | 77.91 | 73.22 | 78.77 | 75.38 | 95.08 | **93.15** | 95.54 | 92.54 | 70.36 | 58.99 | 93.32 | 91.05 | 93.32 | 91.32 |
| Best bias | | 0.00 | 0.00 | -0.32 | 0.67 | 11.28 | 12.36 | 11.99 | 12.25 | -4.79 | -9.11 | 10.00 | 10.63 | 9.72 | 10.56 |
| Last bias | | 0.00 | 0.00 | 0.86 | 2.16 | 17.17 | 19.93 | 17.62 | 19.32 | -7.55 | -14.23 | 15.41 | 17.83 | 15.41 | 18.10 |

\* Best average: the average of Best; Last average: the average of Last; Best bias: the Best bias between Basis and other methods; Last bias: the Last bias between Basis and other methods.

\* The maximum Mean value of each row is bolded.

## PHM2009 with the time domain input

| Task | Loc | Basis | | AdaBN | | MK-MMD | | JMMD | | CORAL | | DANN | | CDAN | |
|---|---|---|---|---|---|---|---|---|---|---|---|---|---|---|---|
| | | Max | Mean | Max | Mean | Max | Mean | Max | Mean | Max | Mean | Max | Mean | Max | Mean |
| 0-1 | Best | 41.60 | 40.70 | 41.35 | 41.04 | 43.27 | 41.41 | 46.79 | **44.42** | 39.74 | 37.50 | 46.47 | 44.04 | 43.59 | 42.82 |
| | Last | 40.06 | 38.31 | 39.87 | 38.32 | 40.06 | 37.37 | 46.79 | **42.63** | 31.73 | 27.69 | 44.55 | 41.28 | 41.67 | 38.91 |
| 0-2 | Best | 45.90 | 42.12 | 46.86 | 45.67 | 49.04 | 45.77 | 47.44 | 45.90 | 39.42 | 34.74 | 49.36 | **46.54** | 47.44 | 44.42 |
| | Last | 38.14 | 36.91 | 43.53 | 42.66 | 47.76 | 42.05 | 45.83 | **43.46** | 30.45 | 26.15 | 45.19 | 42.76 | 42.95 | 40.45 |
| 0-3 | Best | 38.46 | 35.22 | 43.01 | **41.72** | 41.99 | 38.91 | 41.99 | 40.13 | 33.33 | 31.34 | 42.63 | 40.58 | 41.99 | 38.65 |
| | Last | 30.77 | 29.59 | 39.94 | **38.01** | 38.46 | 35.58 | 39.42 | 37.37 | 27.88 | 23.08 | 41.03 | 37.88 | 40.71 | 36.09 |
| 1-0 | Best | 41.92 | 41.20 | 42.69 | 41.27 | 48.40 | 45.96 | 48.08 | **46.99** | 41.99 | 38.14 | 46.79 | 45.38 | 48.08 | 45.77 |
| | Last | 39.17 | 38.01 | 38.78 | 37.68 | 46.47 | 41.47 | 45.19 | **44.36** | 31.73 | 23.46 | 45.19 | 42.12 | 45.83 | 43.65 |
| 1-2 | Best | 53.85 | 53.13 | 54.62 | 53.44 | 58.33 | 56.28 | 57.37 | **56.73** | 53.85 | 51.28 | 58.01 | 56.67 | 58.01 | 55.90 |
| | Last | 51.60 | 50.68 | 53.46 | 50.87 | 55.77 | 51.67 | 56.73 | **54.10** | 43.91 | 28.65 | 54.49 | 52.76 | 55.13 | 52.50 |
| 1-3 | Best | 52.05 | 50.08 | 52.44 | 50.81 | 52.88 | **50.83** | 51.92 | 50.45 | 47.76 | 43.40 | 50.96 | 49.29 | 50.96 | 49.55 |
| | Last | 47.12 | 44.85 | 46.15 | 44.91 | 47.44 | 45.32 | 48.40 | **46.99** | 40.38 | 34.42 | 47.76 | 44.62 | 46.47 | 44.68 |
| 2-0 | Best | 43.14 | 41.94 | 44.49 | 43.41 | 45.51 | 44.81 | 48.08 | **45.26** | 37.50 | 36.09 | 50.00 | 44.94 | 45.51 | 43.85 |
| | Last | 37.18 | 35.79 | 39.81 | 38.11 | 44.23 | 40.70 | 45.83 | 41.28 | 27.88 | 23.85 | 46.79 | **41.34** | 42.31 | 40.00 |
| 2-1 | Best | 53.01 | 52.42 | 53.53 | 51.65 | 58.97 | 54.23 | 55.13 | 53.91 | 45.51 | 42.69 | 56.73 | 53.85 | 58.97 | **54.81** |
| | Last | 50.83 | 48.15 | 50.51 | 48.06 | 57.69 | **51.22** | 50.64 | 49.10 | 37.50 | 33.72 | 54.49 | 49.87 | 55.13 | 48.78 |
| 2-3 | Best | 58.78 | 57.23 | 58.85 | 57.36 | 61.54 | **59.94** | 61.54 | 59.10 | 52.56 | 50.64 | 60.90 | 58.65 | 61.22 | 58.59 |
| | Last | 55.38 | 53.95 | 56.41 | 54.82 | 58.97 | 57.11 | 59.62 | **57.18** | 41.03 | 35.25 | 58.01 | 55.32 | 56.09 | 53.72 |
| 3-0 | Best | 39.23 | 35.95 | 41.92 | **40.13** | 39.10 | 38.14 | 39.74 | 38.65 | 34.29 | 28.91 | 42.31 | 39.93 | 41.67 | 39.55 |
| | Last | 29.10 | 27.31 | 38.85 | **35.69** | 33.97 | 32.37 | 33.01 | 30.07 | 25.00 | 19.30 | 33.97 | 32.63 | 33.65 | 32.63 |
| 3-1 | Best | 51.47 | 48.96 | 51.09 | **50.17** | 50.00 | 46.73 | 51.28 | 48.53 | 43.59 | 42.18 | 50.00 | 49.62 | 50.96 | 49.30 |
| | Last | 43.27 | 41.67 | 47.69 | **45.39** | 44.55 | 42.05 | 47.44 | 44.68 | 38.14 | 35.06 | 45.83 | 42.76 | 46.47 | 44.10 |
| 3-2 | Best | 55.13 | 54.06 | 57.56 | 55.40 | 55.13 | 53.85 | 58.01 | **56.47** | 51.92 | 47.11 | 57.05 | 54.90 | 56.09 | 54.75 |
| | Last | 52.18 | 50.65 | 52.82 | 51.26 | 51.92 | 50.26 | 55.13 | **54.23** | 41.03 | 36.79 | 54.49 | 52.50 | 53.53 | 48.91 |
| Best average | | 47.88 | 46.08 | 49.03 | 47.67 | 50.35 | 48.07 | 50.61 | **48.88** | 43.46 | 40.34 | 50.93 | 48.78 | 50.37 | 48.16 |
| Last average | | 42.90 | 41.32 | 45.65 | 43.82 | 47.27 | 43.93 | 47.84 | **45.45** | 34.72 | 28.95 | 47.65 | 44.65 | 46.66 | 43.70 |
| Best bias | | 0.00 | 0.00 | 1.16 | 1.59 | 2.47 | 1.99 | 2.74 | 2.79 | -4.42 | -5.75 | 3.06 | 2.70 | 2.50 | 2.08 |
| Last bias | | 0.00 | 0.00 | 2.75 | 2.49 | 4.37 | 2.61 | 4.94 | 4.13 | -8.18 | -12.37 | 4.75 | 3.33 | 3.76 | 2.38 |

\* Best average: the average of Best; Last average: the average of Last; Best bias: the Best bias between Basis and other methods; Last bias: the Last bias between Basis and other methods.

\* The maximum Mean value of each row is bolded.

## PHM2009 with the frequency domain input

| Task | Loc | Basis | | AdaBN | | MK-MMD | | JMMD | | CORAL | | DANN | | CDAN | |
|---|---|---|---|---|---|---|---|---|---|---|---|---|---|---|---|
| | | Max | Mean | Max | Mean | Max | Mean | Max | Mean | Max | Mean | Max | Mean | Max | Mean |
| 0-1 | Best | 54.87 | 53.96 | 57.12 | 55.13 | 63.78 | 61.41 | 63.14 | 61.09 | 57.69 | 54.10 | 63.78 | 61.86 | 67.31 | **63.59** |
| | Last | 51.09 | 49.71 | 51.41 | 50.77 | 62.18 | 58.66 | 60.58 | 58.27 | 56.73 | 45.32 | 61.22 | 57.18 | 66.03 | **61.99** |
| 0-2 | Best | 51.67 | 49.13 | 48.40 | 46.96 | 53.53 | 50.71 | 53.21 | 51.35 | 55.13 | 46.86 | 52.24 | 50.26 | 57.37 | **52.82** |
| | Last | 43.97 | 40.39 | 43.14 | 40.92 | 48.72 | 46.09 | 49.68 | 47.95 | 47.12 | 39.04 | 47.44 | 44.81 | 54.49 | **48.40** |
| 0-3 | Best | 45.19 | 43.04 | 44.49 | 41.98 | 48.40 | **46.79** | 48.72 | 47.63 | 44.55 | 43.14 | 46.15 | 45.06 | 50.32 | 46.35 |
| | Last | 32.76 | 28.40 | 38.14 | 36.32 | 43.59 | 40.26 | 44.23 | **42.37** | 41.03 | 38.98 | 42.63 | 40.00 | 47.76 | 40.58 |
| 1-0 | Best | 60.13 | 56.13 | 58.21 | 55.16 | 67.31 | 64.30 | 72.44 | **67.56** | 57.69 | 53.27 | 71.79 | 65.45 | 65.38 | 64.36 |
| | Last | 54.94 | 52.54 | 55.13 | 52.94 | 63.46 | 61.99 | 68.91 | **63.97** | 57.05 | 48.65 | 69.55 | 63.78 | 63.78 | 62.18 |
| 1-2 | Best | 66.99 | 64.27 | 65.00 | 63.54 | 67.31 | 64.49 | 68.91 | 65.00 | 68.59 | 61.47 | 69.55 | **66.92** | 66.99 | 66.48 |
| | Last | 60.00 | 59.58 | 63.08 | 60.28 | 63.78 | 61.28 | 62.82 | 60.64 | 66.03 | 53.85 | 68.27 | 64.74 | 66.67 | **64.81** |
| 1-3 | Best | 62.24 | 60.15 | 67.69 | **64.77** | 58.01 | 56.28 | 61.22 | 59.55 | 58.01 | 53.91 | 60.90 | 59.36 | 66.03 | 61.73 |
| | Last | 57.50 | 54.90 | 64.42 | **61.52** | 55.45 | 51.92 | 57.69 | 54.94 | 53.53 | 48.21 | 58.65 | 56.28 | 61.86 | 58.78 |
| 2-0 | Best | 53.08 | 49.88 | 55.00 | 53.79 | 59.94 | 56.54 | 60.90 | 58.08 | 52.88 | 46.86 | 62.50 | **60.51** | 62.18 | 59.87 |
| | Last | 44.29 | 41.54 | 49.87 | 45.59 | 56.09 | 54.43 | 58.01 | 54.36 | 37.50 | 33.84 | 60.58 | **58.91** | 59.62 | 56.28 |
| 2-1 | Best | 66.99 | 64.73 | 65.13 | 64.31 | 70.51 | 68.27 | 71.79 | 68.97 | 68.27 | 59.04 | 71.15 | **69.74** | 70.51 | 69.61 |
| | Last | 62.44 | 59.89 | 62.44 | 60.62 | 68.27 | 66.28 | 66.99 | 65.58 | 63.14 | 49.17 | 68.91 | **67.89** | 68.59 | 66.86 |
| 2-3 | Best | 75.90 | 74.31 | 74.74 | 71.56 | 78.85 | 76.35 | 81.41 | 77.37 | 72.76 | 70.06 | 82.05 | 77.18 | 80.45 | **78.59** |
| | Last | 71.67 | 70.13 | 70.58 | 69.04 | 76.60 | 73.72 | 80.13 | 74.49 | 68.91 | 59.10 | 77.24 | 73.21 | 78.85 | **75.38** |
| 3-0 | Best | 37.05 | 33.49 | 47.18 | 45.71 | 43.91 | 42.37 | 42.63 | 41.09 | 41.99 | 34.10 | 46.79 | 43.72 | 48.40 | **46.03** |
| | Last | 28.01 | 26.33 | 42.44 | **38.67** | 41.67 | 36.86 | 31.41 | 28.21 | 34.94 | 25.45 | 41.99 | 35.70 | 42.63 | 38.40 |
| 3-1 | Best | 52.56 | 49.35 | 57.31 | 56.30 | 64.42 | 62.63 | 64.74 | **62.88** | 59.94 | 46.60 | 63.46 | 61.60 | 63.78 | 61.99 |
| | Last | 39.74 | 37.10 | 52.95 | 51.23 | 60.26 | 58.91 | 62.18 | 59.62 | 49.36 | 36.73 | 63.14 | **59.68** | 60.90 | 58.59 |
| 3-2 | Best | 71.03 | 68.77 | 72.69 | 71.60 | 74.04 | 70.96 | 73.40 | **71.67** | 75.64 | 64.23 | 71.15 | 70.19 | 73.72 | 71.67 |
| | Last | 62.24 | 56.99 | 67.50 | 65.46 | 71.47 | 68.59 | 70.83 | 67.50 | 70.83 | 60.10 | 69.23 | 67.12 | 69.87 | **69.04** |
| Best average | | 58.14 | 55.60 | 59.41 | 57.57 | 62.50 | 60.09 | 63.54 | 61.02 | 59.43 | 52.80 | 63.46 | 60.99 | 64.37 | **61.92** |
| Last average | | 50.72 | 48.12 | 55.09 | 52.78 | 59.30 | 56.58 | 59.46 | 56.49 | 54.06 | 44.79 | 60.74 | 57.44 | 61.75 | **58.44** |
| Best bias | | 0.00 | 0.00 | 1.27 | 1.97 | 4.36 | 4.49 | 5.32 | 5.39 | 1.29 | -2.80 | 5.32 | 5.39 | 6.23 | 6.32 |
| Last bias | | 0.00 | 0.00 | 4.37 | 4.66 | 8.57 | 8.46 | 10.02 | 9.32 | 3.34 | -3.34 | 10.02 | 9.32 | 11.03 | 10.32 |

\* Best average: the average of Best; Last average: the average of Last; Best bias: the Best bias between Basis and other methods; Last bias: the Last bias between Basis and other methods.

\* The maximum Mean value of each row is bolded.

## SEU with the time domain input

| Task | Loc | Basis | | AdaBN | | MK-MMD | | JMMD | | CORAL | | DANN | | CDAN | |
|---|---|---|---|---|---|---|---|---|---|---|---|---|---|---|---|
| | | Max | Mean | Max | Mean | Max | Mean | Max | Mean | Max | Mean | Max | Mean | Max | Mean |
| 0-1 | Best | 48.91 | 48.33 | 52.93 | 51.59 | 57.04 | **53.81** | 57.04 | 53.14 | 45.45 | 38.80 | 54.84 | 51.32 | 57.77 | 53.40 |
| | Last | 42.70 | 39.18 | 49.82 | 47.69 | 54.40 | **48.65** | 53.81 | 45.98 | 35.78 | 27.57 | 44.72 | 41.61 | 55.28 | 46.48 |
| 1-0 | Best | 61.03 | 58.37 | 58.42 | 56.21 | 68.62 | **62.87** | 67.45 | 62.23 | 56.60 | 54.08 | 66.13 | 59.33 | 63.05 | 60.12 |
| | Last | 59.09 | 53.33 | 53.84 | 50.87 | 67.30 | **61.20** | 65.25 | 59.97 | 50.73 | 33.23 | 63.05 | 55.45 | 57.77 | 53.93 |
| Best average | | 54.97 | 53.35 | 55.68 | 53.90 | 62.83 | **58.34** | 62.25 | 57.68 | 51.03 | 46.44 | 60.49 | 55.32 | 60.41 | 56.76 |
| Last average | | 50.90 | 46.25 | 51.83 | 49.28 | 60.85 | **54.93** | 59.53 | 52.98 | 43.26 | 30.40 | 53.89 | 48.53 | 56.53 | 50.21 |
| Best bias | | 0.00 | 0.00 | 0.71 | 0.55 | 7.86 | 4.99 | 7.28 | 4.33 | -3.94 | -6.91 | 5.52 | 1.97 | 5.44 | 3.41 |
| Last bias | | 0.00 | 0.00 | 0.93 | 3.03 | 9.96 | 8.67 | 8.63 | 6.73 | -7.64 | -15.86 | 2.99 | 2.28 | 5.63 | 3.95 |

\* Best average: the average of Best; Last average: the average of Last; Best bias: the Best bias between Basis and other methods; Last bias: the Last bias between Basis and other methods.

\* The maximum Mean value of each row is bolded.

## SEU with the frequency domain input

| Task | Loc | Basis | | AdaBN | | MK-MMD | | JMMD | | CORAL | | DANN | | CDAN | |
|---|---|---|---|---|---|---|---|---|---|---|---|---|---|---|---|
| | | Max | Mean | Max | Mean | Max | Mean | Max | Mean | Max | Mean | Max | Mean | Max | Mean |
| 0-1 | Best | 37.36 | 34.61 | 41.99 | 36.35 | 42.96 | 40.38 | 46.92 | **44.90** | 53.37 | 38.39 | 47.07 | 43.02 | 49.41 | 41.58 |
| | Last | 27.74 | 22.89 | 24.99 | 20.68 | 39.00 | 33.14 | 43.11 | **36.28** | 30.94 | 25.13 | 44.57 | 34.63 | 41.64 | 32.34 |
| 1-0 | Best | 46.72 | 38.92 | 50.97 | 43.04 | 63.34 | 57.13 | 56.30 | 51.64 | 45.89 | 37.77 | 63.05 | **60.88** | 61.58 | 57.18 |
| | Last | 37.86 | 33.48 | 42.05 | 35.80 | 56.60 | 51.44 | 50.59 | 46.13 | 41.06 | 29.18 | 59.24 | **56.60** | 60.26 | 53.37 |
| Best average | | 42.04 | 36.77 | 46.48 | 39.70 | 53.15 | 48.75 | 51.61 | 48.27 | 49.63 | 38.08 | 55.06 | **51.95** | 55.50 | 49.38 |
| Last average | | 32.80 | 28.19 | 33.52 | 28.24 | 47.80 | 42.29 | 46.85 | 41.20 | 36.00 | 27.16 | 51.91 | **45.62** | 50.95 | 42.86 |
| Best bias | | 0.00 | 0.00 | 4.44 | 2.93 | 11.11 | 11.99 | 9.57 | 11.50 | 7.59 | 1.31 | 13.02 | 15.18 | 13.46 | 12.62 |
| Last bias | | 0.00 | 0.00 | 0.72 | 0.05 | 15.00 | 14.10 | 14.05 | 13.02 | 3.20 | -1.03 | 19.11 | 17.43 | 18.15 | 14.67 |

\* Best average: the average of Best; Last average: the average of Last; Best bias: the Best bias between Basis and other methods; Last bias: the Last bias between Basis and other methods.

\* The maximum Mean value of each row is bolded.

## CWRU with the time and frequency domain inputs for PADA

| Task | Loc | Time domain input | | Frequency domain input | |
|---|---|---|---|---|---|
| | | Max | Mean | Max | Mean |
| 0-1 | Best | 100.00 | **98.31** | 97.32 | 96.63 |
| | Last | 100.00 | **96.63** | 95.40 | 93.49 |
| 0-2 | Best | 100.00 | **97.63** | 97.70 | 97.01 |
| | Last | 100.00 | **96.32** | 96.17 | 95.25 |
| 0-3 | Best | 94.61 | **87.43** | 90.42 | 85.99 |
| | Last | 85.63 | **80.12** | 83.83 | 79.40 |
| 1-0 | Best | 100.00 | **100.00** | 99.40 | 98.80 |
| | Last | 100.00 | **100.00** | 98.80 | 95.42 |
| 1-2 | Best | 100.00 | **100.00** | 100.00 | **100.00** |
| | Last | 100.00 | **97.47** | 100.00 | 94.95 |
| 1-3 | Best | 100.00 | **100.00** | 100.00 | 99.90 |
| | Last | 100.00 | 84.92 | 100.00 | **92.25** |
| 2-0 | Best | 100.00 | **100.00** | 100.00 | **100.00** |
| | Last | 100.00 | 94.95 | 100.00 | **99.79** |
| 2-1 | Best | 100.00 | **100.00** | 100.00 | **100.00** |
| | Last | 100.00 | **100.00** | 100.00 | **100.00** |
| 2-3 | Best | 98.88 | 88.88 | 100.00 | **100.00** |
| | Last | 99.30 | 79.72 | 100.00 | **85.87** |
| 3-0 | Best | 100.00 | **100.00** | 100.00 | **100.00** |
| | Last | 100.00 | 89.90 | 100.00 | **94.53** |
| 3-1 | Best | 100.00 | **89.16** | 85.42 | 72.50 |
| | Last | 91.67 | 35.83 | 47.92 | **45.83** |
| 3-2 | Best | 100.00 | **99.17** | 100.00 | **99.17** |
| | Last | 100.00 | 73.75 | 85.42 | **78.75** |
| Best average | | 99.55 | **97.55** | 97.52 | 95.83 |
| Last average | | 98.05 | 85.80 | 92.30 | **87.96** |

\* Best average: the average of Best; Last average: the average of Last; Best bias: the Best bias between Basis and other methods; Last bias: the Last bias between Basis and other methods.

\* The maximum Mean value of each row is bolded.

## CWRU with the time domain input for OSBP

| Task | Loc | OSBP | | | | | | | | | |
|---|---|---|---|---|---|---|---|---|---|---|---|
| | | ALL* | | UNK | | OS | | ALL | | H-score | |
| | | Max | Mean | Max | Mean | Max | Mean | Max | Mean | Max | Mean |
| 0-1 | Best | 97.32 | 92.03 | 95.74 | 97.87 | 96.25 | 90.13 | 97.08 | 92.92 | 96.52 | 94.82 |
| | Last | 68.20 | 80.53 | 40.43 | 17.02 | 55.24 | 67.31 | 63.96 | 70.84 | 50.76 | 23.74 |
| 0-2 | Best | 97.32 | 91.27 | 95.74 | 97.02 | 96.27 | 89.11 | 97.08 | 92.14 | 96.52 | 93.91 |
| | Last | 85.82 | 84.83 | 42.55 | 23.40 | 76.49 | 73.25 | 79.22 | 75.55 | 56.90 | 34.48 |
| 0-3 | Best | 89.92 | 91.77 | 81.69 | 73.80 | 85.40 | 86.47 | 88.03 | 87.64 | 85.61 | 81.53 |
| | Last | 75.63 | 86.89 | 55.21 | 19.72 | 61.44 | 74.17 | 66.34 | 71.46 | 48.05 | 30.24 |
| 1-0 | Best | 90.00 | 82.42 | 83.10 | 73.52 | 87.99 | 79.25 | 88.12 | 80.00 | 86.41 | 77.10 |
| | Last | 48.42 | 70.84 | 80.28 | 43.38 | 46.43 | 63.92 | 57.09 | 63.37 | 60.41 | 33.66 |
| 1-2 | Best | 95.77 | 91.13 | 83.13 | 78.32 | 93.99 | 89.31 | 88.96 | 84.22 | 89.01 | 83.51 |
| | Last | 97.18 | 87.18 | 40.96 | 21.81 | 89.13 | 77.88 | 66.88 | 51.95 | 57.63 | 32.84 |
| 1-3 | Best | 91.61 | 84.62 | 76.51 | 70.00 | 89.37 | 82.48 | 83.50 | 76.76 | 83.38 | 75.95 |
| | Last | 75.52 | 77.76 | 65.66 | 32.17 | 74.03 | 71.22 | 70.23 | 53.27 | 70.25 | 41.82 |
| 2-0 | Best | 84.94 | 79.88 | 90.53 | 82.11 | 83.77 | 77.56 | 86.97 | 80.69 | 87.64 | 80.08 |
| | Last | 90.36 | 75.18 | 37.89 | 43.79 | 81.58 | 67.20 | 71.26 | 63.75 | 53.40 | 44.84 |
| 2-1 | Best | 79.34 | 80.56 | 92.63 | 77.89 | 72.76 | 71.99 | 83.44 | 79.74 | 85.47 | 77.43 |
| | Last | 55.87 | 86.67 | 75.79 | 17.05 | 39.99 | 71.08 | 62.01 | 65.19 | 64.32 | 16.48 |
| 2-3 | Best | 96.84 | 90.63 | 88.24 | 86.05 | 93.81 | 85.58 | 93.53 | 88.87 | 92.34 | 88.05 |
| | Last | 98.95 | 84.21 | 39.50 | 9.41 | 88.53 | 64.04 | 76.05 | 55.40 | 56.46 | 14.16 |
| 3-0 | Best | 87.32 | 83.24 | 87.39 | 79.83 | 85.07 | 79.74 | 87.36 | 81.69 | 87.36 | 80.96 |
| | Last | 64.08 | 76.20 | 49.58 | 26.39 | 55.52 | 63.67 | 57.47 | 53.49 | 55.91 | 35.76 |
| 3-1 | Best | 67.61 | 48.45 | 67.93 | 57.81 | 66.98 | 50.49 | 67.86 | 55.65 | 67.77 | 50.69 |
| | Last | 29.58 | 50.70 | 27.00 | 14.85 | 28.94 | 41.64 | 27.60 | 23.12 | 28.23 | 20.29 |
| 3-2 | Best | 64.79 | 67.61 | 78.90 | 59.91 | 68.19 | 65.80 | 75.65 | 61.69 | 71.15 | 59.58 |
| | Last | 87.32 | 83.10 | 24.05 | 9.11 | 71.50 | 64.44 | 38.64 | 26.17 | 37.71 | 15.55 |
| Best average | | 86.90 | 81.97 | 85.13 | 77.84 | 84.99 | 78.99 | 86.47 | 80.17 | 85.77 | 78.63 |
| Last average | | 73.08 | 78.67 | 46.58 | 23.18 | 64.07 | 66.65 | 61.40 | 56.12 | 53.34 | 28.66 |

* Best average: the average of Best; Last average: the average of Last; Best bias: the Best bias between Basis and other methods; Last bias: the Last bias between Basis and other methods.

## CWRU with the frequency domain input for OSBP

| Task | Loc | OSBP | | | | | | | | | |
|---|---|---|---|---|---|---|---|---|---|---|---|
| | | ALL* | | UNK | | OS | | ALL | | H-score | |
| | | Max | Mean | Max | Mean | Max | Mean | Max | Mean | Max | Mean |
| 0-1 | Best | 80.84 | 84.29 | 91.49 | 78.72 | 75.91 | 78.81 | 82.47 | 83.44 | 85.84 | 80.84 |
| | Last | 89.66 | 87.05 | 55.32 | 42.13 | 82.86 | 77.99 | 84.42 | 80.20 | 68.42 | 56.03 |
| 0-2 | Best | 86.97 | 85.21 | 97.87 | 80.85 | 84.02 | 79.87 | 88.64 | 84.54 | 92.10 | 81.88 |
| | Last | 82.76 | 83.91 | 55.32 | 25.11 | 79.03 | 73.42 | 78.57 | 74.93 | 66.31 | 34.95 |
| 0-3 | Best | 89.50 | 84.03 | 64.79 | 57.46 | 83.06 | 74.99 | 83.82 | 77.93 | 75.16 | 68.23 |
| | Last | 86.97 | 85.63 | 22.54 | 12.12 | 73.92 | 71.33 | 72.17 | 68.74 | 35.80 | 20.65 |
| 1-0 | Best | 78.95 | 79.79 | 83.10 | 71.55 | 76.71 | 76.31 | 80.08 | 77.55 | 80.97 | 74.59 |
| | Last | 41.58 | 79.26 | 92.96 | 23.94 | 40.97 | 69.84 | 55.56 | 64.21 | 57.46 | 21.34 |
| 1-2 | Best | 91.55 | 82.39 | 75.30 | 72.77 | 89.30 | 81.08 | 82.79 | 77.21 | 82.63 | 76.46 |
| | Last | 76.76 | 90.28 | 37.95 | 26.02 | 71.29 | 81.10 | 55.84 | 55.65 | 50.79 | 39.61 |
| 1-3 | Best | 77.62 | 74.54 | 74.10 | 69.04 | 76.99 | 73.74 | 75.73 | 71.58 | 75.82 | 71.47 |
| | Last | 59.44 | 74.82 | 40.96 | 21.69 | 56.65 | 67.08 | 49.51 | 46.28 | 48.50 | 31.74 |
| 2-0 | Best | 83.13 | 76.87 | 67.37 | 68.42 | 78.67 | 72.63 | 77.39 | 73.79 | 74.42 | 71.22 |
| | Last | 89.16 | 81.57 | 31.58 | 18.11 | 79.51 | 70.08 | 68.20 | 58.47 | 46.64 | 28.34 |
| 2-1 | Best | 76.06 | 71.45 | 84.21 | 78.32 | 66.79 | 60.21 | 78.57 | 73.57 | 79.93 | 72.64 |
| | Last | 51.17 | 62.25 | 100.00 | 41.26 | 37.50 | 47.43 | 66.23 | 55.78 | 67.70 | 31.96 |
| 2-3 | Best | 93.68 | 85.47 | 90.76 | 82.69 | 90.01 | 77.76 | 92.56 | 84.40 | 92.20 | 83.38 |
| | Last | 77.37 | 90.74 | 38.66 | 26.22 | 59.22 | 75.28 | 62.46 | 65.89 | 51.55 | 39.83 |
| 3-0 | Best | 92.25 | 74.65 | 78.99 | 78.15 | 88.86 | 71.09 | 86.21 | 76.25 | 85.11 | 76.08 |
| | Last | 80.99 | 68.73 | 21.01 | 16.98 | 68.02 | 54.91 | 53.44 | 45.13 | 33.36 | 26.98 |
| 3-1 | Best | 61.97 | 55.21 | 71.31 | 60.68 | 63.66 | 56.01 | 69.16 | 59.42 | 66.31 | 55.10 |
| | Last | 63.38 | 49.01 | 36.71 | 14.09 | 56.14 | 39.84 | 42.86 | 22.14 | 46.49 | 18.86 |
| 3-2 | Best | 73.24 | 62.82 | 67.93 | 62.19 | 71.83 | 62.64 | 69.16 | 62.34 | 70.49 | 61.81 |
| | Last | 74.65 | 61.12 | 35.02 | 7.51 | 64.55 | 47.66 | 44.16 | 18.87 | 47.68 | 10.52 |
| Best average | | 82.15 | 76.39 | 78.94 | 71.74 | 78.82 | 72.10 | 80.55 | 75.17 | 81.03 | 72.81 |
| Last average | | 72.82 | 76.20 | 47.34 | 22.93 | 64.14 | 64.66 | 61.14 | 54.77 | 51.73 | 30.07 |

* Best average: the average of Best; Last average: the average of Last; Best bias: the Best bias between Basis and other methods; Last bias: the Last bias between Basis and other methods.

## CWRU with the time domain input for UAN

| Task | Loc | UAN | | | | | | | | | |
|---|---|---|---|---|---|---|---|---|---|---|---|
| | | ALL* | | UNK | | OS | | ALL | | H-score | |
| | | Max | Mean | Max | Mean | Max | Mean | Max | Mean | Max | Mean |
| 0-1 | Best | 98.19 | 95.38 | 100.00 | 100.00 | 98.44 | 95.97 | 98.42 | 95.98 | 99.09 | 97.62 |
| | Last | 92.09 | 85.04 | 95.65 | 89.67 | 92.53 | 85.65 | 92.59 | 85.68 | 93.84 | 87.19 |
| 0-2 | Best | 95.21 | 92.70 | 100.00 | 97.39 | 95.83 | 93.31 | 95.79 | 93.26 | 97.55 | 94.83 |
| | Last | 76.43 | 75.43 | 100.00 | 80.09 | 80.13 | 77.97 | 79.63 | 77.29 | 86.64 | 81.52 |
| 0-3 | Best | 96.62 | 89.28 | 95.83 | 98.33 | 96.85 | 87.66 | 96.49 | 90.81 | 96.23 | 93.51 |
| | Last | 90.30 | 79.91 | 95.83 | 92.08 | 88.55 | 78.81 | 91.23 | 81.96 | 92.98 | 85.51 |
| 1-0 | Best | 98.43 | 96.23 | 97.87 | 99.15 | 98.43 | 96.53 | 98.32 | 96.81 | 98.15 | 97.65 |
| | Last | 95.00 | 88.12 | 100.00 | 88.89 | 94.80 | 87.29 | 96.10 | 88.29 | 97.44 | 88.41 |
| 1-2 | Best | 97.92 | 98.34 | 100.00 | 98.30 | 98.61 | 98.32 | 98.95 | 98.31 | 98.95 | 98.31 |
| | Last | 93.18 | 86.82 | 93.48 | 83.91 | 93.66 | 85.97 | 93.33 | 85.33 | 93.33 | 84.86 |
| 1-3 | Best | 97.92 | 93.33 | 100.00 | 95.00 | 98.61 | 93.89 | 98.96 | 94.17 | 98.95 | 94.15 |
| | Last | 95.65 | 71.30 | 97.87 | 89.36 | 96.39 | 76.78 | 96.77 | 80.43 | 96.75 | 78.39 |
| 2-0 | Best | 97.89 | 97.26 | 97.48 | 97.65 | 97.83 | 97.35 | 97.66 | 97.48 | 97.69 | 97.45 |
| | Last | 90.53 | 84.63 | 95.80 | 85.38 | 91.66 | 84.76 | 93.46 | 85.05 | 93.09 | 84.98 |
| 2-1 | Best | 100.00 | 98.31 | 97.59 | 95.30 | 99.52 | 97.70 | 98.47 | 96.40 | 98.78 | 96.78 |
| | Last | 94.74 | 85.68 | 95.78 | 87.83 | 94.88 | 86.07 | 95.40 | 87.05 | 95.26 | 86.61 |
| 2-3 | Best | 100.00 | 98.67 | 97.87 | 97.02 | 99.29 | 98.13 | 98.91 | 97.83 | 98.92 | 97.81 |
| | Last | 91.11 | 87.56 | 95.74 | 87.66 | 92.63 | 87.55 | 93.48 | 87.61 | 93.37 | 87.52 |
| 3-0 | Best | 100.00 | 100.00 | 97.83 | 97.83 | 99.28 | 99.28 | 98.89 | 98.89 | 98.90 | 98.90 |
| | Last | 100.00 | 83.64 | 95.65 | 83.04 | 98.55 | 83.29 | 97.78 | 83.33 | 97.78 | 83.32 |
| 3-1 | Best | 98.32 | 98.32 | 100.00 | 100.00 | 99.30 | 99.30 | 98.59 | 98.59 | 99.15 | 99.15 |
| | Last | 97.00 | 93.00 | 95.45 | 90.91 | 95.08 | 92.55 | 96.72 | 92.62 | 96.22 | 91.93 |
| 3-2 | Best | 98.32 | 98.32 | 100.00 | 100.00 | 97.22 | 98.47 | 98.59 | 98.59 | 99.15 | 99.15 |
| | Last | 94.17 | 92.62 | 95.65 | 93.91 | 93.75 | 92.75 | 94.44 | 92.86 | 94.91 | 93.26 |
| Best average | | 98.24 | 96.35 | 98.71 | 98.00 | 98.27 | 96.33 | 98.17 | 96.43 | 98.46 | 97.11 |
| Last average | | 92.52 | 84.48 | 96.41 | 88.47 | 92.79 | 84.95 | 93.41 | 85.63 | 94.30 | 86.12 |

* Best average: the average of Best; Last average: the average of Last; Best bias: the Best bias between Basis and other methods; Last bias: the Last bias between Basis and other methods.

## CWRU with the frequency domain input for UAN

| Task | Loc | UAN | | | | | | | | | |
|---|---|---|---|---|---|---|---|---|---|---|---|
| | | ALL* | | UNK | | OS | | ALL | | H-score | |
| | | Max | Mean | Max | Mean | Max | Mean | Max | Mean | Max | Mean |
| 0-1 | Best | 93.98 | 90.84 | 100.00 | 100.00 | 94.70 | 91.93 | 94.74 | 92.00 | 96.89 | 95.19 |
| | Last | 85.61 | 83.45 | 95.65 | 89.56 | 87.04 | 84.55 | 87.04 | 84.32 | 90.35 | 86.38 |
| 0-2 | Best | 90.42 | 87.90 | 100.00 | 100.00 | 91.64 | 89.46 | 91.58 | 89.37 | 94.97 | 93.55 |
| | Last | 82.14 | 80.00 | 95.45 | 90.00 | 84.41 | 81.45 | 83.95 | 81.36 | 88.30 | 84.58 |
| 0-3 | Best | 86.92 | 85.15 | 100.00 | 98.75 | 85.53 | 82.78 | 89.12 | 87.44 | 93.00 | 91.44 |
| | Last | 77.64 | 74.68 | 89.58 | 86.25 | 73.61 | 70.45 | 79.65 | 76.63 | 83.18 | 79.98 |
| 1-0 | Best | 92.15 | 90.25 | 100.00 | 98.70 | 92.66 | 90.30 | 93.70 | 91.94 | 95.91 | 94.27 |
| | Last | 86.25 | 80.75 | 97.78 | 86.67 | 85.67 | 79.91 | 88.78 | 82.05 | 91.65 | 83.49 |
| 1-2 | Best | 97.92 | 95.68 | 95.74 | 96.57 | 97.19 | 95.91 | 96.84 | 96.14 | 96.92 | 96.09 |
| | Last | 90.91 | 88.18 | 91.30 | 87.82 | 90.71 | 87.83 | 91.11 | 88.00 | 91.11 | 87.96 |
| 1-3 | Best | 95.83 | 95.81 | 100.00 | 97.07 | 97.22 | 96.22 | 97.92 | 96.45 | 97.87 | 96.43 |
| | Last | 89.13 | 86.96 | 95.74 | 90.64 | 91.38 | 88.17 | 92.47 | 88.82 | 92.32 | 88.66 |
| 2-0 | Best | 97.89 | 93.68 | 94.96 | 96.47 | 97.32 | 94.25 | 96.26 | 95.23 | 96.40 | 95.03 |
| | Last | 90.53 | 85.90 | 89.08 | 86.22 | 90.24 | 85.93 | 89.72 | 86.07 | 89.80 | 86.01 |
| 2-1 | Best | 96.84 | 93.68 | 98.80 | 94.82 | 97.22 | 94.02 | 98.08 | 94.48 | 97.81 | 94.35 |
| | Last | 91.58 | 81.68 | 91.57 | 84.94 | 91.54 | 82.29 | 91.57 | 83.76 | 91.57 | 83.26 |
| 2-3 | Best | 97.78 | 95.11 | 95.74 | 97.02 | 96.99 | 95.71 | 96.74 | 96.09 | 96.75 | 96.01 |
| | Last | 86.67 | 80.89 | 93.62 | 91.06 | 88.75 | 83.97 | 90.22 | 86.09 | 90.01 | 85.51 |
| 3-0 | Best | 100.00 | 98.64 | 97.83 | 98.26 | 99.28 | 98.48 | 98.89 | 98.45 | 98.90 | 98.43 |
| | Last | 100.00 | 91.36 | 93.48 | 90.00 | 97.83 | 91.00 | 96.67 | 90.67 | 96.63 | 90.64 |
| 3-1 | Best | 96.00 | 92.00 | 100.00 | 100.00 | 95.83 | 90.42 | 96.72 | 93.44 | 95.76 | 95.81 |
| | Last | 94.00 | 86.40 | 100.00 | 93.64 | 93.75 | 82.96 | 95.08 | 87.71 | 96.91 | 89.84 |
| 3-2 | Best | 98.32 | 97.98 | 100.00 | 100.00 | 99.30 | 98.33 | 98.59 | 98.31 | 99.15 | 98.98 |
| | Last | 95.15 | 83.59 | 95.65 | 93.04 | 95.34 | 93.58 | 95.24 | 93.49 | 95.40 | 93.28 |
| Best average | | 95.34 | 93.08 | 98.59 | 98.14 | 94.11 | 95.41 | 95.71 | 94.11 | 96.87 | 95.46 |
| Last average | | 93.08 | 84.49 | 94.08 | 89.15 | 89.19 | 84.34 | 90.13 | 85.75 | 91.44 | 86.63 |

* Best average: the average of Best; Last average: the average of Last; Best bias: the Best bias between Basis and other methods; Last bias: the Last bias between Basis and other methods.

## CWRU with the time and frequency domain inputs for multi-domain transfer

| Task | Loc | Time Basis Max | Time Basis Mean | Time IAN Max | Time IAN Mean | Time MS-UADA Max | Time MS-UADA Mean | Freq Basis Max | Freq Basis Mean | Freq IAN Max | Freq IAN Mean | Freq MS-UADA Max | Freq MS-UADA Mean |
|---|---|---|---|---|---|---|---|---|---|---|---|---|---|
| 012-3 | Best | 100.00 | **100.00** | 100.00 | **100.00** | 100.00 | **100.00** | 98.71 | 97.35 | 98.06 | 96.51 | 100.00 | **100.00** |
| | Last | 99.03 | 97.35 | 100.00 | 98.12 | 100.00 | **99.55** | 95.47 | 93.14 | 87.38 | 84.79 | 100.00 | **99.55** |
| 013-2 | Best | 100.00 | **100.00** | 100.00 | **100.00** | 100.00 | **100.00** | 100.00 | **100.00** | 100.00 | 99.81 | 100.00 | **100.00** |
| | Last | 100.00 | **100.00** | 100.00 | **100.00** | 100.00 | **100.00** | 99.68 | 98.31 | 100.00 | 97.27 | 100.00 | **100.00** |
| 023-1 | Best | 100.00 | **100.00** | 100.00 | **100.00** | 100.00 | **100.00** | 100.00 | **100.00** | 100.00 | **100.00** | 100.00 | **100.00** |
| | Last | 100.00 | 99.94 | 100.00 | 99.94 | 100.00 | **100.00** | 99.68 | 99.35 | 100.00 | 99.22 | 100.00 | **100.00** |
| 123-0 | Best | 100.00 | **100.00** | 100.00 | **100.00** | 100.00 | **100.00** | 99.23 | 98.62 | 99.23 | 98.77 | 100.00 | 98.47 |
| | Last | 100.00 | **100.00** | 100.00 | 99.85 | 100.00 | 99.92 | 99.47 | 97.55 | 98.47 | 97.16 | 98.85 | 98.47 |
| Best average | | 100.00 | **100.00** | 100.00 | **100.00** | 100.00 | **100.00** | 99.49 | 98.99 | 99.32 | 98.77 | 100.00 | 99.83 |
| Last average | | 99.76 | 99.32 | 100.00 | 99.48 | 100.00 | **99.98** | 98.33 | 97.09 | 96.46 | 94.61 | 99.71 | 99.50 |
| Best bias | | 0.00 | 0.00 | 0.00 | 0.00 | 0.00 | 0.00 | 0.00 | 0.00 | -0.16 | -0.22 | 0.52 | 0.84 |
| Last bias | | 0.00 | 0.00 | 0.24 | 0.15 | 0.24 | 0.66 | 0.00 | 0.00 | -1.86 | -2.48 | 1.39 | 2.42 |

\* Best average: the average of Best; Last average: the average of Last; Best bias: the Best bias between Basis and other methods; Last bias: the Last bias between Basis and other methods.

\* The maximum Mean value of each row is bolded.

## PU-Types with the time domain input

| Task | Loc | Basis Max | Basis Mean | AdaBN Max | AdaBN Mean | MK-MMD Max | MK-MMD Mean | JMMD Max | JMMD Mean | CORAL Max | CORAL Mean | DANN Max | DANN Mean | CDAN Max | CDAN Mean |
|---|---|---|---|---|---|---|---|---|---|---|---|---|---|---|---|
| 0-1 | Best | 60.34 | 57.12 | 46.65 | 41.76 | 55.22 | 46.86 | 56.72 | 53.13 | 64.68 | **58.41** | 53.73 | 50.75 | 56.22 | 50.65 |
| | Last | 55.54 | **48.01** | 39.06 | 38.10 | 41.79 | 33.73 | 32.84 | 28.66 | 49.75 | 34.63 | 35.82 | 33.43 | 36.32 | 34.53 |
| 1-0 | Best | 50.40 | 46.32 | 74.27 | **69.95** | 49.33 | 44.53 | 58.67 | 54.67 | 54.00 | 42.13 | 57.33 | 54.40 | 58.67 | 56.27 |
| | Last | 34.13 | 33.79 | 68.40 | **61.71** | 36.67 | 32.13 | 47.33 | 37.73 | 33.33 | 33.33 | 40.67 | 35.20 | 38.67 | 35.47 |
| Best average | | 55.37 | 51.72 | 60.46 | **55.85** | 52.28 | 45.70 | 57.70 | 53.90 | 59.34 | 50.27 | 55.53 | 52.57 | 57.45 | 53.46 |
| Last average | | 44.84 | 40.90 | 53.73 | **49.90** | 39.23 | 32.93 | 40.09 | 33.20 | 41.54 | 33.98 | 38.25 | 34.32 | 37.50 | 35.00 |
| Best bias | | 0.00 | 0.00 | 5.09 | 4.13 | -3.10 | -6.02 | 2.33 | 2.18 | 3.97 | -1.45 | 0.16 | 0.85 | 2.07 | 1.74 |
| Last bias | | 0.00 | 0.00 | 8.90 | 9.00 | -5.61 | -7.97 | -4.75 | -7.70 | -3.30 | -6.92 | -6.59 | -6.58 | -7.34 | -5.90 |

\* Best average: the average of Best; Last average: the average of Last; Best bias: the Best bias between Basis and other methods; Last bias: the Last bias between Basis and other methods.

\* The maximum Mean value of each row is bolded.

## PU-Types with the frequency domain input

| Task | Loc | Basis Max | Basis Mean | AdaBN Max | AdaBN Mean | MK-MMD Max | MK-MMD Mean | JMMD Max | JMMD Mean | CORAL Max | CORAL Mean | DANN Max | DANN Mean | CDAN Max | CDAN Mean |
|---|---|---|---|---|---|---|---|---|---|---|---|---|---|---|---|
| 0-1 | Best | 62.54 | 44.78 | 49.95 | 41.56 | 46.77 | 46.57 | 45.77 | 41.29 | 67.66 | **59.50** | 54.73 | 49.35 | 44.78 | 42.79 |
| | Last | 36.86 | 33.63 | 38.26 | 36.80 | 45.27 | 42.19 | 33.83 | 33.04 | 39.30 | 35.02 | 45.77 | **45.47** | 44.78 | 37.51 |
| 1-0 | Best | 71.60 | 66.56 | 70.53 | 66.53 | 63.33 | 61.33 | 68.67 | 64.93 | 65.33 | 57.60 | 69.33 | 66.40 | 74.00 | **68.80** |
| | Last | 62.27 | 59.79 | 66.80 | 61.41 | 59.33 | 56.13 | 62.67 | 58.00 | 38.00 | 34.53 | 65.33 | 63.46 | 73.33 | **64.27** |
| Best average | | 67.07 | 55.67 | 60.24 | 54.05 | 55.05 | 53.95 | 57.22 | 53.11 | 66.50 | **58.55** | 62.03 | 57.88 | 59.39 | 55.79 |
| Last average | | 49.57 | 46.71 | 52.53 | 49.11 | 52.30 | 49.16 | 48.25 | 45.52 | 38.65 | 34.78 | 55.55 | **54.47** | 59.06 | 50.89 |
| Best bias | | 0.00 | 0.00 | -6.83 | -1.62 | -12.02 | -1.72 | -9.85 | -2.55 | -0.58 | 2.88 | -5.04 | 2.21 | -7.68 | 0.12 |
| Last bias | | 0.00 | 0.00 | 2.96 | 2.40 | 2.74 | 2.45 | -1.32 | -1.19 | -10.92 | -11.93 | 5.99 | 7.76 | 9.49 | 4.18 |

\* Best average: the average of Best; Last average: the average of Last; Best bias: the Best bias between Basis and other methods; Last bias: the Last bias between Basis and other methods.

\* The maximum Mean value of each row is bolded.



\end{document}